%% file: main.tex
\documentclass[letterpaper,twocolumn,10pt]{article}
\usepackage{usenix-2020-09}

\usepackage[normalem]{ulem}

\usepackage{epsfig,endnotes}
\usepackage{subfig}
\usepackage{comment}
\usepackage{amsmath}
\usepackage{graphicx}
\usepackage{array}
\usepackage{multirow}
\usepackage{color}
\usepackage{booktabs}
\usepackage{algorithm}
\usepackage{caption}
\usepackage{comment}
\usepackage{filecontents}
\usepackage{lipsum}
\usepackage{url}

\usepackage{tabularx}

\usepackage[noend]{algpseudocode}
\algrenewcommand\algorithmicindent{1.0em}

\newcommand{\SYSNAME}{BiDiFilter}

\begin{document}	

\date{}

\title{\Large \bf Multilevel Bidirectional Cache Filter}


\author{
	Ohad Eytan \ \ \ \ \ \ \ \ Roy Friedman\\
	Computer Science Department\\
	Technion\\
	\texttt{\{ohadey,roy\}@cs.technion.ac.il}
}

\maketitle

\begin{abstract}
	Modern caches are often required to handle a massive amount of data, which exceeds the amount of available memory; thus, hybrid caches, specifically DRAM/SSD combination, become more and more prevalent. In such environments, in addition to the classical hit-ratio target, saving writes to the second-level cache is a dominant factor to avoid write amplification and wear out, two notorious phenomena of SSD. 
	
	This paper presents BiDiFilter, a novel multilevel caching scheme that controls demotions and promotions between cache levels using a frequency sketch filter. 
	Further, it splits the higher cache level into two areas to keep the most recent and the most frequent items close to the user. 
	
	We conduct an extensive evaluation over real-world traces, comparing to previous multilevel policies. 
	We show that using our mechanism yields an $x10$ saving of writes in almost all cases and often improving latencies by up to $20\%$.
	
\end{abstract}

\input{introduction}

\input{related}

\input{algorithms}

\input{evaluation}

\input{discussion}

\clearpage
\bibliographystyle{plain}
\bibliography{refs}

\end{document}

%% file: introduction.tex
\section{Introduction}

Distributed caching systems, such as Redis~\cite{redis} and Memcached~\cite{MemCacheD}, improve the access latency perceived by users by storing parts of their applications data closer to the application or end user.
In principle, the more storage area allocated for caching, the easier it becomes to ensure that most requested data can be served from cache.

Yet, the cost vs. performance tradeoff of common storage technologies, as illustrated in Table~\ref{tab:tradeoff}, poses the following challenges to such caching systems' designers.
DRAM memory is the fastest, and therefore provides users with the best response time.
However, it is also the most expensive, meaning that it is also relatively limited in size.
SSD or flash technology is an order of magnitude slower, but also two orders of magnitude cheaper.
Further, SSDs can only accommodate a limited amount of writes in each of their memory cells, and also suffer from the write amplification problem, which means that writes becomes slower over time and cause expedite the degradation of the medium.
Finally, HDDs are yet another two orders of magnitude slower (at least for random access) and another order of magnitude cheaper.
For this reason, many caching solutions are multi-level or hierarchical.

\begin{table}[h]
	\begin{center}
		\begin{tabular}{l|c|c}
			\textbf{Storage} & \textbf{Latency} & \textbf{Price} \\
			& & $\$/GB$ \\
			\hline
			DRAM & \textasciitilde60--100ns & \textasciitilde5--10 \\
			\hline
			SSD & \textasciitilde10-200us & \textasciitilde0.1--1 \\
			\hline
			HDD & \textasciitilde2-3ms & \textasciitilde0.02 \\
		\end{tabular}
	\end{center}
	\normalsize
	\vspace{0.0cm}
	\caption{Price vs. performance trade-off. Based on \cite{NHC, Latency, CSCILecture}}
	\label{tab:tradeoff}
\end{table}

Given the significant access time differences between these technologies, often the higher level cache resides in DRAM, the next level in SSD, and if needed, a lower level is placed in HDDs.
Sometimes, the combination of DRAM and SSDs serve as a cache for the local HDD.
It is common wisdom that the items whom are most likely to be accessed should be placed in the first level of the cache (DRAM), the next set of items in terms of access likelihood in the second level of the cache, etc.
This minimizes the overall expected data access time.
Due to the price performance trade-off illustrated in Table~\ref{tab:tradeoff}, SSD is indeed the preferred second level cache in many deployed settings~\cite{netflixcache,netflixcache2,CacheLib,Flashield,LRB}.

Alas, SSDs suffer from two important limitations when serving as an underlying cache technology.
First, each page in an SSD device can endure a limited number of writes~\cite{BD10,HKAA17,F2FS}.
In other words, frequent writes to SSDs shorten their lifetime.
Second, SSDs are prone to the write amplification problem~\cite{HKAA17}.
This is due to the fact that the minimal erasable unit of an SSD is a complete page~\cite{Pannier,Flashield,RIPQ}, typically at least 4KB.
For these reasons, keeping the ``hot'' data in DRAM reduces the number of writes to SSD, thereby improving the SSD's lifetime and performance.

The choice of which item should be placed in each cache is governed by a cache management policy.
When considering each cache level in isolation, it is known that the best management policy in terms of maximizing the hit-ratio depends on that cache's workload.
Interestingly, most systems manage each of the cache levels in isolation.
We argue that by adding a frequency based filtering mechanism, it is possible to improve the expected access time as well as significantly reduce the number of writes to the second level cache (SSD).

\paragraph*{Contributions:} 
We propose a novel scheme for exclusive multi-level caching, in which we place a frequency based filter (with aging) between cache levels, as illustrated in Figure~\ref{fig:biditinylfu}.
That is, whenever an item $x$ is evacuated from a high level cache $L_i$, it is only inserted into the lower level cache $L_{i+1}$ if the corresponding would-be-victim at $L_{i+1}$, according to $L_{i+1}$'s management policy, is less frequent than $x$.
Similarly, we promote an accessed item $x$ from $L_{i+1}$ into $L_i$ only if $x$ is more frequent than the would-be-victim of $L_i$ according to $L_i$'s management policy.
The intuition behind this is that by keeping the more frequently accessed items at higher levels, which offer faster access times, the overall expected access time is shortened.
Further, this reduces the amount of items being switched between levels, which reduces the number of writes to the low levels.
In particular, when $L_2$ (or a lower level) is implemented in SSD, reducing writes is important as discussed~above.

Further, we split $L_1$'s total area in two: a \emph{Window space} used to store newly arriving items, and a \emph{Veterans space} that holds items that were promoted from lower levels.
The reason for separating between new items and veterans in $L_1$ is that we need to give new items some time to build their reference counts and other statistics before we can decide whether it makes sense to continue caching them or not.
This is obtained by placing a new item in the Window space, where it does not compete with the veterans until it becomes a victim candidate among the Window space items.

Our next contribution is an extensive performance evaluation study of our scheme against previous methods~\cite{Demote, Promote}. 
We show that our approach yields a saving of at least $90\%$ of the level-two writes among various traces and improves the expected latency, by up to $20\%$ 
percents on some of them, depending on the measurement technique. 
We also released our JAVA implementation of all schemes as part of the Caffeine Simulator~\cite{CaffeineProject}.

\paragraph*{Paper road-map:} The rest of this paper is structured as follows: 
In section~\ref{multi:sec:algorithms} we describe our scheme design. Section~\ref{multi:sec:eval} presents the evaluation and measurements on various traces. Finally, section~\ref{multi:sec:discussion} concludes with a results discussion and possible extensions.

%% file: related.tex
\section{Related Work}
\label{sec:related}

The task of managing a multi-level cache includes the management policy within each layer coupled with a mechanism to decide how to combine the layers.
The problem of managing a single layer cache has been extensively studied, and many policies have been suggested to address this~\cite{CAR,LHD,AdaptSize,Cloud1,Cloud2,TinyLFU,AdaptiveTinyLFU,ClockPro,2Q,ARC,LRB,AdaptiveCaches,MiniSim}.
Below, we discuss in more detail approaches for combining multiple cache~levels.

CHOPT~\cite{CHOPT} is an optimal offline algorithm for data placement in a multi-level cache where a data item can be accessed directly from any level of the cache.
The authors of~\cite{CHOPT} further present a sampling based approximation of CHOPT and demonstrate that on a wide variety of real-world traces it obtained an average access time reduction of $8.2\%-44.8\%$ over other state of the art approaches.

Karma~\cite{Karma} uses application hints to partition blocks according to their expected access patterns.
Each such partition is placed in a cache level and managed by a policy that are likely to be best for that partition based on the assumed access pattern.
Karma is most suited for environments in which such hints can be obtained, such as databases.
In contrast, our scheme \SYSNAME{} does not rely on any application hints.

DEMOTE~\cite{Demote} tries to mimic a single LRU policy for a two level cache and avoid duplication (exclusive caching) as follows:
When an item is demoted from $L_1$ to $L_2$, it is placed at the tail of the LRU list of $L_2$ (as if it was just accessed in $L_2$).
Conversely, when an item is promoted from $L_2$ to $L_1$, it is placed at the head of $L_2$' LRU list (as if it is the LRU item of $L_2$).
This reduces the level of replication between the cache levels, thereby improving the overall hit-rate.

A shortcoming of DEMOTE is that it generates a significant amount of I/O between the cache levels.
This is addressed by PROMOTE~\cite{Promote}, in which cache layers choose probabilistically whether to store an accessed item or not.
The effectiveness of exclusive caching and of combining eviction decisions between the cache levels has been also explored in~\cite{Sink}, where a holistic multi-level approach to ARC~\cite{ARC} has been designed and evaluated.

NHC~\cite{NHC} optimizes the overall multi-level cache throughput performance when they are under heavy load by directing some of the excess load to the non-top layer of the cache.
This is due to the observation that in many modern settings, latency differences between cache levels are not dramatic.
Hence, if the top layer is running at full capacity, it makes sense to enable deeper levels to serve requests directly, thereby improving the throughput of the system.

The multi-level cache simulator described in~\cite{desperately} accounts for several aspects of evaluating multi-level caches beyond hit-ratio.
This includes response time analysis, inter-level I/O, and impact on SSD lifetime. 
Unfortunately, up to the time of this writing, it has not been published.

Kangaroo~\cite{Kangaroo} aims to support caching of billions of tiny objects on flash.
Its goal is to reduce the write amplification of the flash storage, by splitting the flash cache into a small log-structured cache and a larger set associative cache.
Kangaroo also employs a small DRAM cache in front of the flash, but admits almost all objects from the DRAM cache to the flash.
The main focus of Kangaroo is to reduce the number of writes from the on-flash log structured region to the on-flash set associative cache.
In contrast, our work aims to reduce the number of objects that are transferred from the L1 cache (typically DRAM), to the lower level (typically flash).
Hence, Kangaroo and our approach are orthogonal, and in principle can be combined.

%% file: algorithms.tex
\section{\SYSNAME}
\label{multi:sec:algorithms}



\subsection{Design}
In this section, we present the core components of our system, named \SYSNAME.
The main objectives of our design are to:
\begin{enumerate}
	\item \label{obj:hr} Maintain a high overall hit-ratio (i.e., reduce complete misses).
	\item \label{obj:l1} Serve a large portion of the requests from L1 (i.e., increase L1 hits).
	\item \label{obj:writes} Reduce writes and bandwidth usage to the lower levels of the cache.  
\end{enumerate}

One can easily argue that if the access distribution were i.i.d. and constant over time, then keeping the most frequent items at the higher possible level would benefit objectives~\ref{obj:hr} and~\ref{obj:l1}. 
Unfortunately, it is well known~\cite{TinyLFU,AdaptiveTinyLFU,ARC} that access patterns typically exhibit a mix of recency and frequency characteristics, and specific items' relative frequency tends to change over time even in frequency biased workloads.
To deal with both issues, we split the highest level of the cache (L1) into two parts: a \textit{Window} space to which items that are new to the cache are inserted, and a \textit{Veteran} space intended for keeping the most frequent items that were also recently~accessed.

\begin{figure}[!h]
	\begin{center}
		\includegraphics[width=0.80\columnwidth]{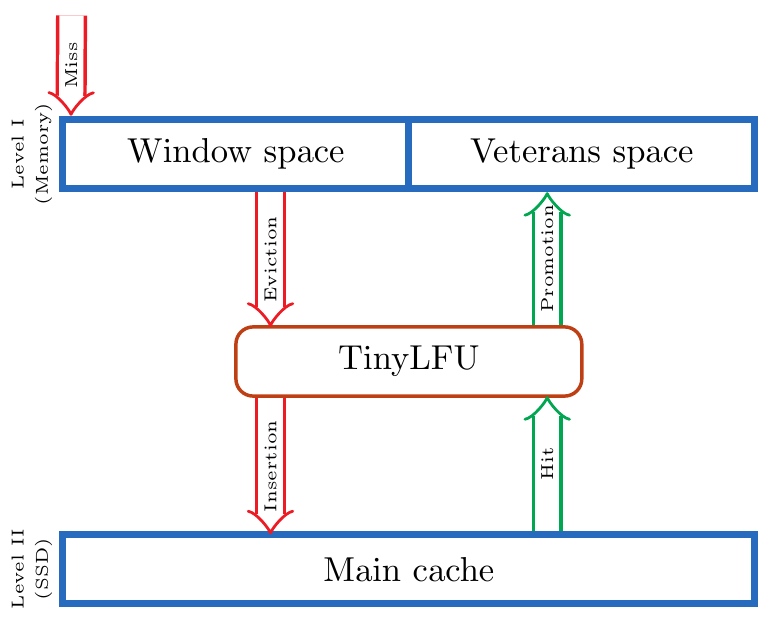}
	\end{center}	
	\caption{\normalfont{\SYSNAME{} scheme}}
	\label{fig:biditinylfu}
\end{figure}

Additionally, we measure the frequency of items and filter admissions between the layers of the cache based on it.
Filtering is performed in both directions: 
\begin{enumerate}
	\item An item evicted from the Window space (at L1) becomes a candidate to be admitted into the lower cache level (i.e., L2).
	Its frequency is compared to a victim item from L2 and the one with the higher frequency stays or is inserted depending on the comparison result.
	This filtering is meant to address objectives~\ref{obj:writes} and~\ref{obj:hr}.
	\item An item at L2 that has been accessed (L2 hit) becomes a candidate to be promoted to the higher cache level (i.e., L1).
	Its frequency is compared to a victim item from the Veterans space and the one with the higher frequency stays or is inserted depending on the result.
	The goal of this filtering is to achieve objective~\ref{obj:l1}.
\end{enumerate}
This mechanism results in a flow where new and recent items live in the Window space, other frequent items are admitted into the lower cache, while the most frequent items are promoted back into the Veterans space. 
Notice that each space has its own internal caching mechanism. 
The outline of the policy is detailed in Algorithm~\ref{alg:biditinylfu} and is illustrated in Figure~\ref{fig:biditinylfu}. 
For brevity, we omit edge cases, e.g., the cache is still not~full.

To implement the filtering mechanism, we rely on a frequency sketch with aging, similar to the TinyLFU mechanism~\cite{TinyLFU}.
That is, to enable keeping track of a very large number of items' frequencies in a space efficient manner, we employ a sketch such as Count-Min Sketch~\cite{CMSketch}.
Moreover, to accomodate for frequency changes, and prevent once popular items from polluting the cache for a long time, we employ periodic halving of the sketch counters.
To further expedite this process and reduce memory consumption, we also cap the size of the counters to $\lceil W/C\rceil$, where $W$ is the maximal number of items we are willing to track and $C$ is the number of items in the cache.
More detailed motivation for this approach can be found in~\cite{TinyLFU}.

\begin{algorithm*}[h]
	\caption{\SYSNAME{}} \label{alg:biditinylfu}
	\begin{algorithmic}[1]
		\Procedure{OnMiss}{$newItem$}
		\State insert $newItem$ into $Window$ space 
		\State $candidate$ $\gets$ a victim from $Window$ space \color{blue}\Comment{${\text{\normalsize $candidate$ is a potential} \atop \text{\normalsize admission for L2}}$}\color{black}
		\State $victim$ $\gets$ a victim from $Main$ cache
		\If{frequency($candidate$) $\ge$ frequency($victim$)} 
		\State admit $candidate$ into $Main$ cache
		\State evict $victim$ from $Main$ cache
		\Else
		\State evict $candidate$ from $Window$ space 
		\EndIf
		\EndProcedure
		\\
		\Procedure{OnHit}{$item$}
		\If {$item$ $\in$ L1}
			\State handle $item$ according to the internal cache policy
		\Else \color{blue}\Comment{$item$ $\in$ L2 is a potential for promotion}\color{black}
			\State $victim$ $\gets$ a victim from $Veterans$ space
			\If{frequency($item$) $\ge$ frequency($victim$)} 
				\State promote $item$ into $Veterans$ space
				\State evict $victim$ into $Main$ cache
			\Else
				\State handle $item$ according to the internal cache policy
			\EndIf 
		\EndIf
		\EndProcedure
	\end{algorithmic}
\end{algorithm*}

\subsection{A Multi-Level Extension of \SYSNAME}

\begin{figure}[!h]
	\begin{center}
		\includegraphics[width=0.6\columnwidth]{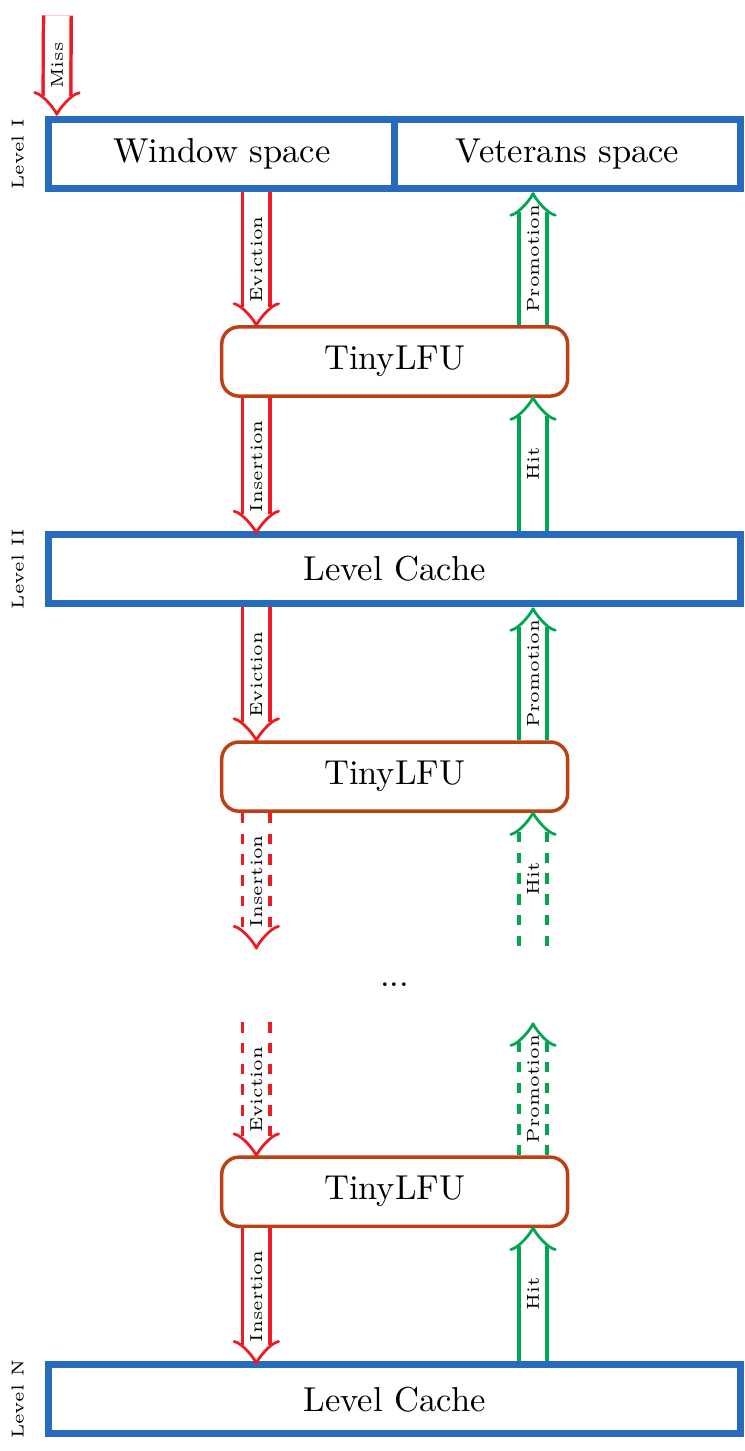}
	\end{center}	
	\caption{\normalfont{N-dimensional \SYSNAME{} scheme}}
	\label{fig:biditinylfundim}
\end{figure}

An extension of \SYSNAME{} to $N$ levels is presented in Figure~\ref{fig:biditinylfundim}.
Here, we place a TinyLFU filter between each pair of levels.
As before, when an item becomes the victim of level $L_i$, it is admitted to $L_{i+1}$ if its frequency estimate is higher than the would be cache victim of level $L_{i+1}$.
In contrast, when an item at level $L_i$ (for $i > 1$) is accessed, it is promoted to $L_{i-1}$ if its frequency estimate is higher than the frequency estimate of the would be victim of layer $L_{i-1}$.

 
\subsection{Synthetic Motivating Example}
To exemplify the benefit of two areas in the first-level cache, we produced synthetic traces, mixing frequency and recency biases. Given a \textit{skew} and \textit{recency} parameters in $(0,1)$, each access of a trace is generated by picking a recent item with a probability of the recency argument and an item from a Zipf distribution with the skew argument otherwise. The creation process of the traces is described in Algorithm~\ref{alg:traces-alg}.

\begin{figure}[!h]
		\centering \includegraphics[width=0.8\columnwidth]{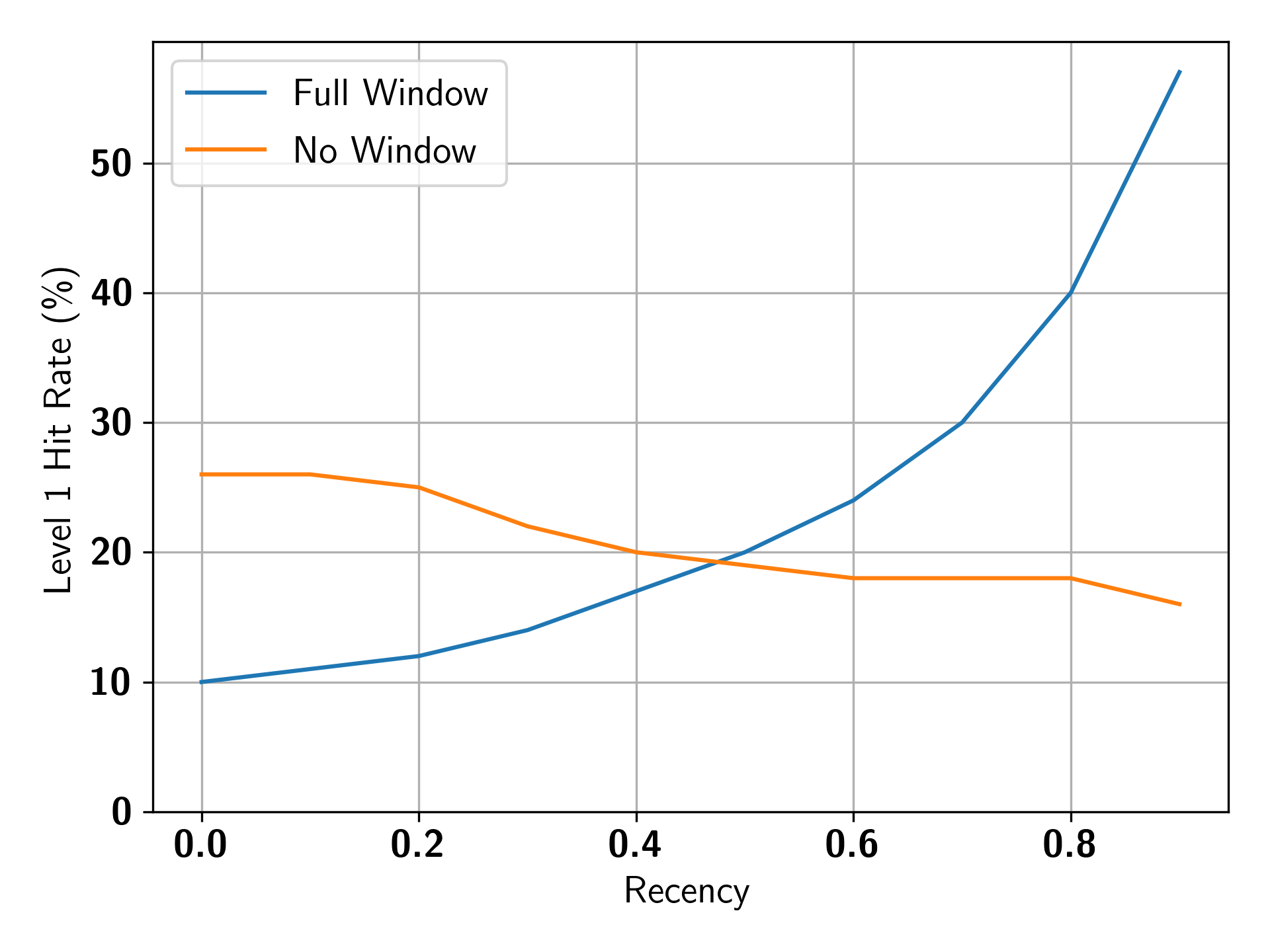}
		\label{eva:zipf:l1}
	\caption{Results for a synthetic zipf trace of length 10 million accesses, with $0.5$ skewness parameter and different recencies, where level 2 cache size is 50\% of the unique items and the ratio between level 1 and level 2 sizes is 1:10.}
	\label{alg:zipf}
\end{figure}

Using a skew parameter of $0.5$ and recency parameters ranging from $0$ to $1$, we ran versions of \SYSNAME{} with Full Window area (i.e., no Veterans area) and with No Window area (i.e., full Veterans area). 
The results for those runs are shown in Figure~\ref{alg:zipf}. 
As can be expected, when the recency parameter is close to $0$, meaning it is a frequency biased trace, the Full Window version outperforms the No Window version.
On the contrary, as the recency increases, the Full Window version becomes better and better. 
This motivates us to split the first-level cache into two areas to benefit from both~worlds.

\begin{algorithm*}[h]
	\caption{Creation of Synthetic Traces} \label{alg:traces-alg}
	\begin{algorithmic}[1]
		\Procedure{Create Trace}{$length$, $groundSet$, $skew$, $recency$}
		\State $trace$ $\gets$ emptyList()
		\While{$trace$ size $<$ $length$}
		\If{rand() < $recency$} \color{blue}\Comment{rand() is uniformly distributed $\in (0,1)$}\color{black}
		\State $trace$.append(one of the $10$ recent items) \color{blue}\Comment{${\text{\normalsize Notice this amplifies both} \atop \text{\normalsize recency \textbf{and skewness}}}$}\color{black}
		\Else
		\State $trace$.append(sample an item from a Zipf distribution with $skew$ and $groundSet$ parameters)
		\EndIf
		\EndWhile
		\State \Return $trace$
		\EndProcedure
	\end{algorithmic}
\end{algorithm*}

\subsection{The Tiebreak Dilemma}
During our evaluation (described thoroughly in section~\ref{multi:sec:eval} below), we observed that a substantial amount of the comparisons done by the sketching mechanism results in equal estimated frequencies.
Hence, choosing whether to admit or reject in these cases could impact the performance significantly.
Choosing to admit biases the policy towards recency and increases the number of writes to lower levels, while choosing to reject biases the policy towards frequency and decreases the number of writes.
Below we presents the performance of both options. 
Since the overall mechanism is frequency oriented due to the frequency filter, and since the number of writes is reduced considerably, we chose to admit whenever there is such a tie.
However, this can be a configurable option to allow more strict~filtering.

\subsection{Implementation}
We implemented \SYSNAME{} in Java and used the Caffeine Simulator~\cite{CaffeineProject} to test it.
Our configuration includes Count-Min Sketch with 4 bits per counter, and we tested different split ratios of L1 between the Window space and the Veterans space. 
For simplicity, LRU is the internal cache management policy for each of the spaces in L1, and similarly to the main cache in~\cite{TinyLFU}, SLRU is employed in the internal cache of L2.


%% file: evaluation.tex
\section{Evaluation}
\label{multi:sec:eval}
In this section, we describe our evaluation process, conducted in a simulated environment using four real-world modern traces.
First, we evaluated multiple configurations of our system to study the consequences of choosing one over another.
Then, we compare our chosen version with state-of-the-art competing methods.

We measured hits, misses, and the number of writes on all cache levels as well as average latency.
From an algorithmic point of view, evictions and writes happen in a sequential manner and hold the following access to the system. Nevertheless, on many production systems, writes are done by a secondary process (often in batches) while reads continue to be served. Obviously, if the system is under pressure, the writes can still delay the reads.
To reflect this, we present latencies with and without considering writes.
We denote $T_{li}$ the access time to serve an item from layer $i$, $H_{li}$ the hit ratio of layer $i$, $T_{miss}$ is time to serve an item that it not in any of the cache levels (miss penalty), while $M$ is the complete miss rate.
We combine these to the following formulation:
\begin{align}
	& {Avg\ Read \atop Latency} = \frac{T_{l1} \cdot H_{l1} + T_{l2} \cdot H_{l2} + T_{miss} \cdot M}{\text{number of requests}}
\end{align}
\begin{multline}
	{Avg\ Read\&Write \atop Latency} = \\ \frac{T_{l1} \cdot H_{l1} + T_{l2} \cdot H_{l2} + T_{miss} \cdot M + T_{l1} \cdot W_{l1} + T_{l2} \cdot W_{l2}}{\text{number of requests}}
\end{multline}
Based on the values in Table~\ref{tab:tradeoff}, we set $T_{l1}=2ns$, $T_{l2}=200us$, $T_{miss}=100ns$, $H=Hits$, $M=Misses$ and $W=Writes$.
Keep in mind that the actual latency is in-between Avg\ Read Latency and Avg\ Read\&Write Latency, and is heavily affected by engineering optimization.
We used a version of the Caffeine Simulator~\cite{CaffeineProject} modified to handle multilevel caches to run all of our experiments.

\subsection{Traces}
\label{multi:sec:eval:traces}
To conduct our research, we used four real-world modern traces that were published in recent years. SYSTOR~\cite{Systor} is a trace of accesses to a storage of a VDI system. 
CDN~\cite{PracticalBounds} is a trace of Wikipedia CDN deployment. TENCENT~\cite{Tencent} is a trace of a large-scale photo service. TWITTER~\cite{Twitter} is a trace of anonymized cache requests from a Twitter production cluster. 
Table~\ref{multi:tab:traces} lists the number of accesses and uniques items in each trace. 
Notice that to avoid the complications of size-aware cache policies, we split oversized items into chunks of 4kb, and treated all items as equal-sized. 
We ran the traces for different L2 cache sizes as a percentage of the unique items number, ranging from $10\%$ to $100\%$ percent, and for different ratios between L1 and L2, ranging from $1:10$ to $1:100$. 
Due to the tremendous number of items in the TENCENT trace, we were able to run it with L2 size only up to $50\%$ of the unique items. 

\begin{table}[h]
	\caption{Summary of Traces}
	\label{multi:tab:traces}		
	\begin{center}
		\begin{tabular}{l|c|c|c} 
			\textbf{Name} & \textbf{Accesses} & \textbf{Unique Objects} & \\
			& $millions$ & $millions$ \\
			\hline
			SYSTOR1 \cite{Systor} & 527 & 452 & \\
			CDN1 \cite{PracticalBounds} & 4,541 & 628 \\
			TENCENT1 \cite{Tencent} & 2,797 & 1,110 \\
			TWITTER1 \cite{Twitter} & 939 & 44 \\
		\end{tabular} \\
		\vspace{0.2cm}
		Total number of accesses and unique objects for each trace. 
	\end{center}
	\vspace{-0.5cm}
	\normalsize
\end{table}

\vspace{-1cm}
\subsection{Comparison of different configurations}
As mentioned above, our system has few available configurations, mainly the ratio we split the level-one cache and whether we admit or reject in a frequency tie.  
To get some insights into these configurations, we ran BiDiFilter with a window space of $1\%$, $50\%$, and $99\%$, as well as with admission and rejection on a tie. 
Combining those options, we get six different versions.
We name them BiDiFilterXY, where X is the window size and Y is T or F for admission or rejection, respectively.
Additionally, we ran a version where the level-one cache is united (i.e., window and veterans share together all the space) and we admit on a tie, and name it BiDiFilterLRU.
The results of these simulations are presented in figures~\ref{eva:between:writes}~and~\ref{eva:between:latency}.

The most eye-catching phenomenon is the large difference in the number of writes between tie-admission configurations and tie-rejection configurations. Figure~\ref{eva:between:writes} shows the number of writes on a log scale, and the difference between those two sets of configurations is about two orders of magnitude on all traces and cache sizes. 
This indicates that a significant part of the potential admissions are tied; probably most of them are with low frequencies. Notice that among the tie-admission group -- the $1\%$ window usually has fewer writes, but among the tie-rejection group -- the $99\%$ is better. However, the split ratio seems less important here.

Regarding the latencies in figure~\ref{eva:between:latency}, BiDiFilterLRU is noticeably the poorest option. 
For TENCENT1 and TWITTER1 traces, the tie-rejection group is considerably worst, sometimes by an $x2$ factor, apparently because they are very recency oriented with fresh items arriving constantly. 
On the contrary, in SYSTOR1 and CDN1, the differences are much smaller, and there is no clear split between the groups. 
There is also no notable impact if we measure the latency with or without writes.
In all cases, BiDiFilter50T performs well; hence we continue to test this configuration against other existing~methods.

\newcommand{\Height}{3.5cm}
\begin{figure*}[h]
	\centering SYSTOR1 \\
	\subfloat[\normalfont{$L1:L2 = 1:10$ }]{
		\includegraphics[trim=0 0 0 10, clip, height=\Height]{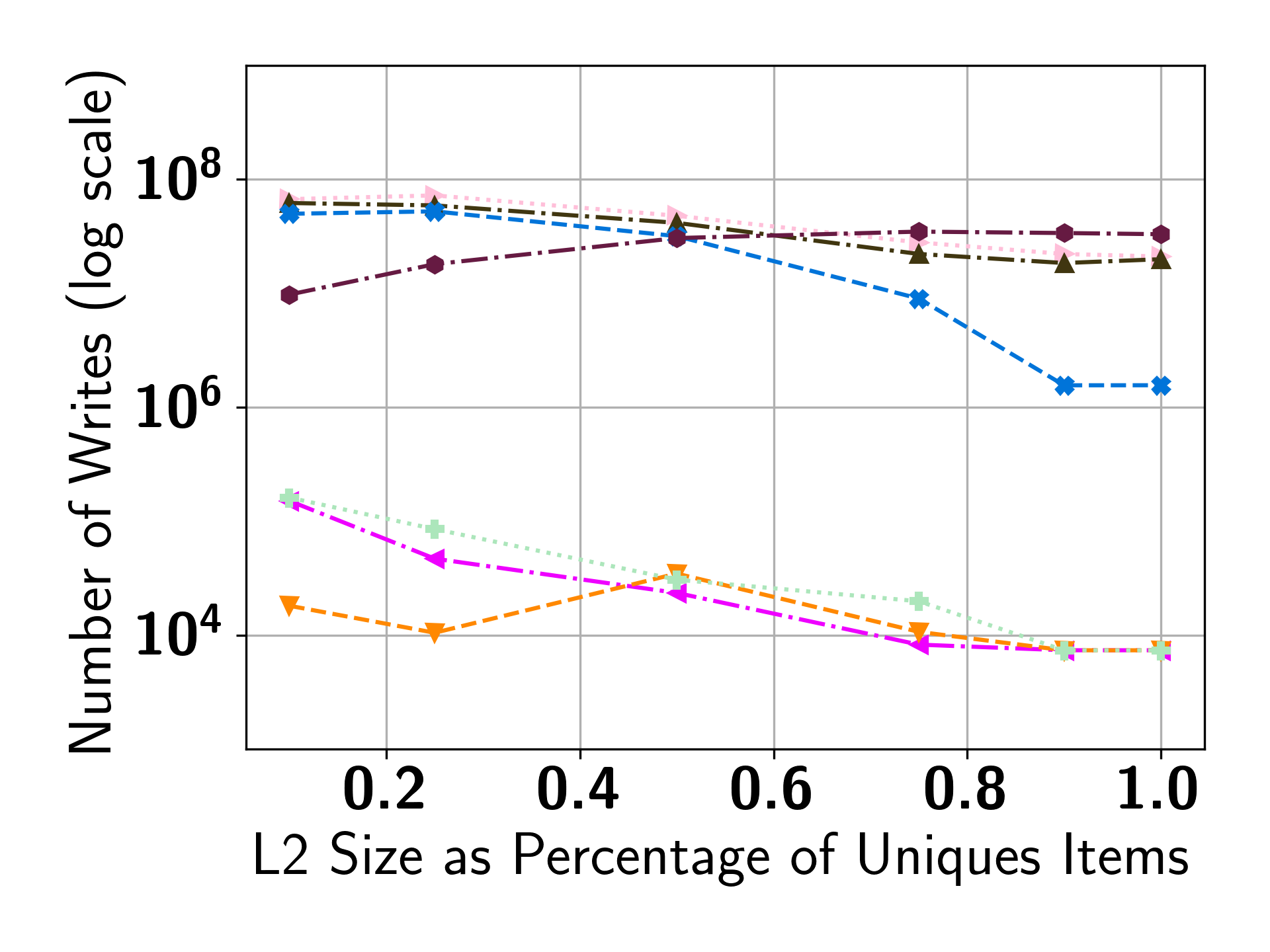}
	}
	\subfloat[\normalfont{$L1:L2 = 1:20$ }]{
		\includegraphics[trim=85 0 0 10, clip, height=\Height]{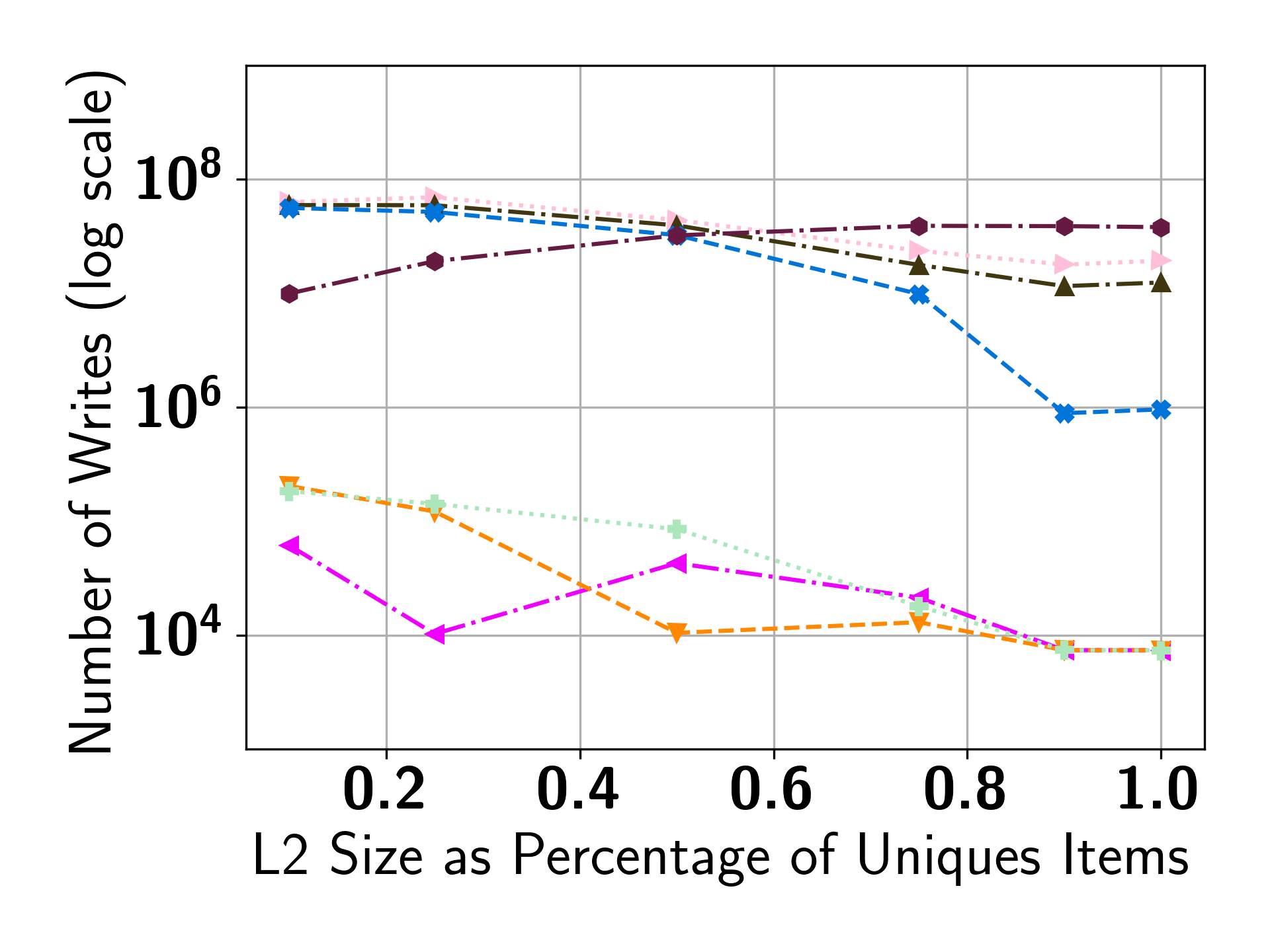}
	}
	\subfloat[\normalfont{$L1:L2 = 1:50$ }]{
		\includegraphics[trim=85 0 0 10, clip, height=\Height]{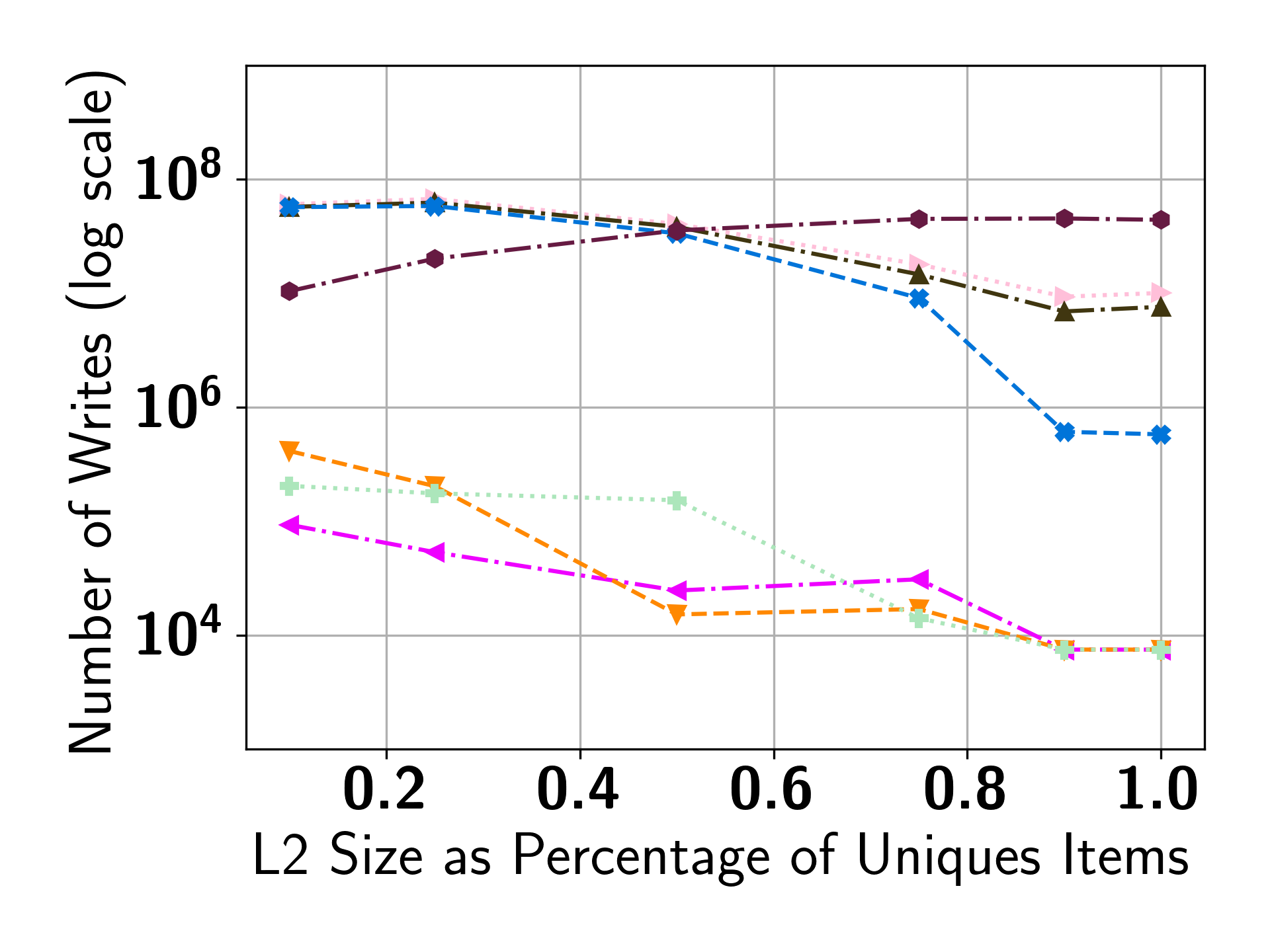}
	}
	\subfloat[\normalfont{$L1:L2 = 1:100$ }]{
		\includegraphics[trim=85 0 0 10, clip, height=\Height]{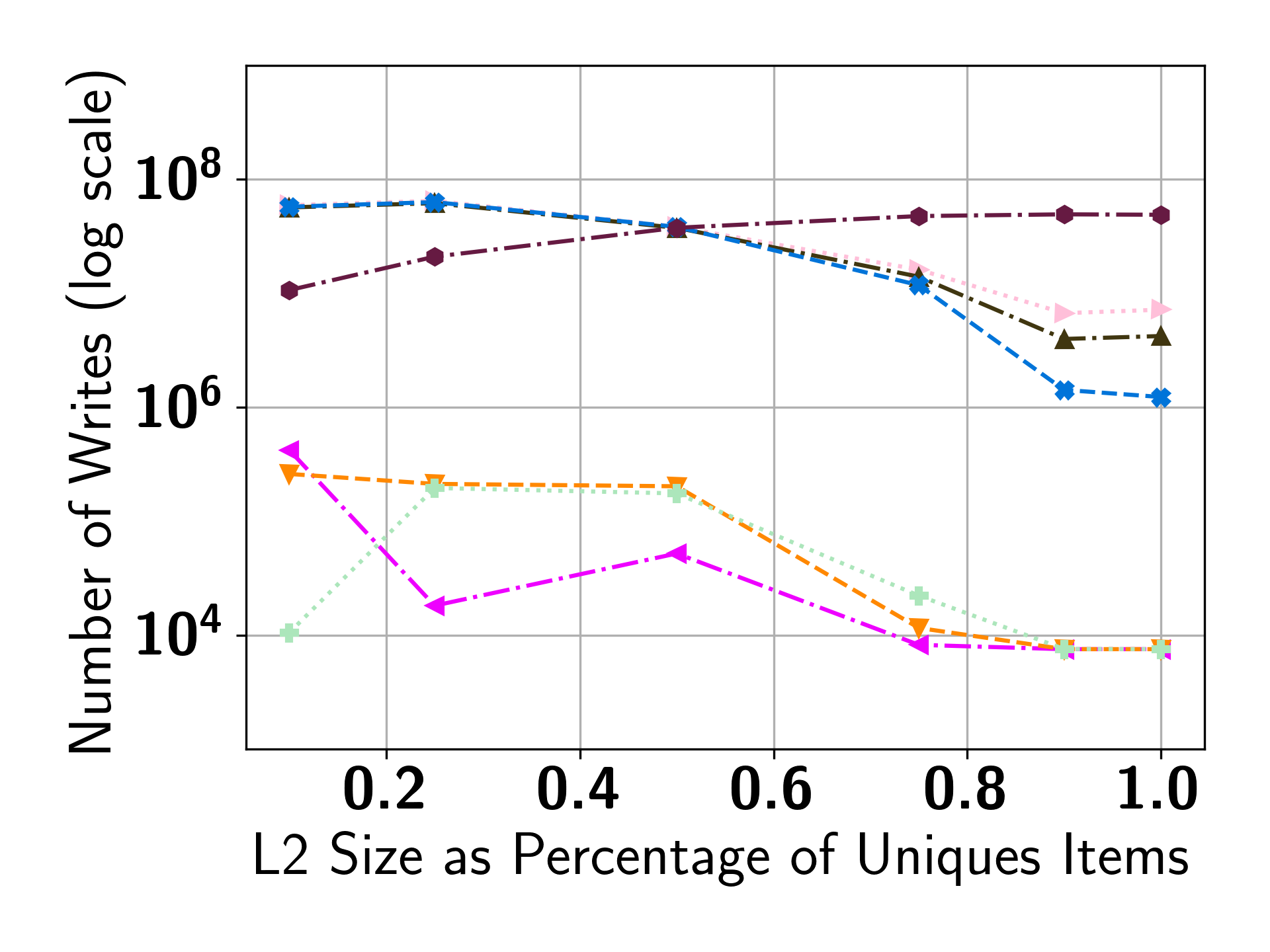}
	}
	\\
	\vspace{0.2cm}
	\centering CDN1 \\ 
	\vspace{-0.2cm}
	\subfloat[\normalfont{$L1:L2 = 1:10$ }]{
		\includegraphics[trim=0 0 0 10, clip, height=\Height]{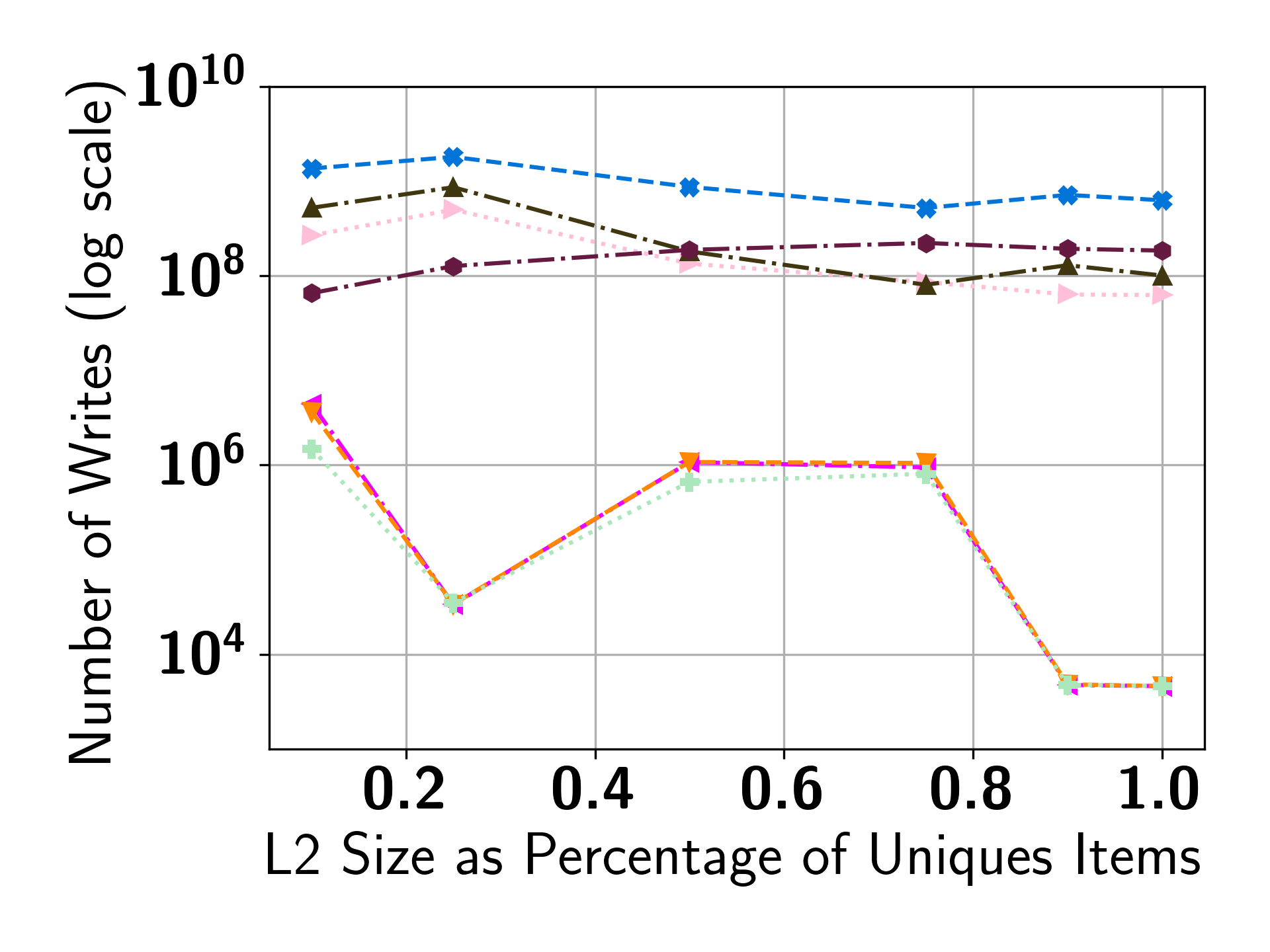}
	}
	\subfloat[\normalfont{$L1:L2 = 1:20$ }]{
		\includegraphics[trim=86 0 0 10, clip, height=\Height]{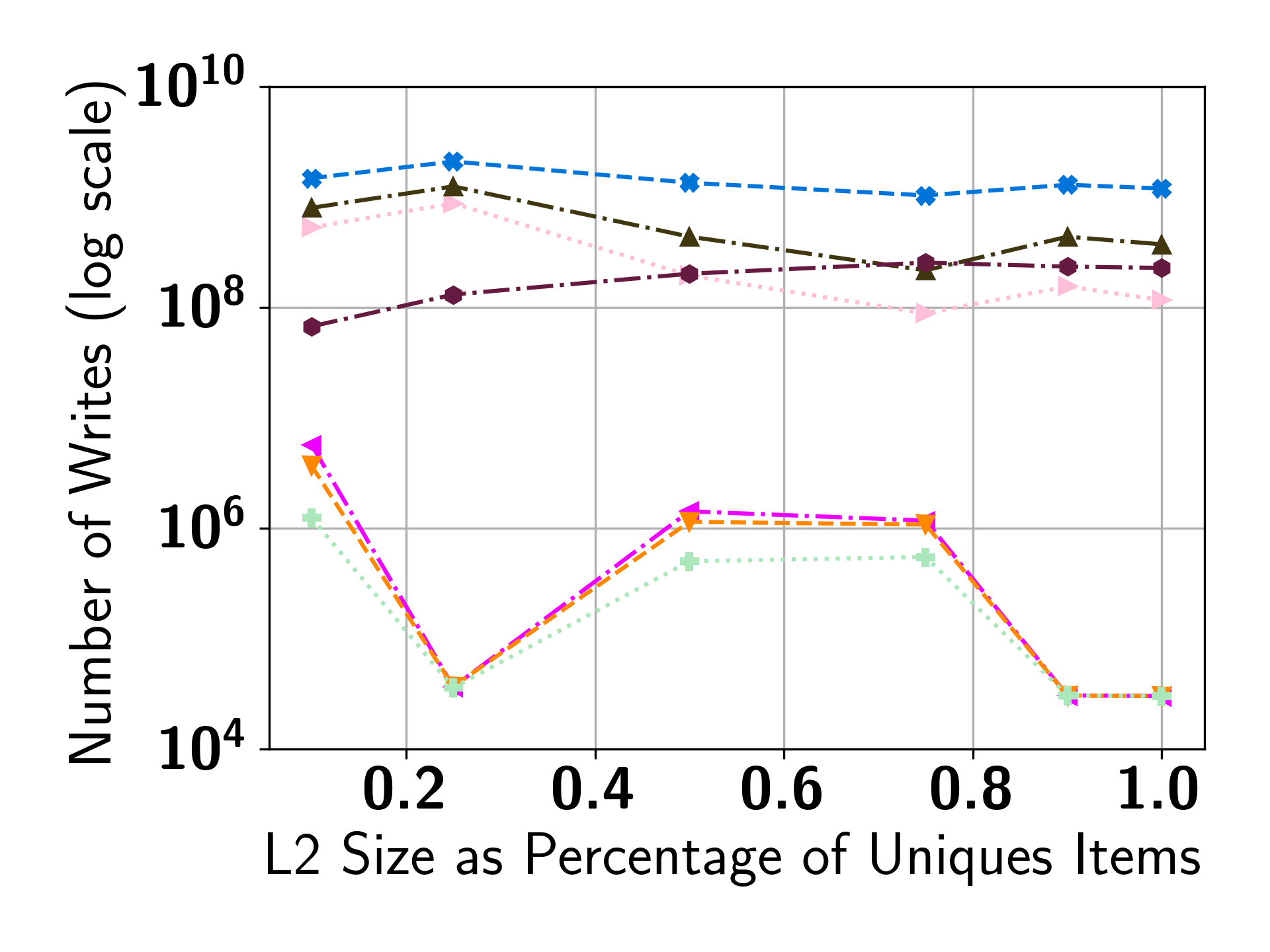}
	}
	\subfloat[\normalfont{$L1:L2 = 1:50$ }]{
		\includegraphics[trim=86 0 0 10, clip, height=\Height]{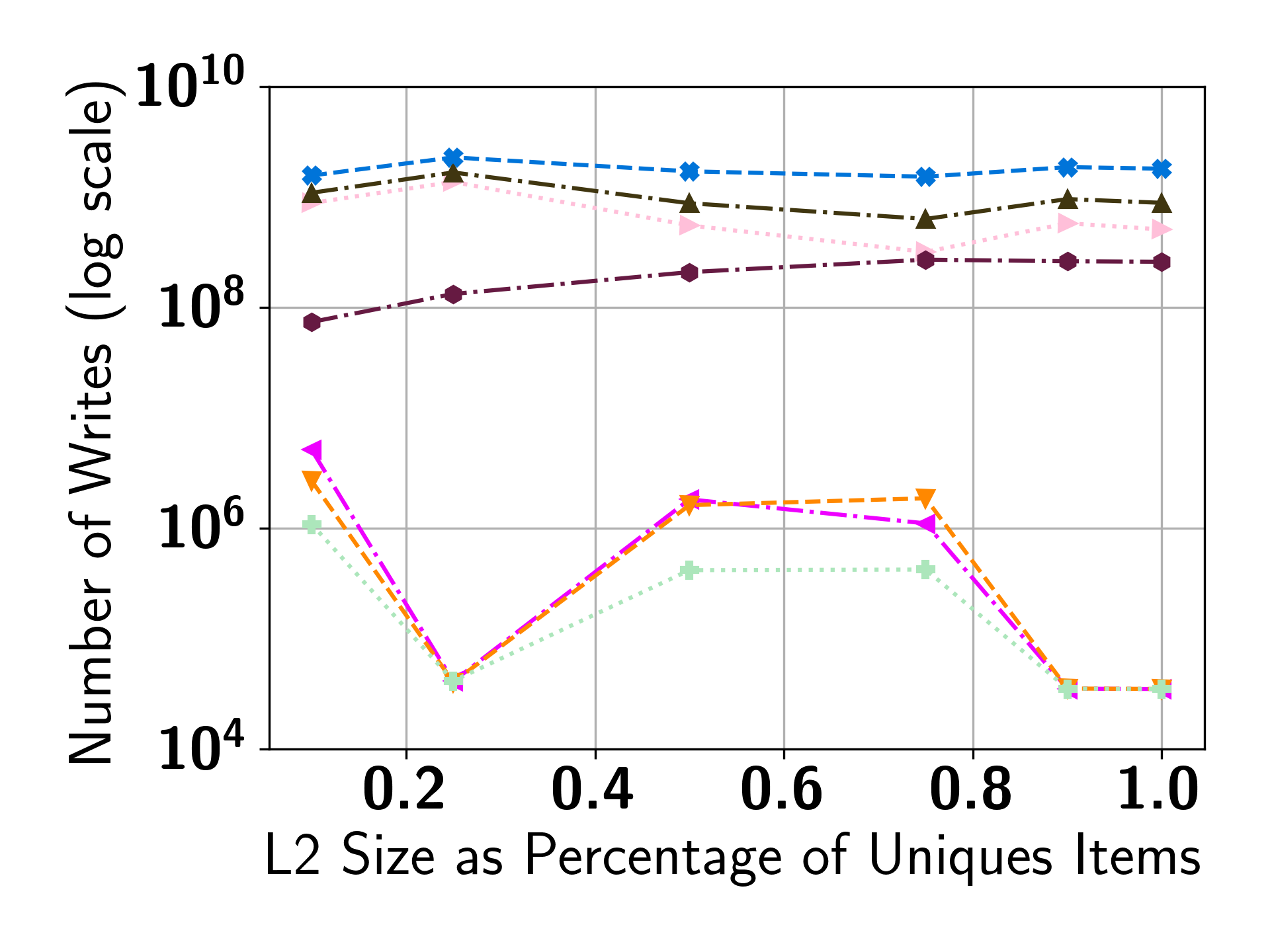}
	}
	\subfloat[\normalfont{$L1:L2 = 1:100$ }]{
		\includegraphics[trim=86 0 0 10, clip, height=\Height]{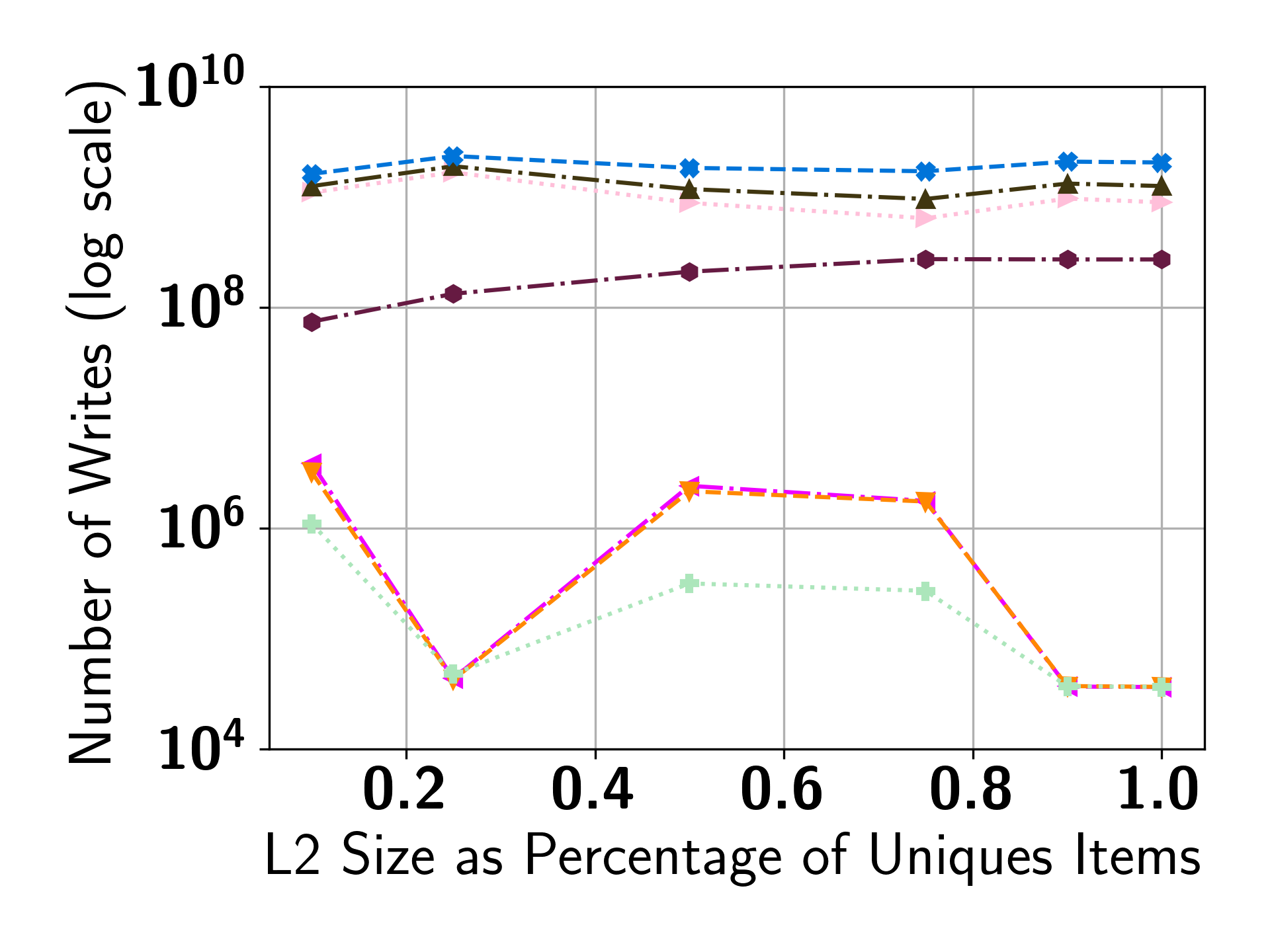}
	}
	\\
	\vspace{0.2cm}
	\centering TWITTER1 \\ 
	\vspace{-0.2cm}
	\subfloat[\normalfont{$L1:L2 = 1:10$ }]{
		\includegraphics[trim=0 0 0 10, clip, height=\Height]{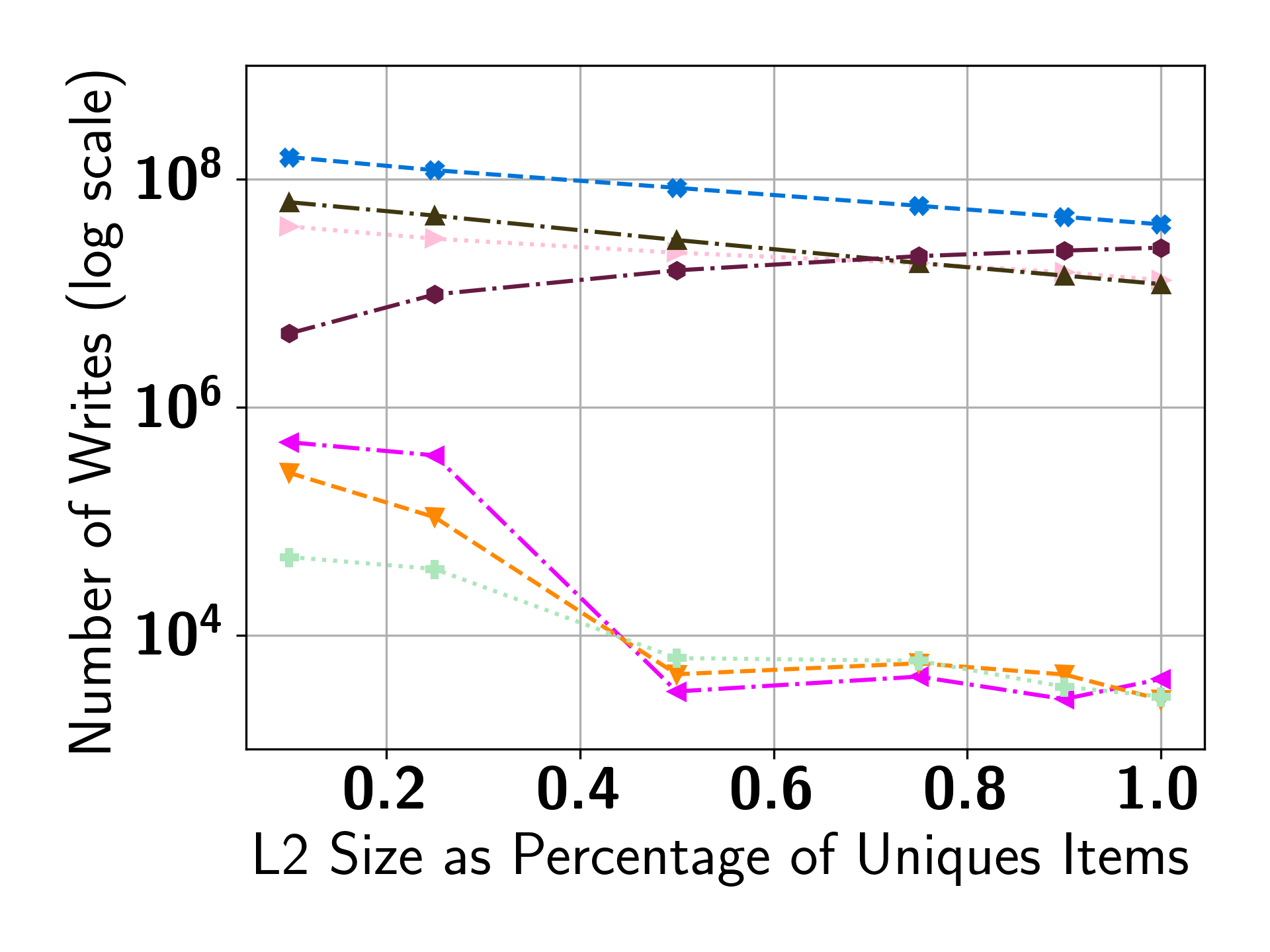}
	}
	\subfloat[\normalfont{$L1:L2 = 1:20$ }]{
		\includegraphics[trim=86 0 0 10, clip, height=\Height]{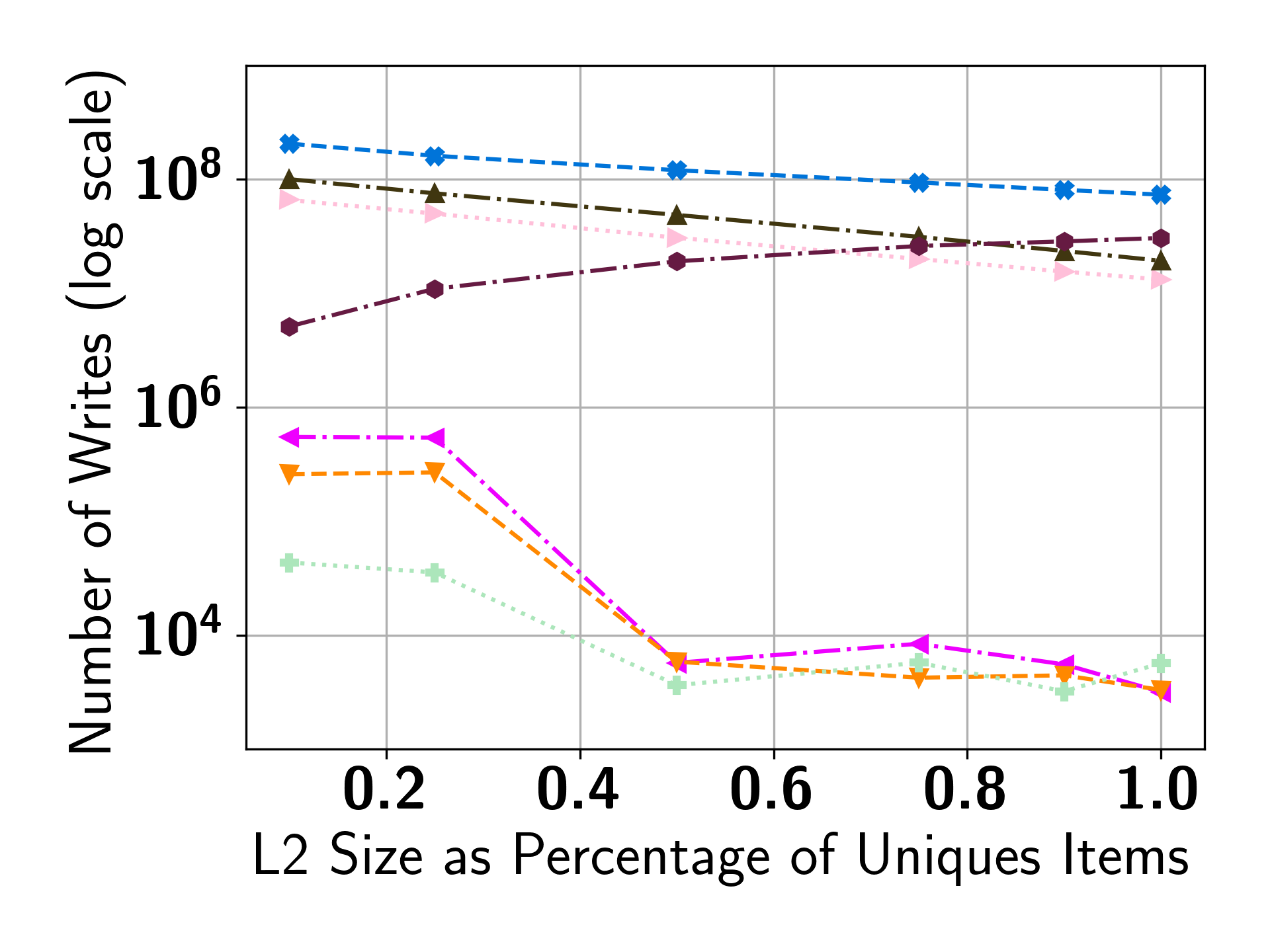}
	}
	\subfloat[\normalfont{$L1:L2 = 1:50$ }]{
		\includegraphics[trim=86 0 0 10, clip, height=\Height]{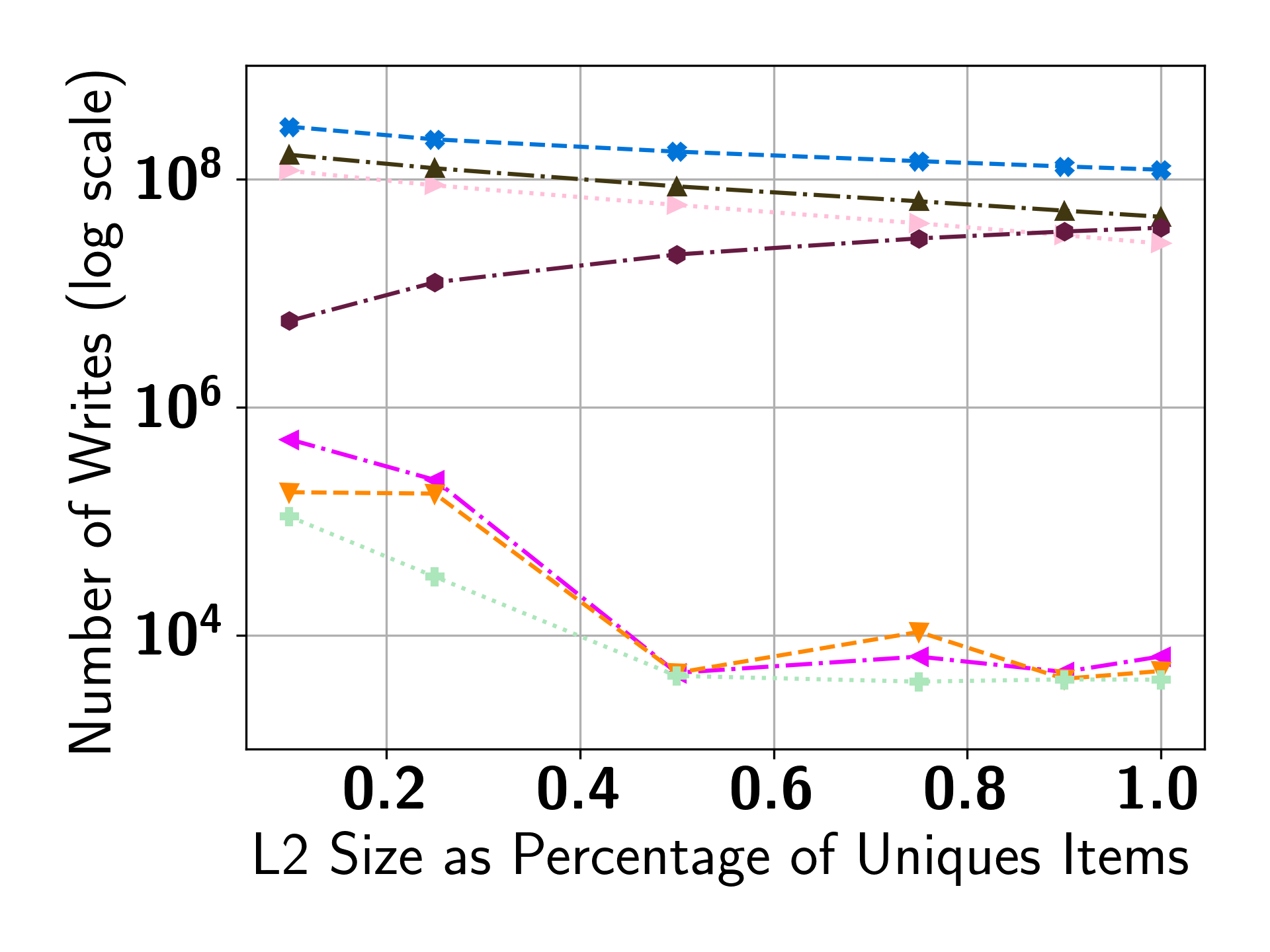}
	}
	\subfloat[\normalfont{$L1:L2 = 1:100$ }]{
		\includegraphics[trim=86 0 0 10, clip, height=\Height]{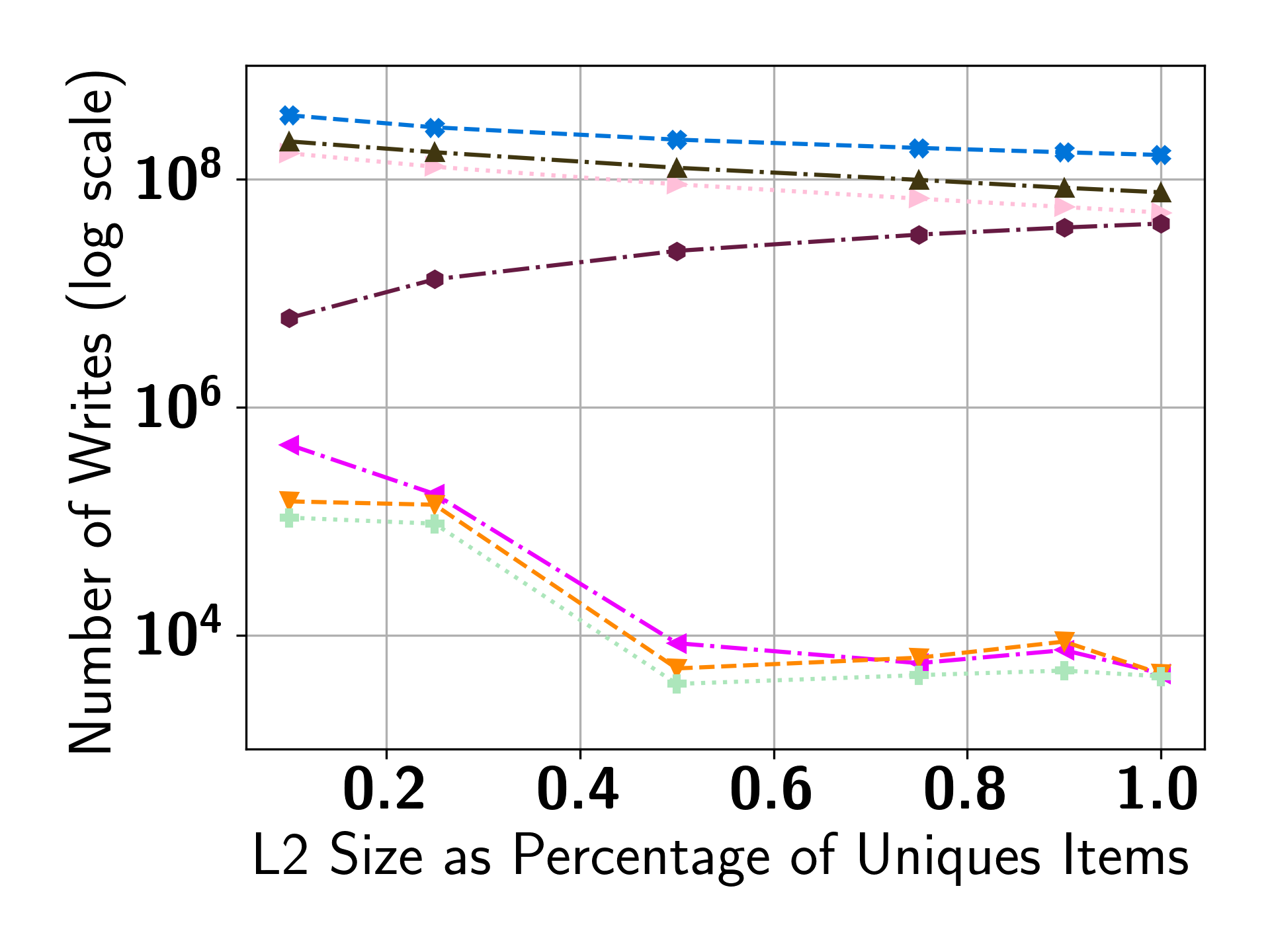}
	}
	\\
	\vspace{0.2cm}
	\centering TENCENT1 \\ 
	\vspace{-0.2cm}
	\subfloat[\normalfont{$L1:L2 = 1:10$ }]{
		\includegraphics[trim=0 0 0 10, clip, height=\Height]{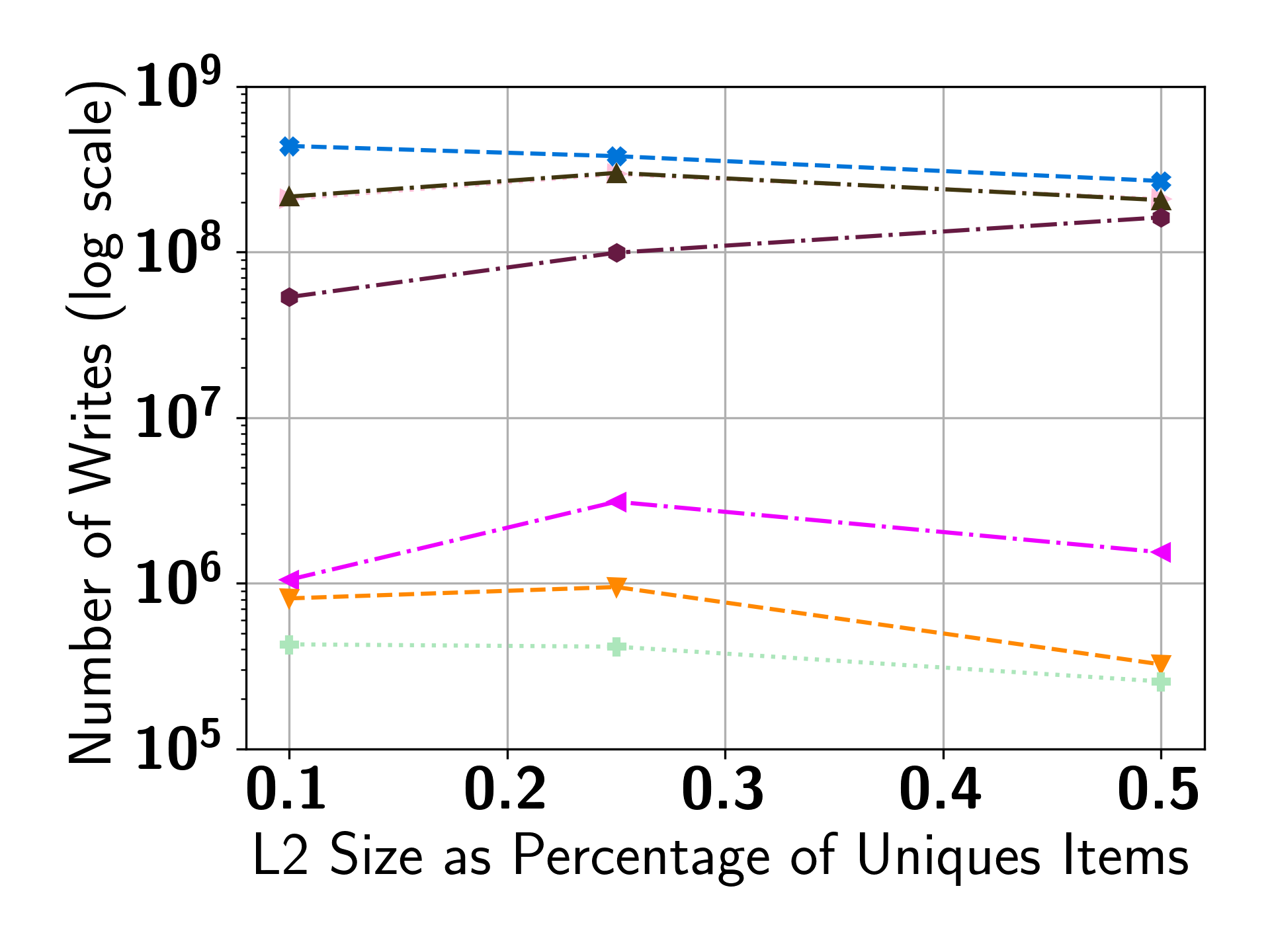}
	}
	\subfloat[\normalfont{$L1:L2 = 1:20$ }]{
		\includegraphics[trim=86 0 0 10, clip, height=\Height]{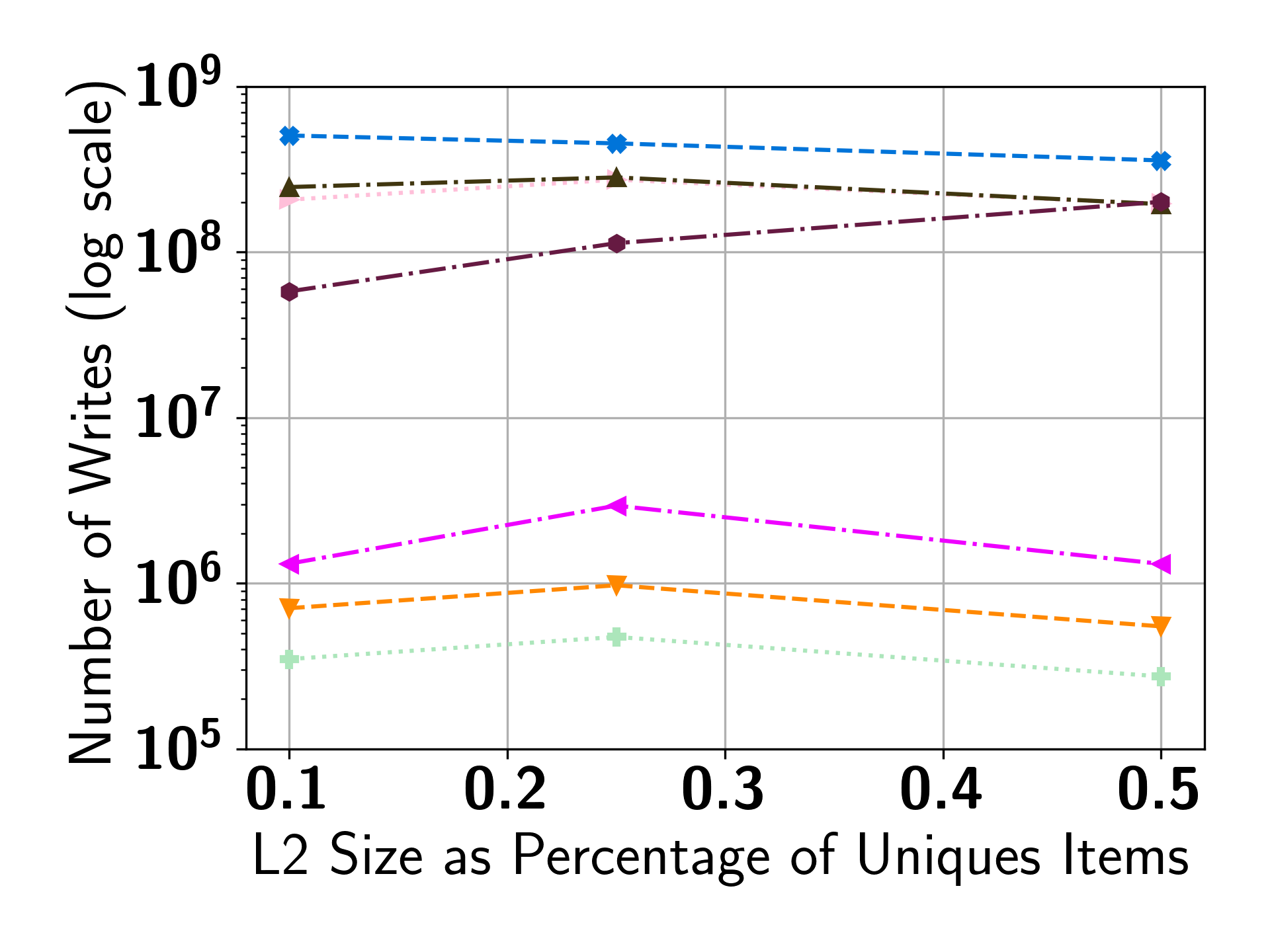}
	}
	\subfloat[\normalfont{$L1:L2 = 1:50$ }]{
		\includegraphics[trim=86 0 0 10, clip, height=\Height]{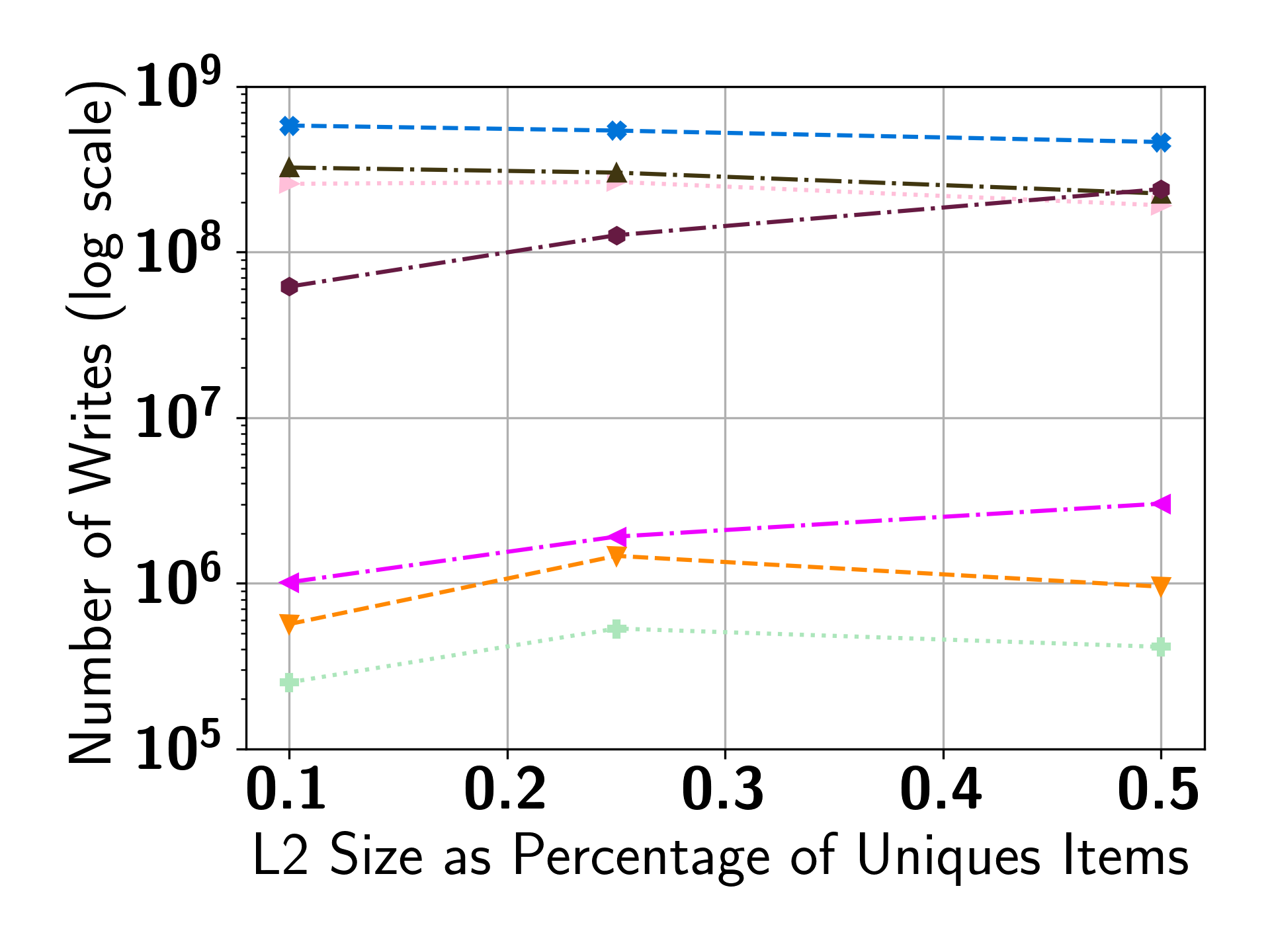}
	}
	\subfloat[\normalfont{$L1:L2 = 1:100$ }]{
		\includegraphics[trim=86 0 0 10, clip, height=\Height]{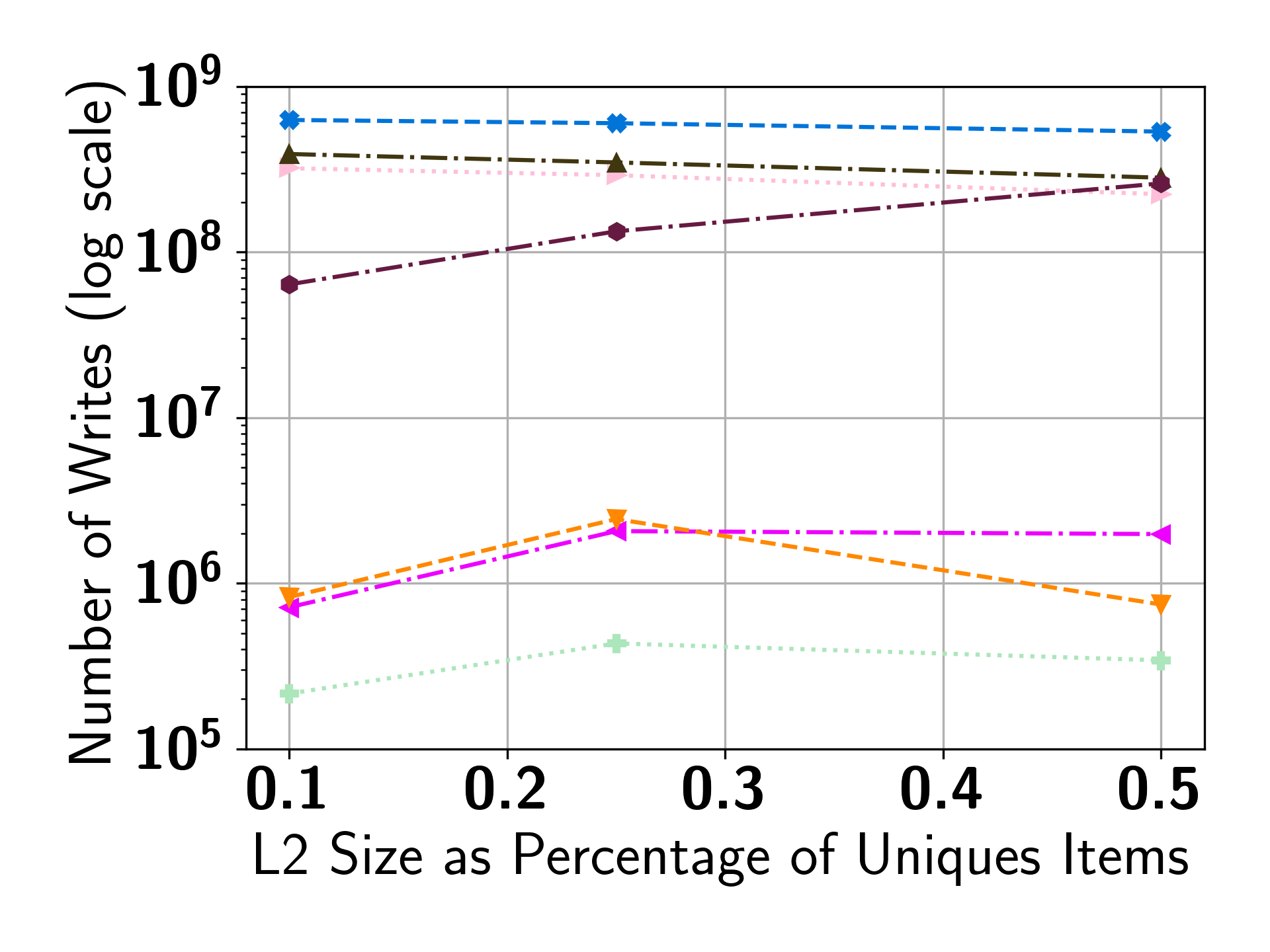}
	}
	\\
	\centering \includegraphics[height=1.3cm]{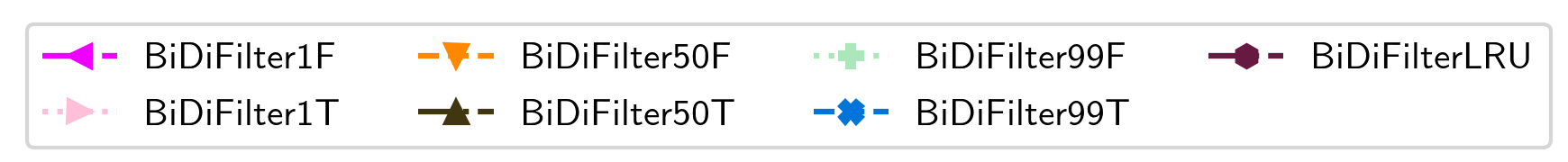}

	\caption{Number of writes to L2 for multiple traces and multiple ratios between L1 and L2}
	\label{eva:between:writes}
\end{figure*}

\newcommand{\HeightL}{3.5cm}
\begin{figure*}[h]
	\begin{tabularx}{\textwidth}{ >{\centering\arraybackslash}X >{\centering\arraybackslash}X  }
		SYSTOR1 & CDN1 \\
	 	Hit-Ratio ranges from 7\%--23\% & Hit-Ratio ranges from 20\%--86\%	
	\end{tabularx} \\ 
	\subfloat[\normalfont{$L1:L2 = 1:10$ }]{ \begin{tabular}[b]{c}%
		\includegraphics[trim=0 0 0 10, clip, height=\HeightL]{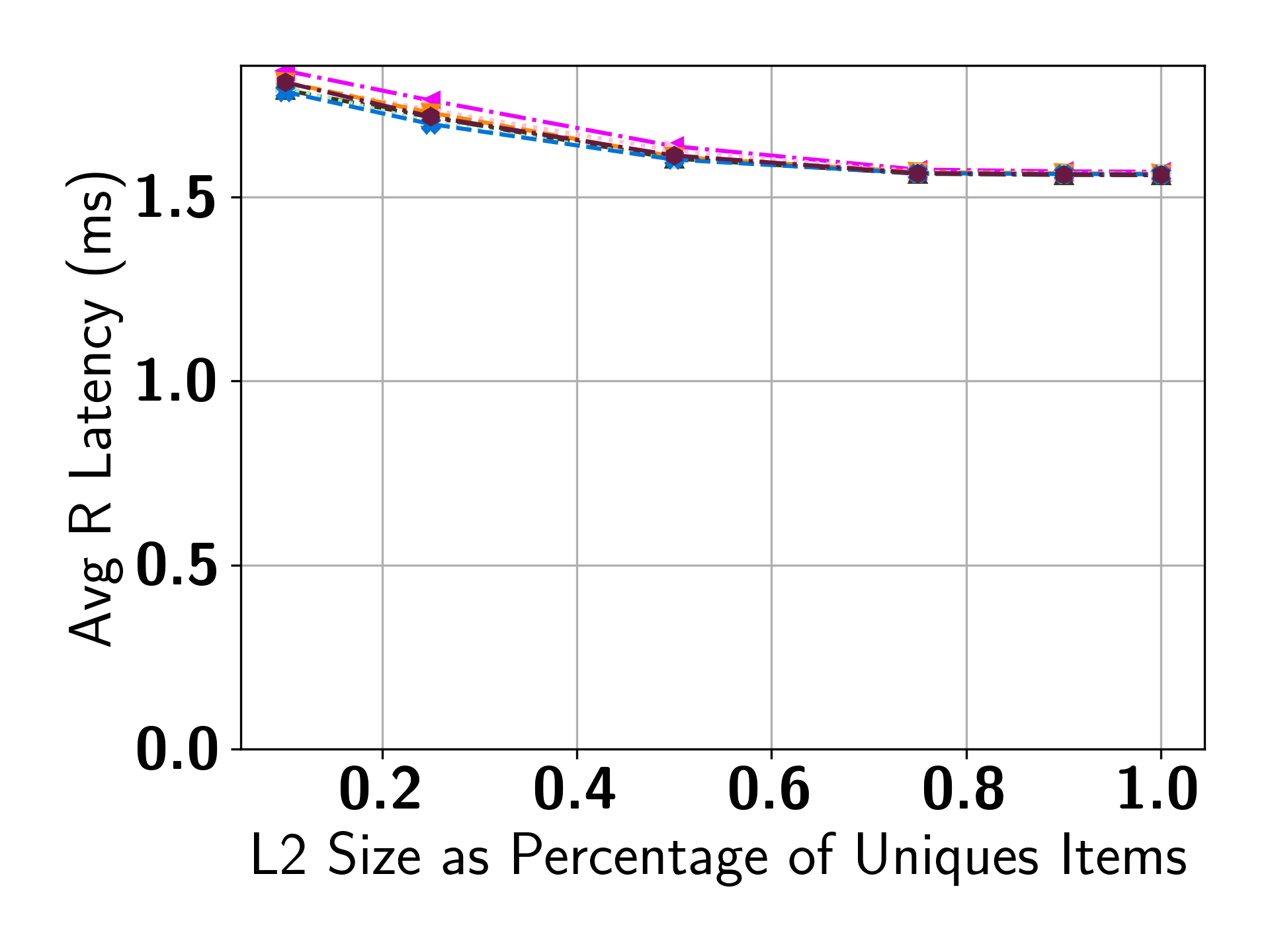} \\
		\includegraphics[trim=0 0 0 10, clip, height=\HeightL]{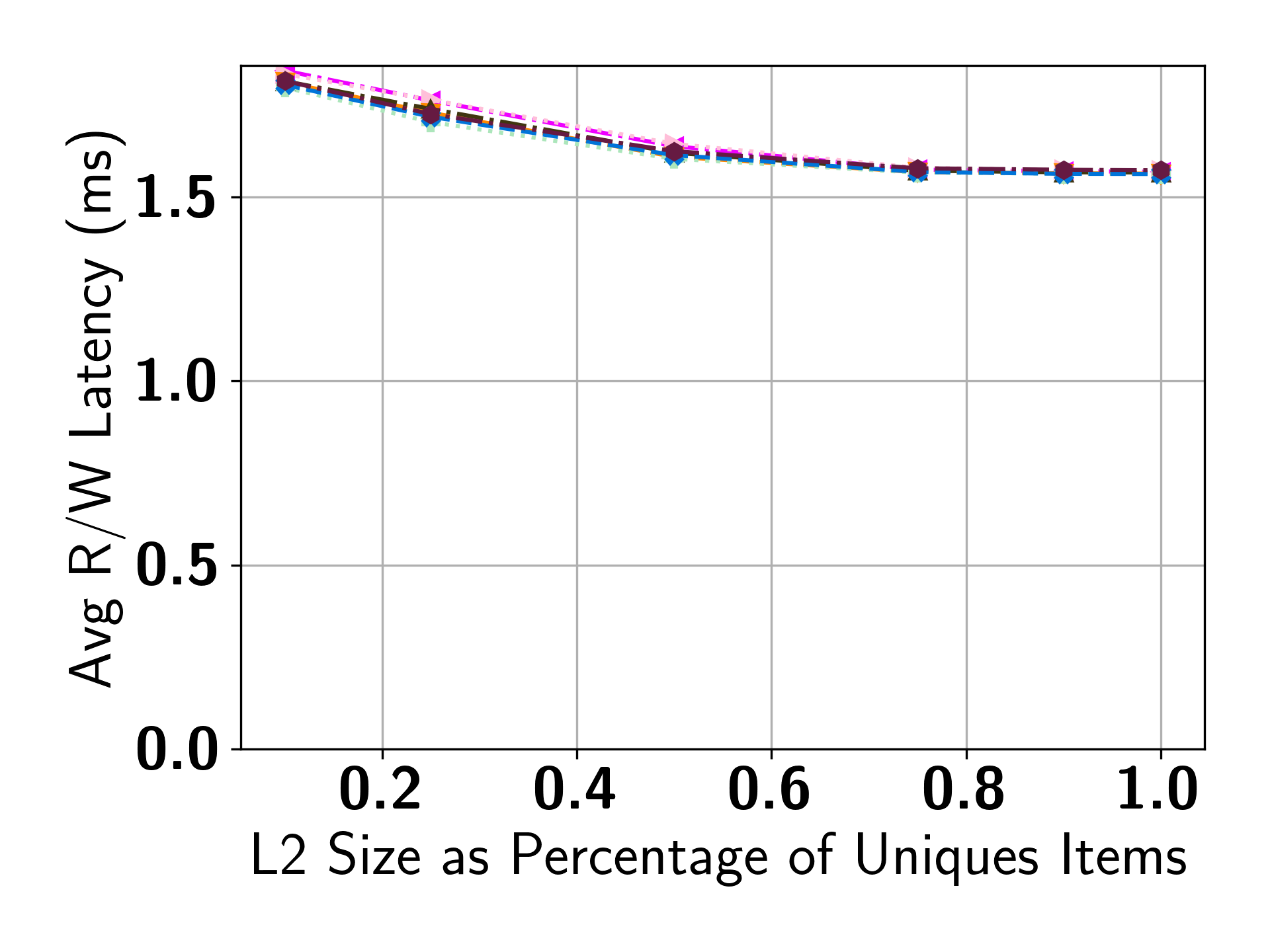}
	\end{tabular} \hspace{-0.7cm}
	}
	\subfloat[\normalfont{$L1:L2 = 1:100$ }]{ \begin{tabular}[b]{c}%
		\includegraphics[trim=84 0 0 10, clip, height=\HeightL]{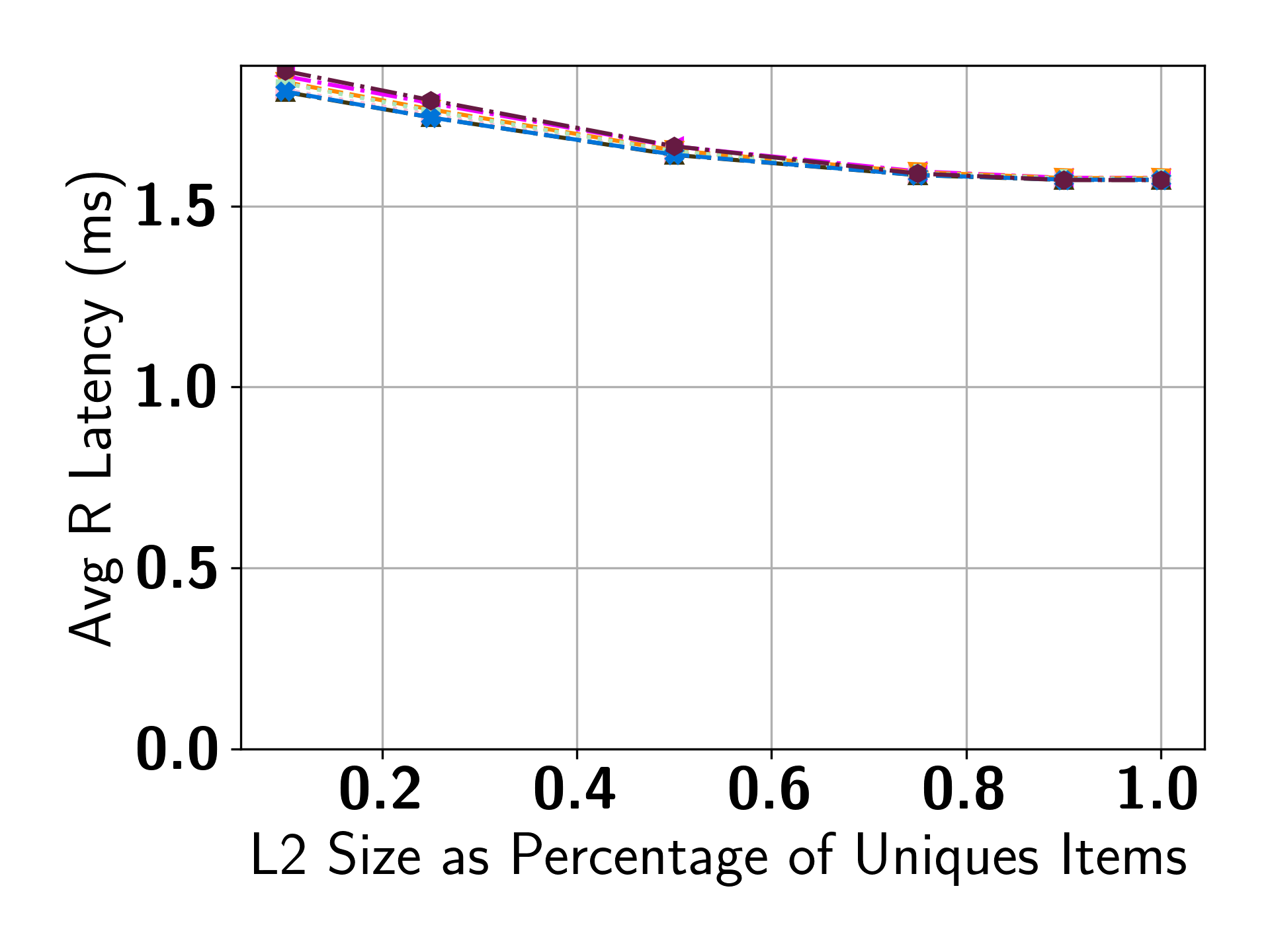} \\
		\includegraphics[trim=84 0 0 10, clip, height=\HeightL]{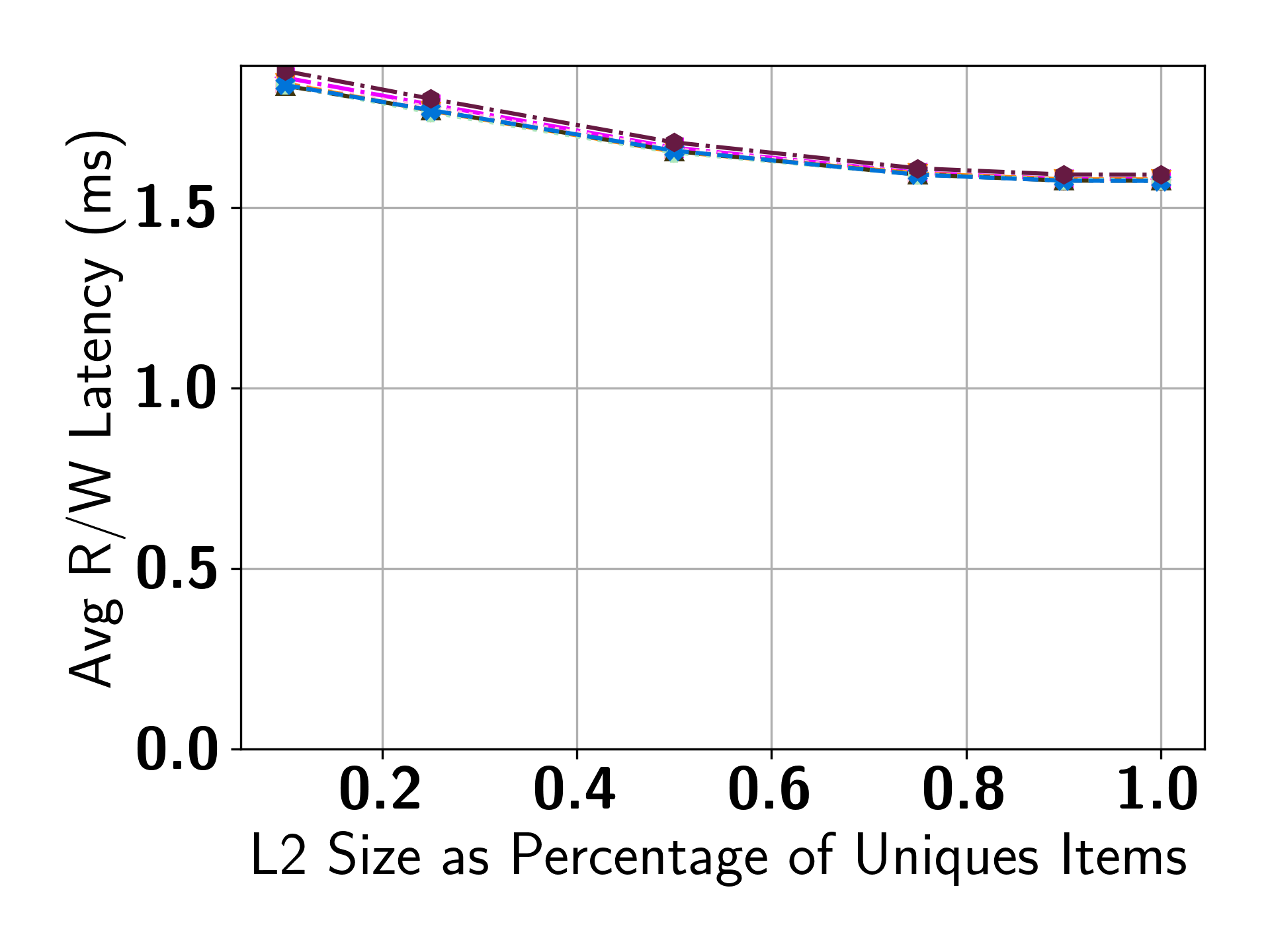}
	\end{tabular} \hspace{-0.7cm}
	}
	\subfloat[\normalfont{$L1:L2 = 1:10$ }]{ \begin{tabular}[b]{c}%
		\includegraphics[trim=50 0 0 10, clip, height=\HeightL]{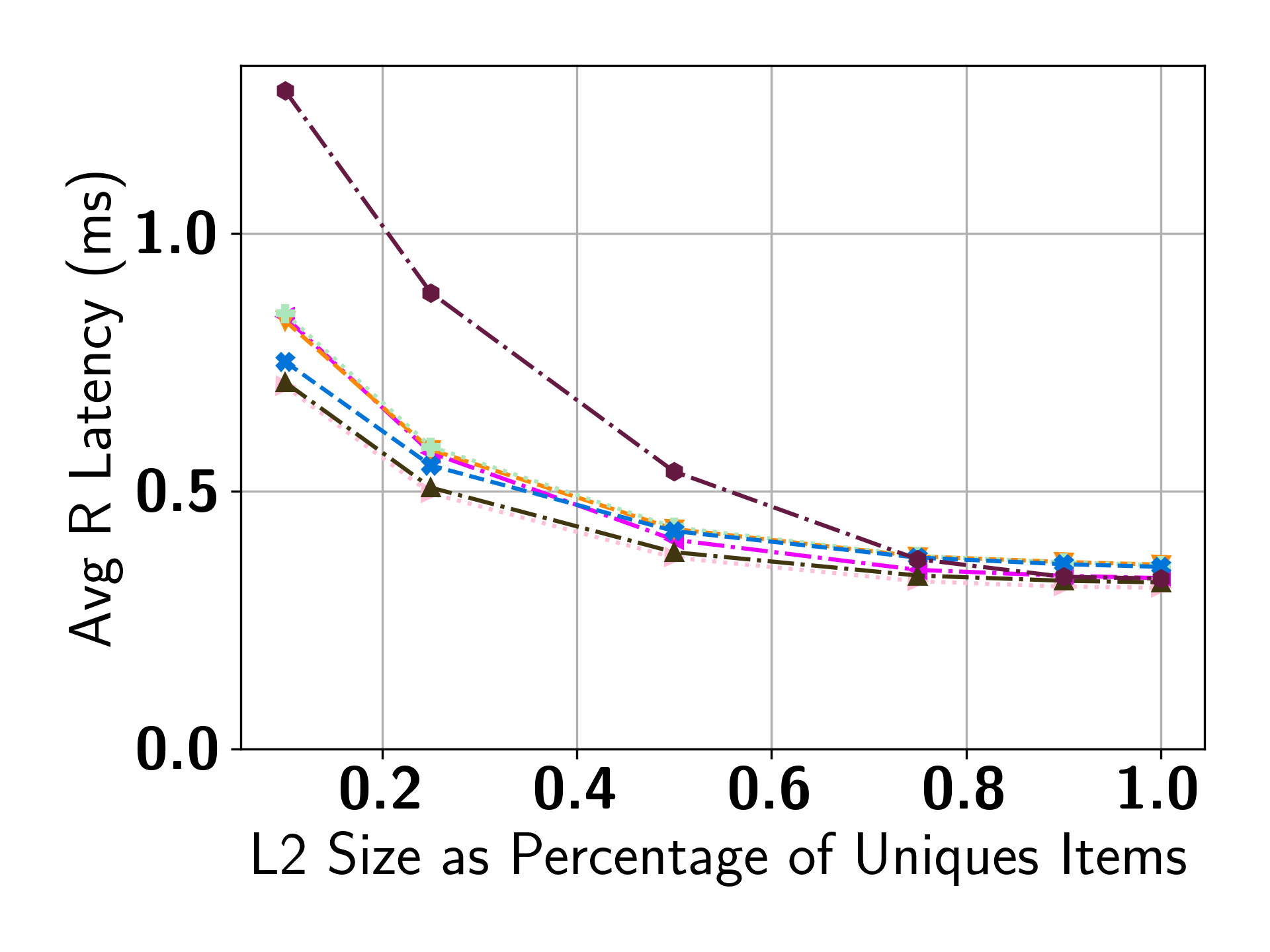} \\
		\includegraphics[trim=50 0 0 10, clip, height=\HeightL]{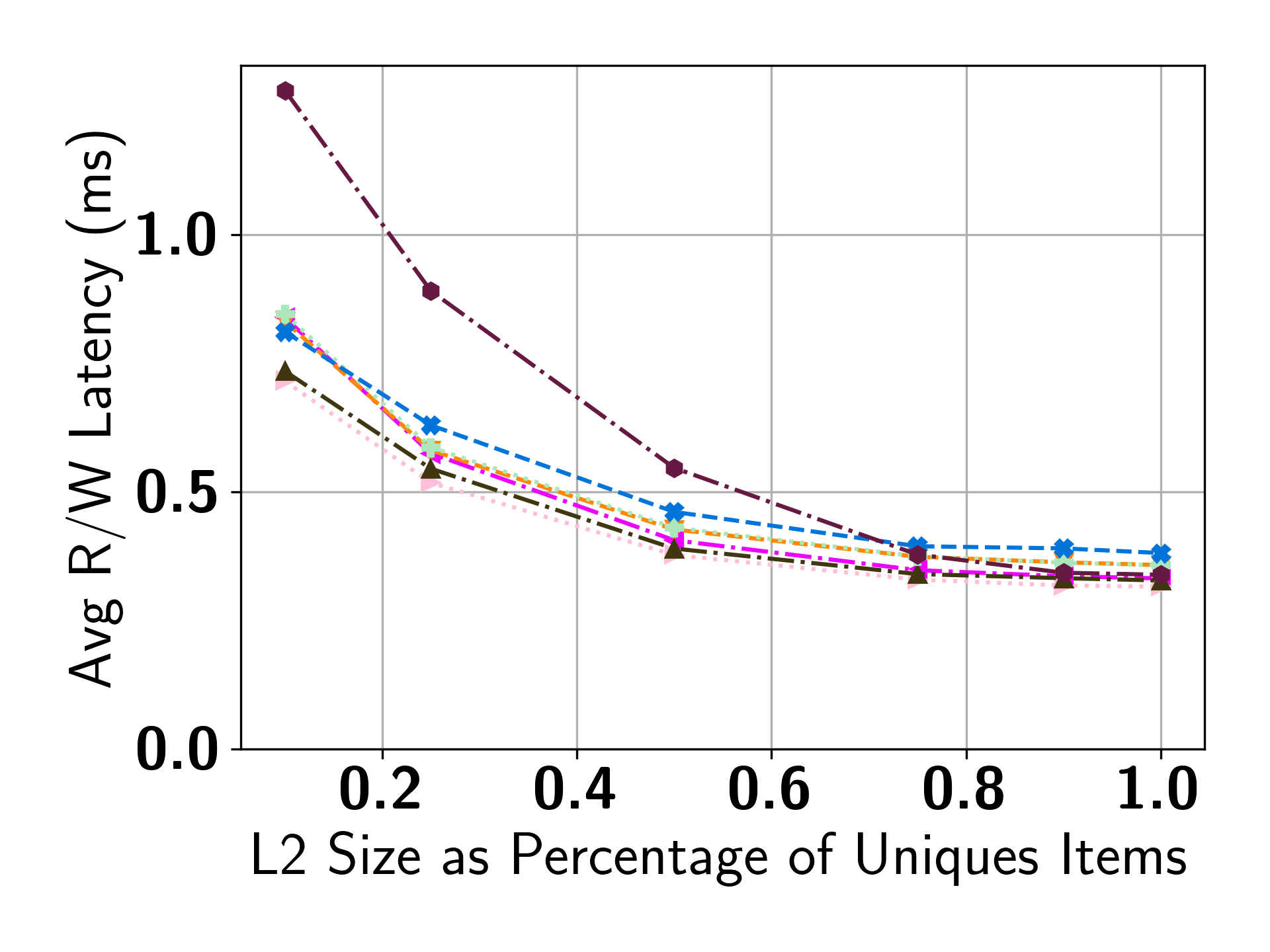}
	\end{tabular} \hspace{-0.7cm}
	}
	\subfloat[\normalfont{$L1:L2 = 1:100$ }]{ \begin{tabular}[b]{c}%
		\includegraphics[trim=50 0 0 10, clip, height=\HeightL]{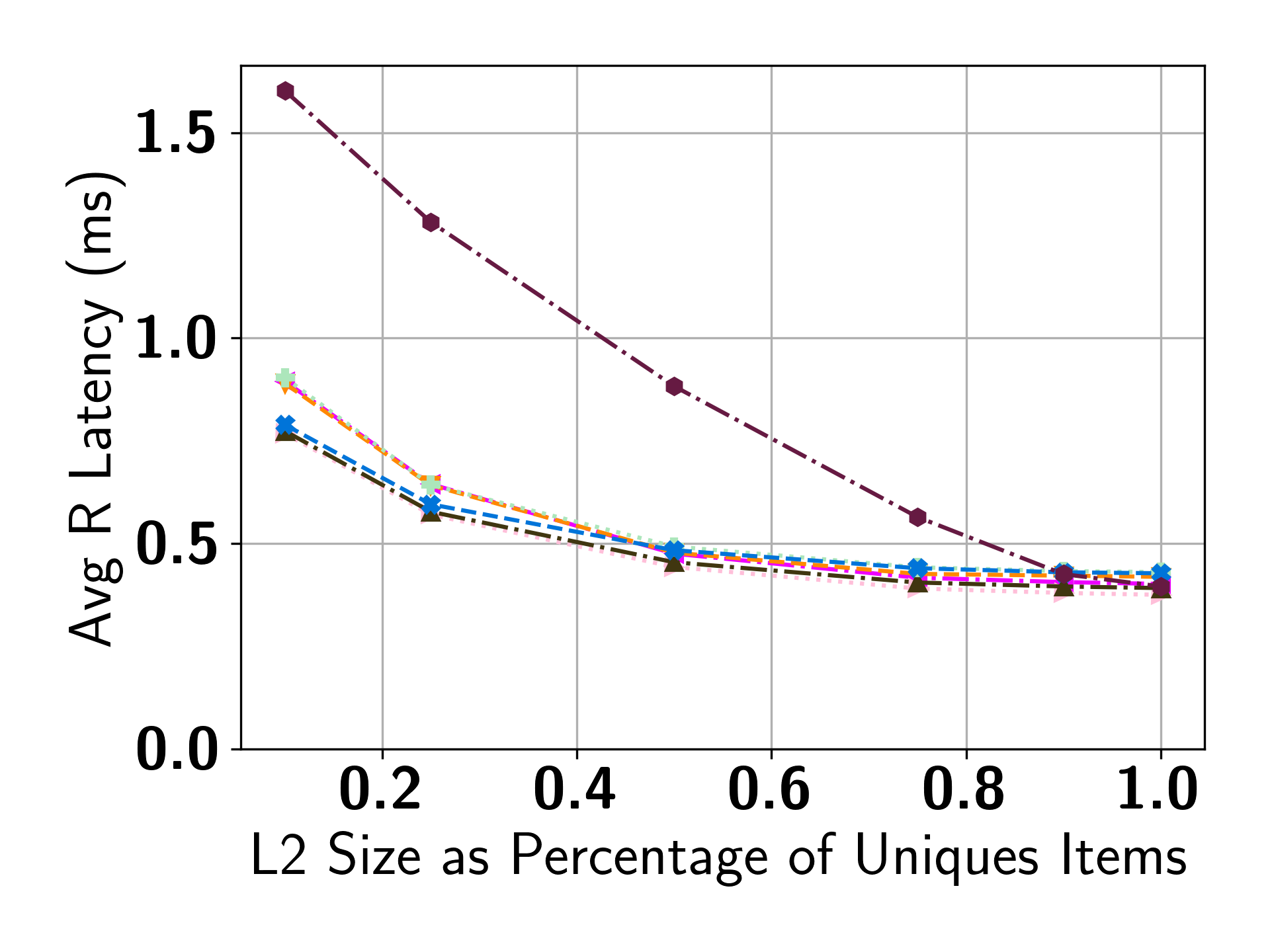} \\
		\includegraphics[trim=50 0 0 10, clip, height=\HeightL]{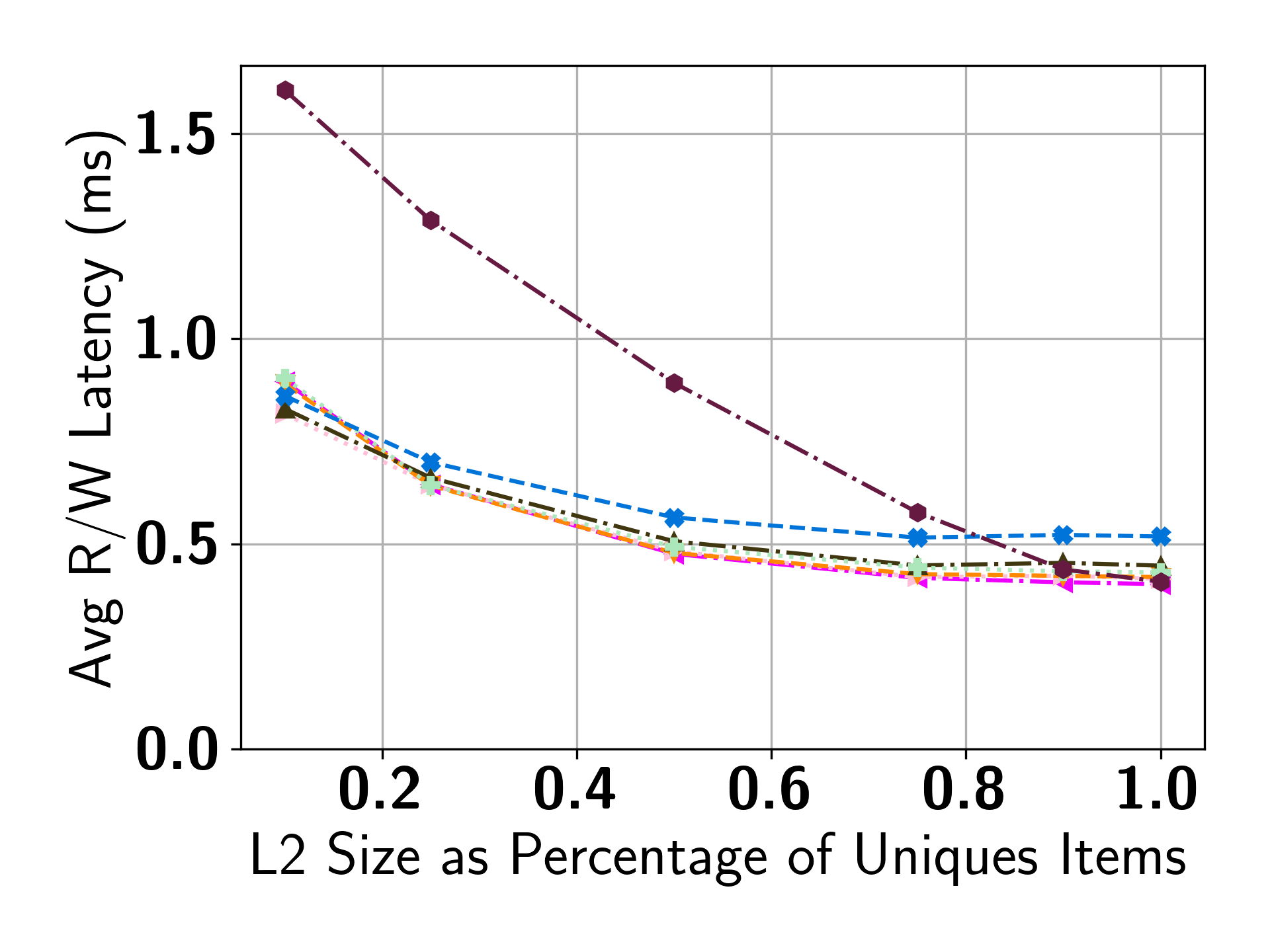}
	\end{tabular} \hspace{-0.7cm}
	}
	\\
	\vspace{0.2cm}
	\begin{tabularx}{\textwidth}{ >{\centering\arraybackslash}X >{\centering\arraybackslash}X  }
	TWITTER1 & TENCENT1 \\
	Hit-Ratio ranges from 65\%--93\% & Hit-Ratio ranges from 18\%--59\%	
	\end{tabularx} \\ 
	\subfloat[\normalfont{$L1:L2 = 1:10$ }]{ \begin{tabular}[b]{c}%
		\includegraphics[trim=0 0 0 10, clip, height=\HeightL]{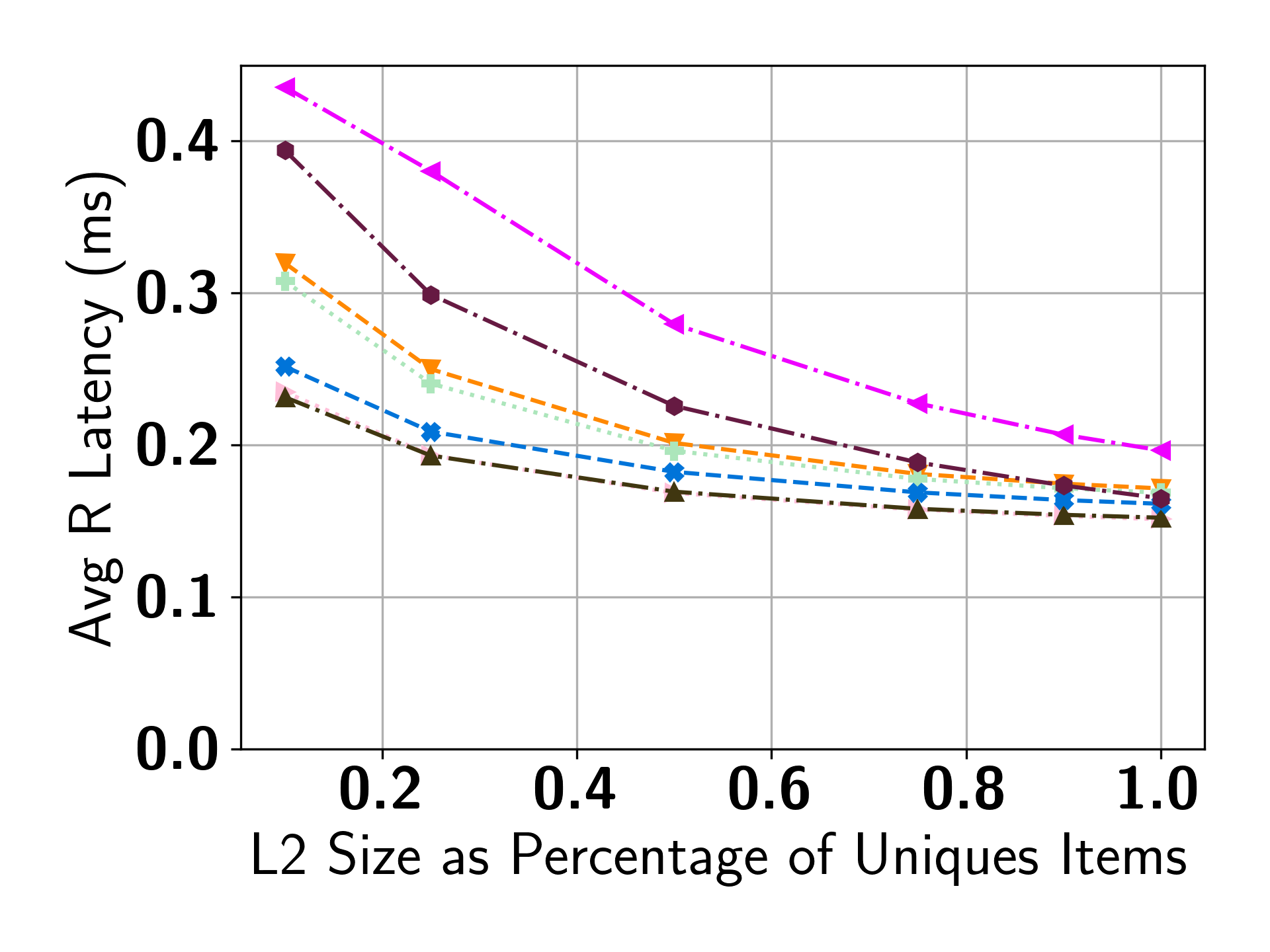} \\
		\includegraphics[trim=0 0 0 10, clip, height=\HeightL]{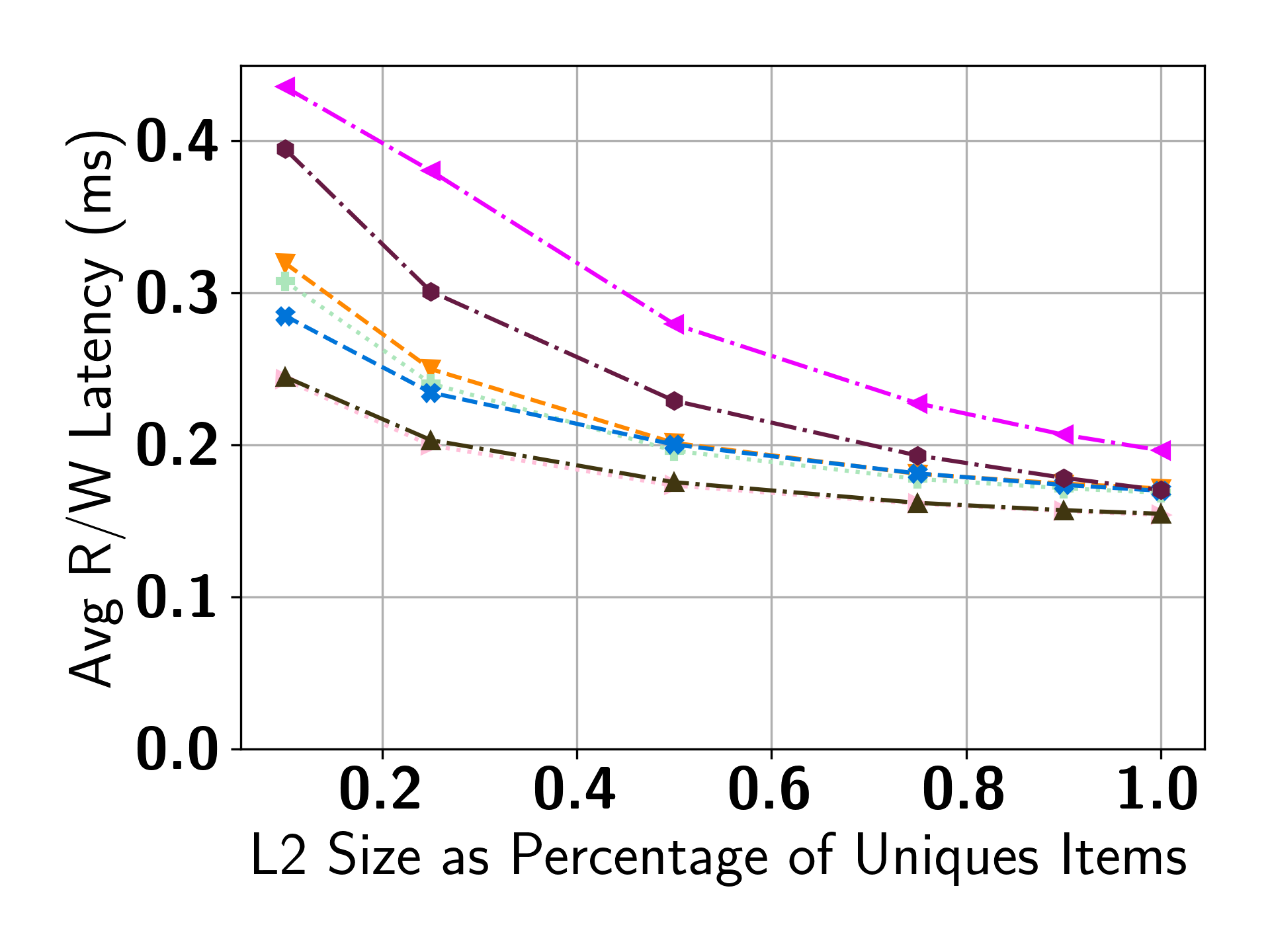}
	\end{tabular} \hspace{-0.7cm}
	}
	\subfloat[\normalfont{$L1:L2 = 1:100$ }]{ \begin{tabular}[b]{c}%
		\includegraphics[trim=50 0 0 10, clip, height=\HeightL]{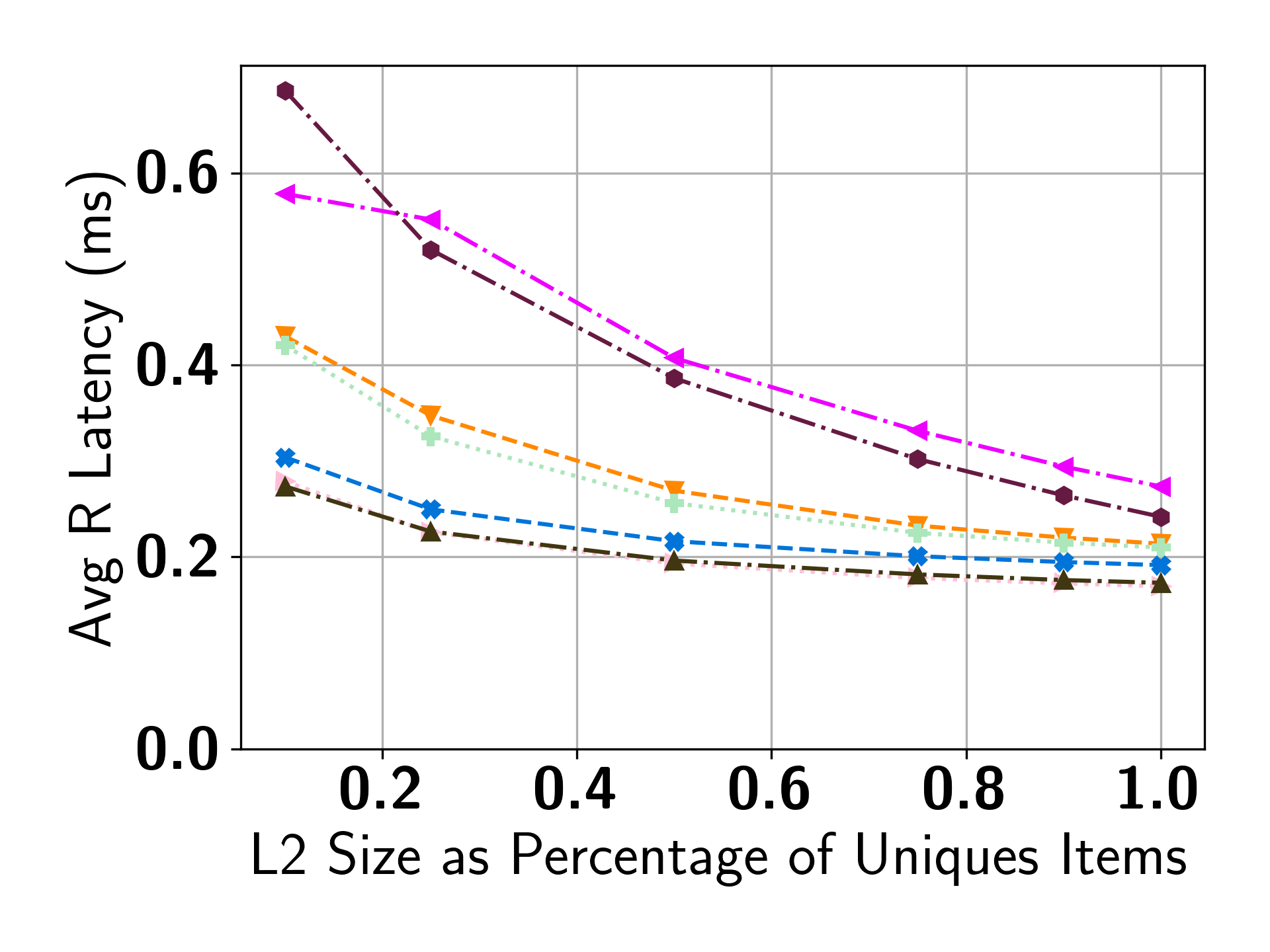} \\
		\includegraphics[trim=50 0 0 10, clip, height=\HeightL]{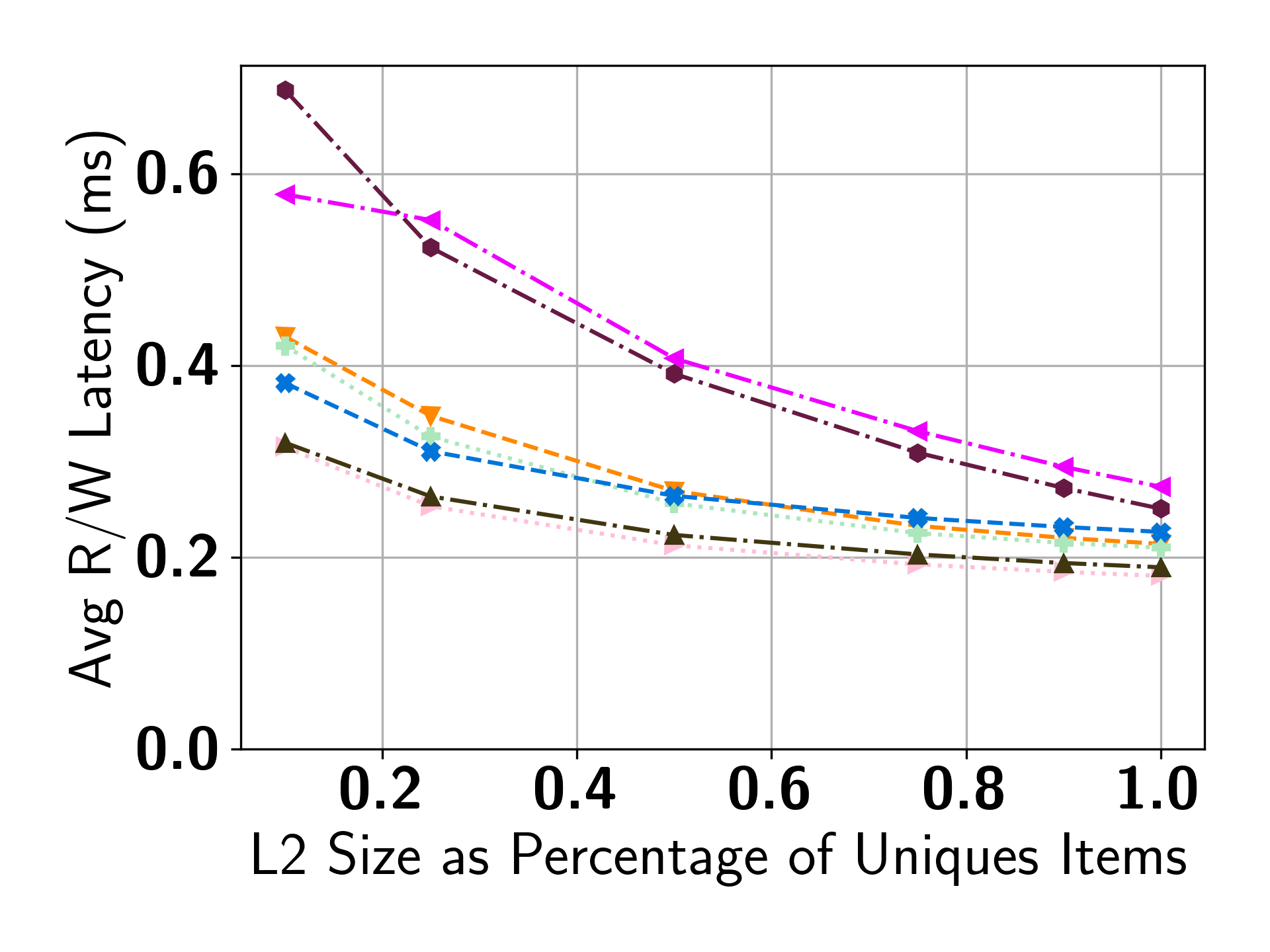}
	\end{tabular} \hspace{-0.7cm}
	}
	\subfloat[\normalfont{$L1:L2 = 1:10$ }]{ \begin{tabular}[b]{c}%
		\includegraphics[trim=50 0 0 10, clip, height=\HeightL]{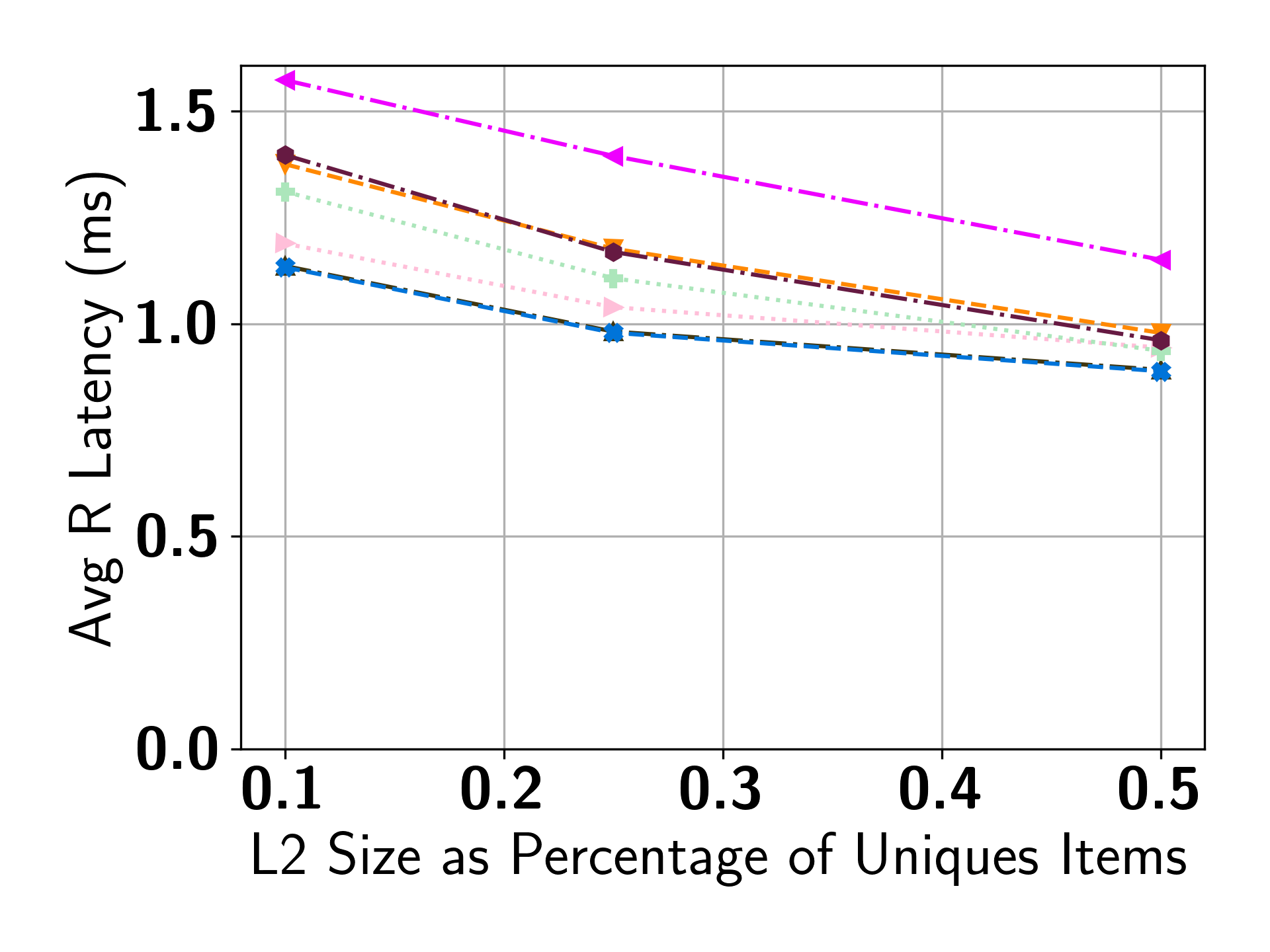} \\
		\includegraphics[trim=50 0 0 10, clip, height=\HeightL]{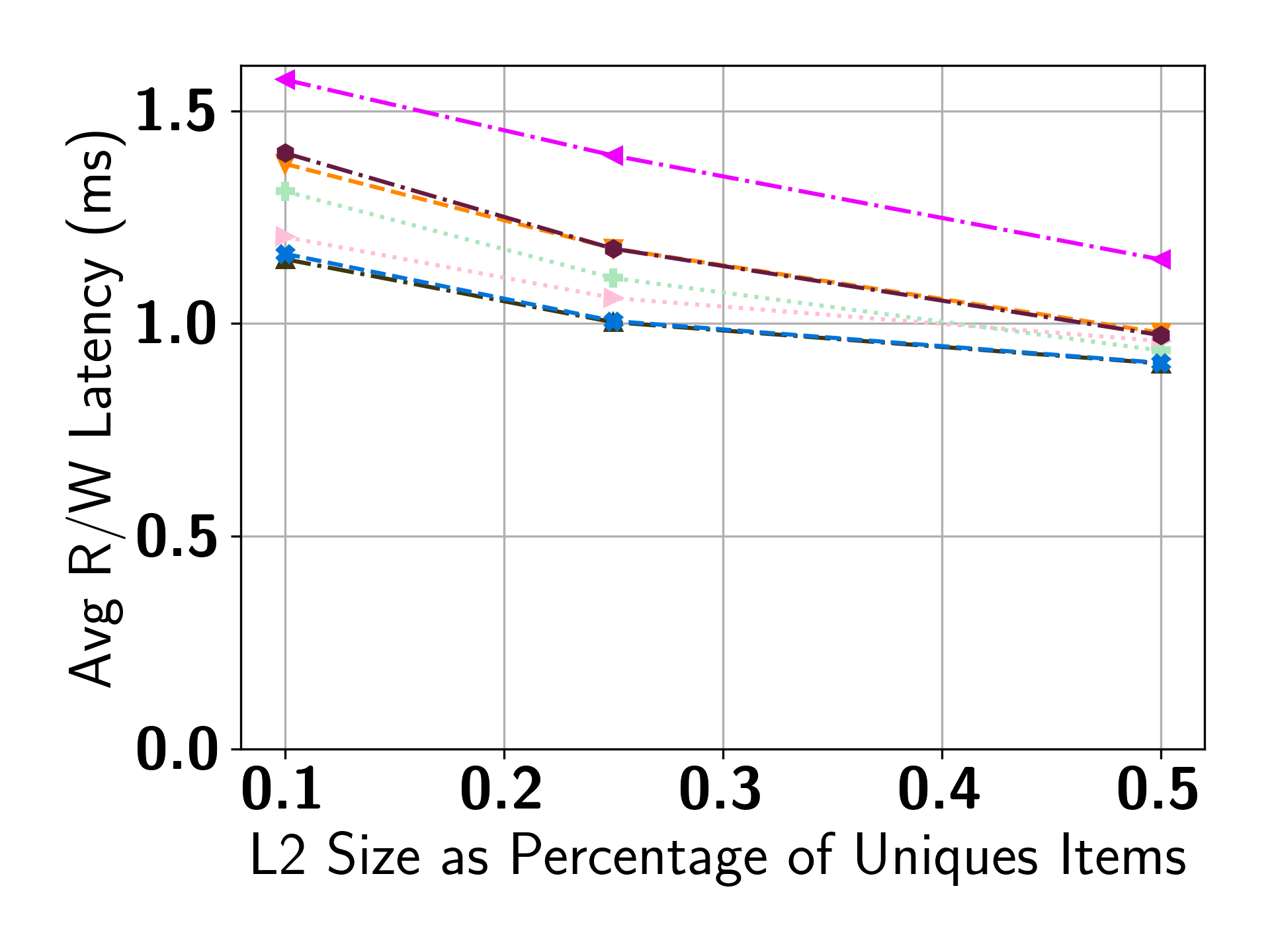}
	\end{tabular} \hspace{-0.7cm}
	}
	\subfloat[\normalfont{$L1:L2 = 1:100$ }]{ \begin{tabular}[b]{c}%
		\includegraphics[trim=84 0 0 10, clip, height=\HeightL]{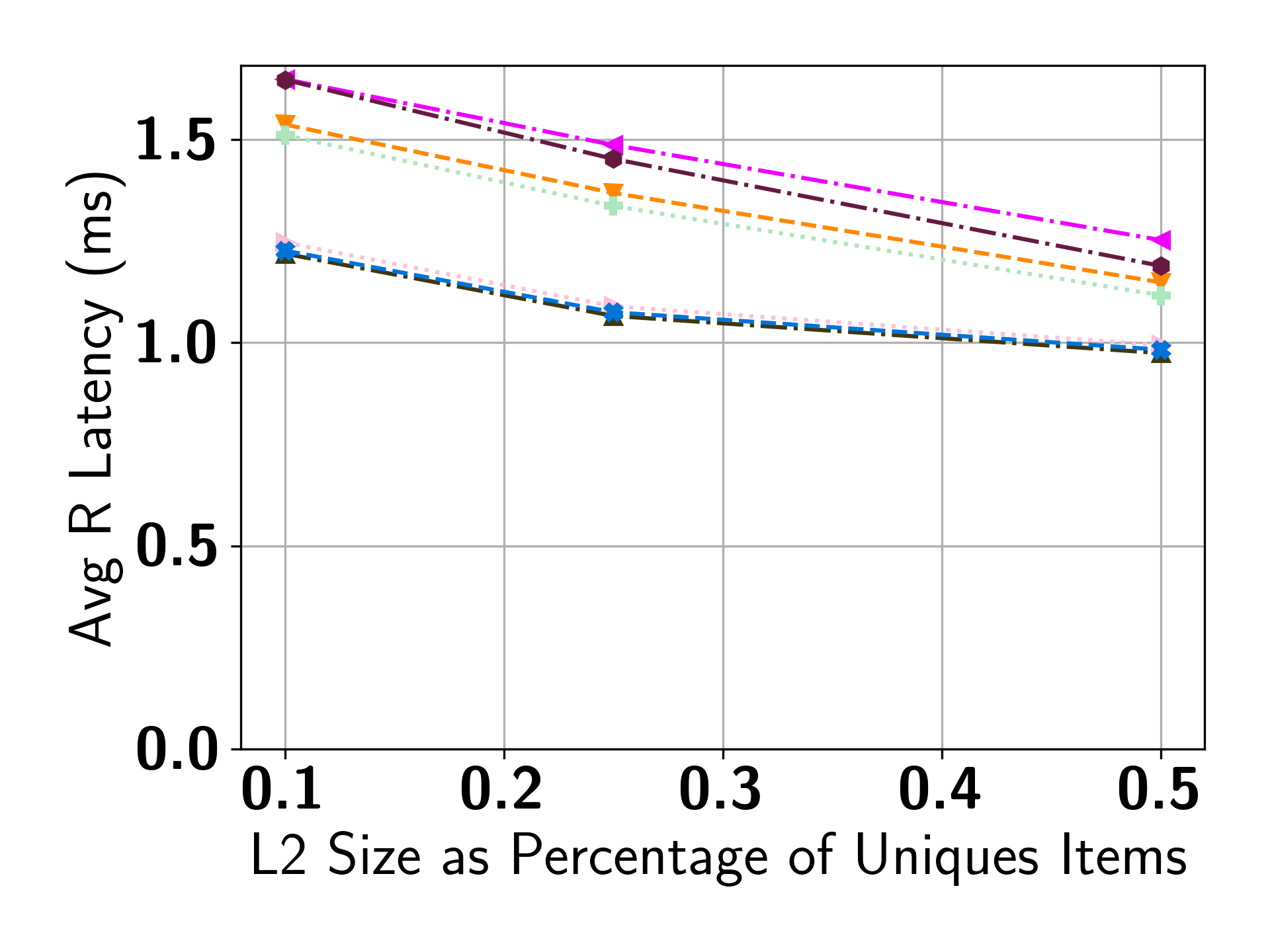} \\
		\includegraphics[trim=84 0 0 10, clip, height=\HeightL]{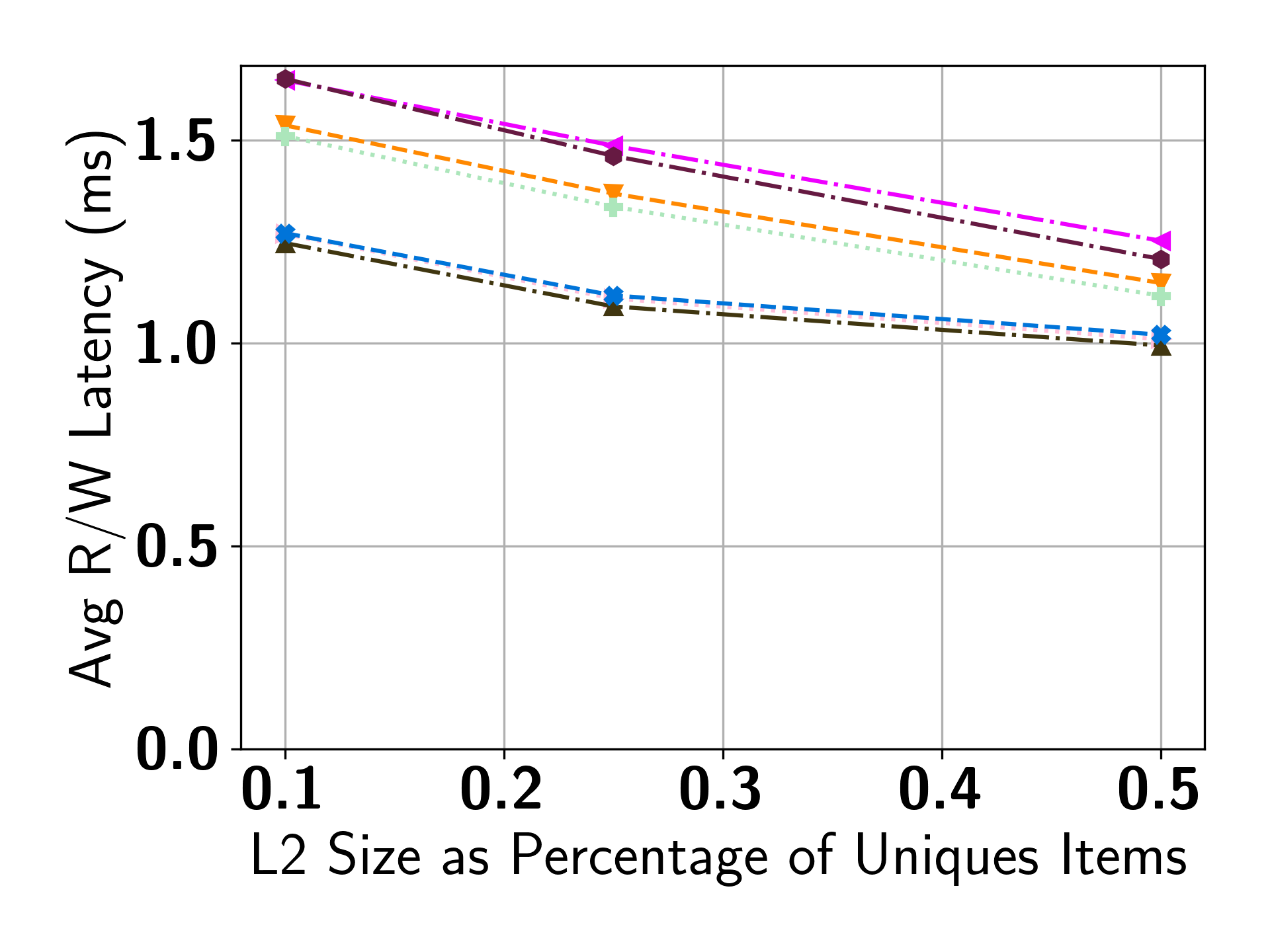}
	\end{tabular} \hspace{-0.7cm}
	}
	\\
	\centering \includegraphics[height=1.3cm]{Plots/our-legend-lines.png}
	\caption{Average latency per request, assuming latencies of 2ms, 200us and 100ns for disk, L2, and L1 respectively, for multiple traces and multiple ratios between L1 and L2}
	\label{eva:between:latency}
\end{figure*}

\subsection{Comparison with other policies}
\begin{figure*}[h]
	\centering SYSTOR1 \\
	\subfloat[\normalfont{$L1:L2 = 1:10$ }]{
		\includegraphics[trim=0 0 0 10, clip, height=\Height]{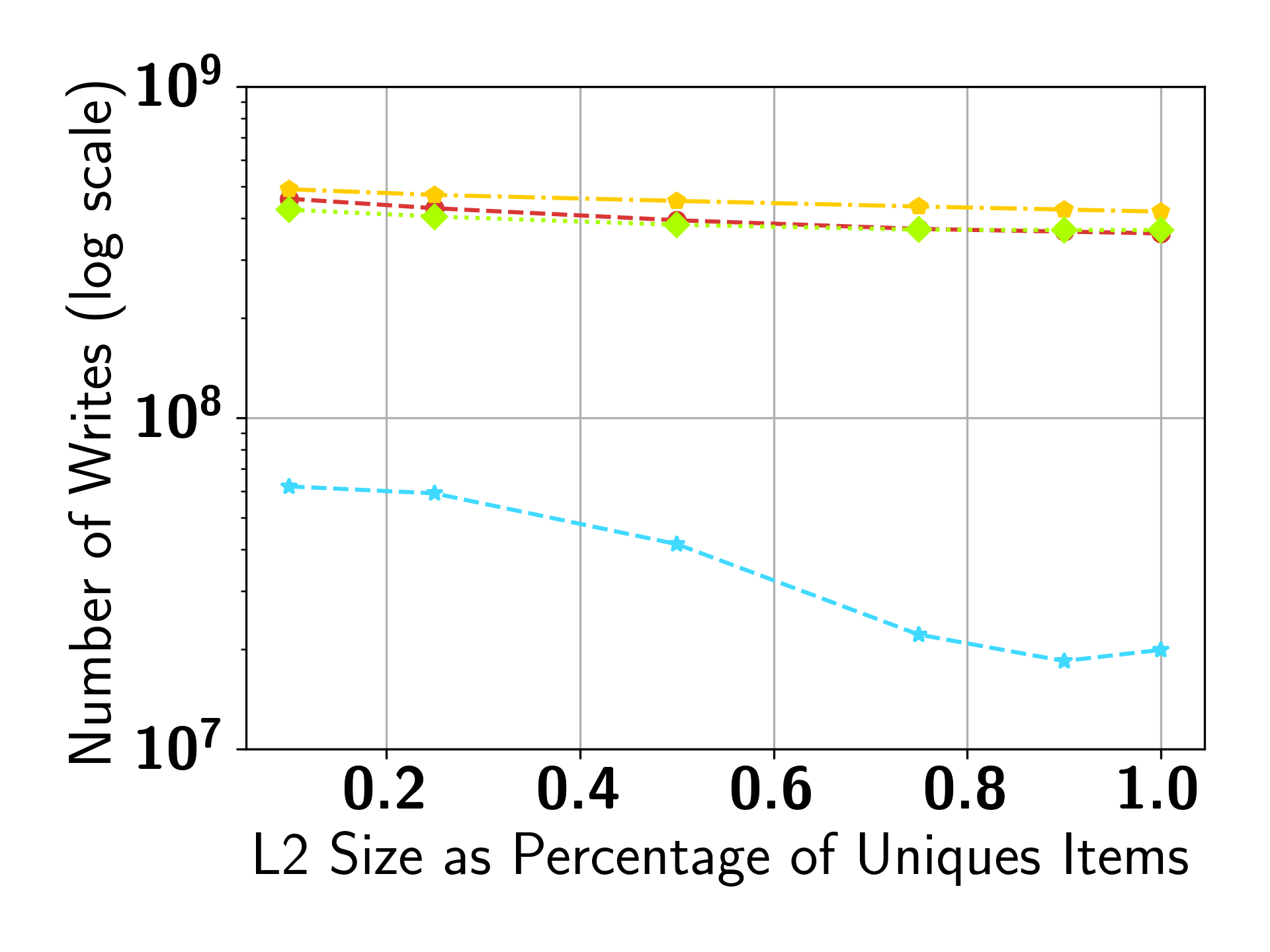}
	}
	\subfloat[\normalfont{$L1:L2 = 1:20$ }]{
		\includegraphics[trim=50 0 0 10, clip, height=\Height]{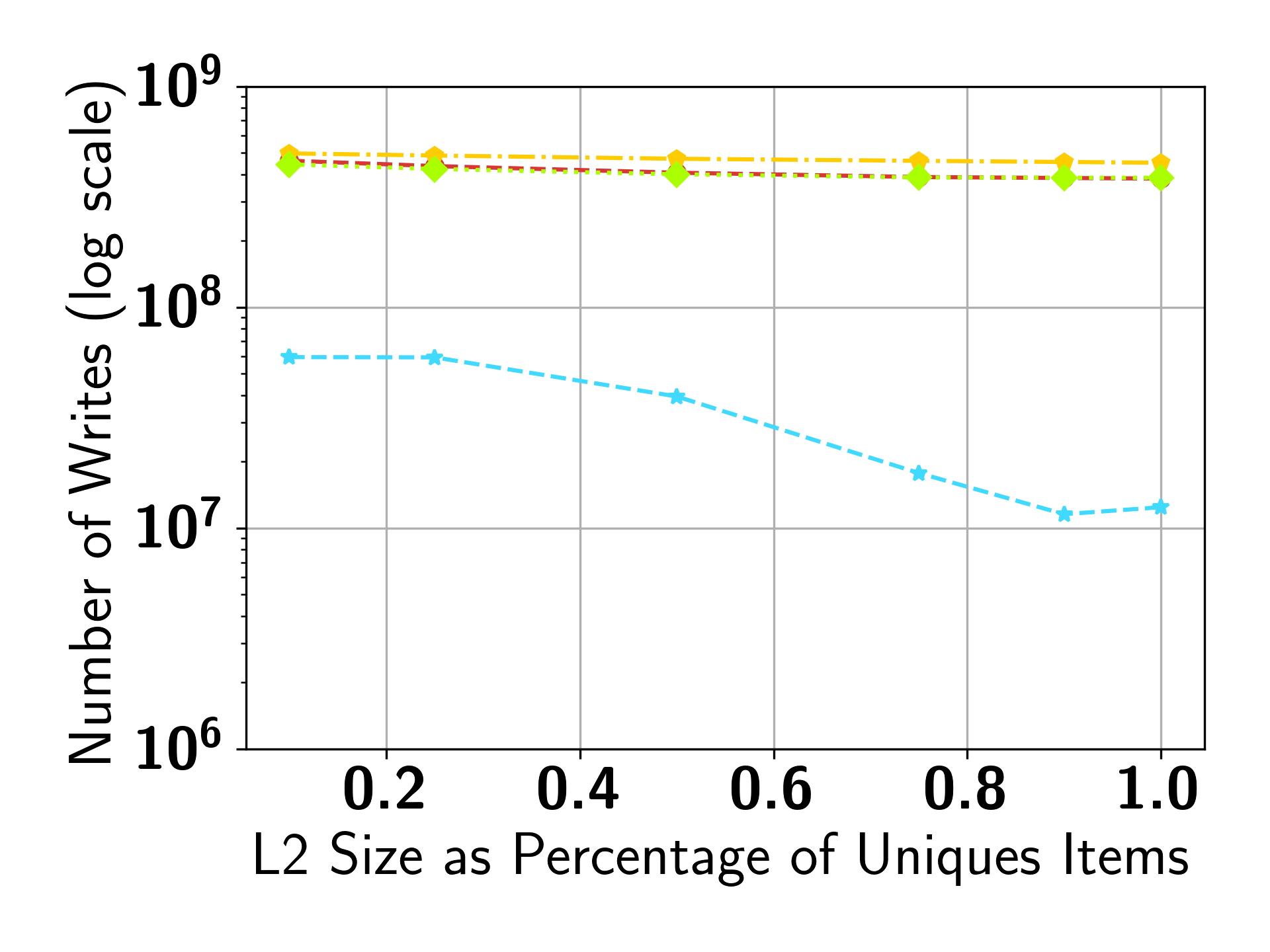}
	}
	\subfloat[\normalfont{$L1:L2 = 1:50$ }]{
		\includegraphics[trim=86 0 0 10, clip, height=\Height]{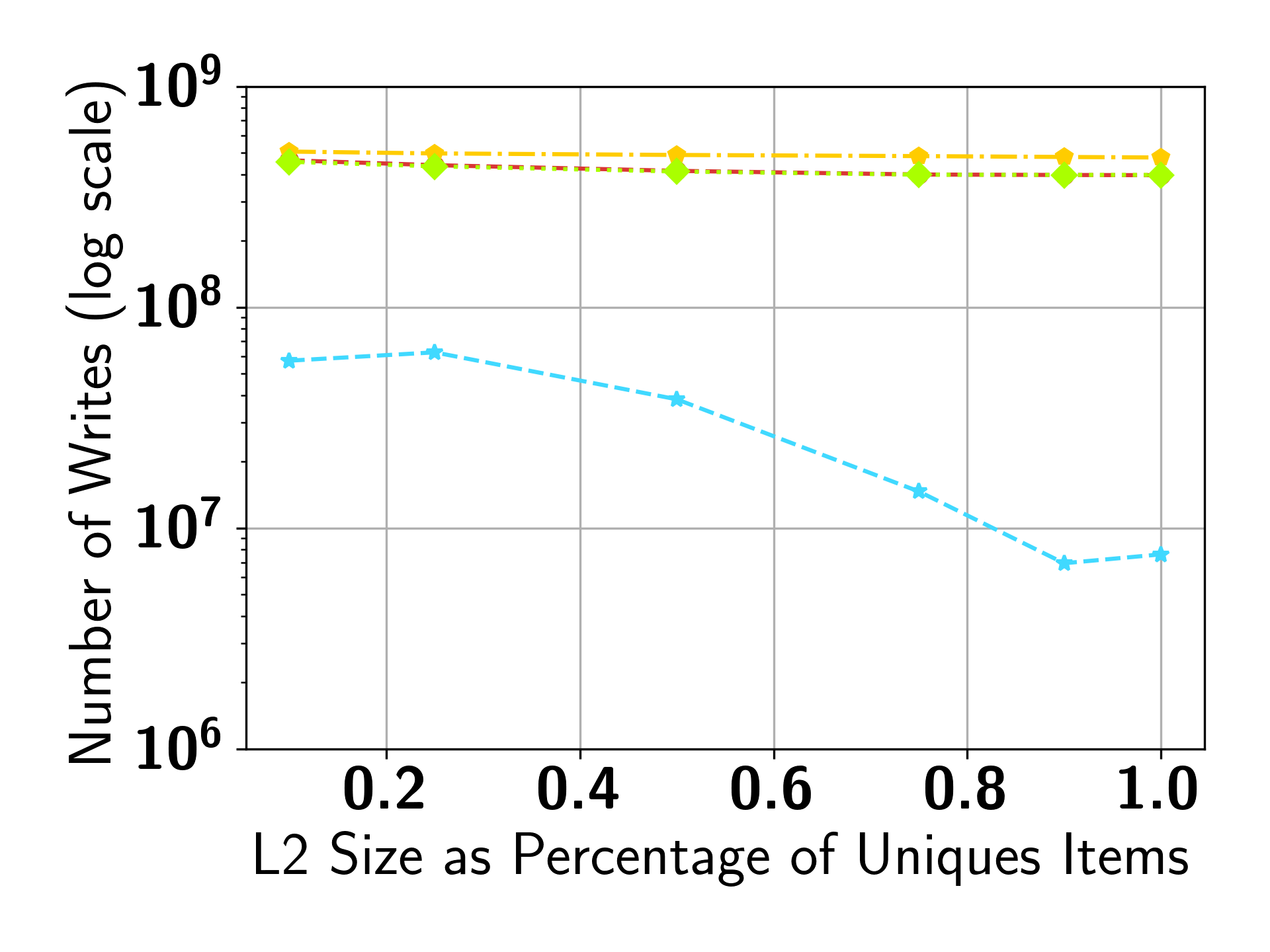}
	}
	\subfloat[\normalfont{$L1:L2 = 1:100$ }]{
		\includegraphics[trim=86 0 0 10, clip, height=\Height]{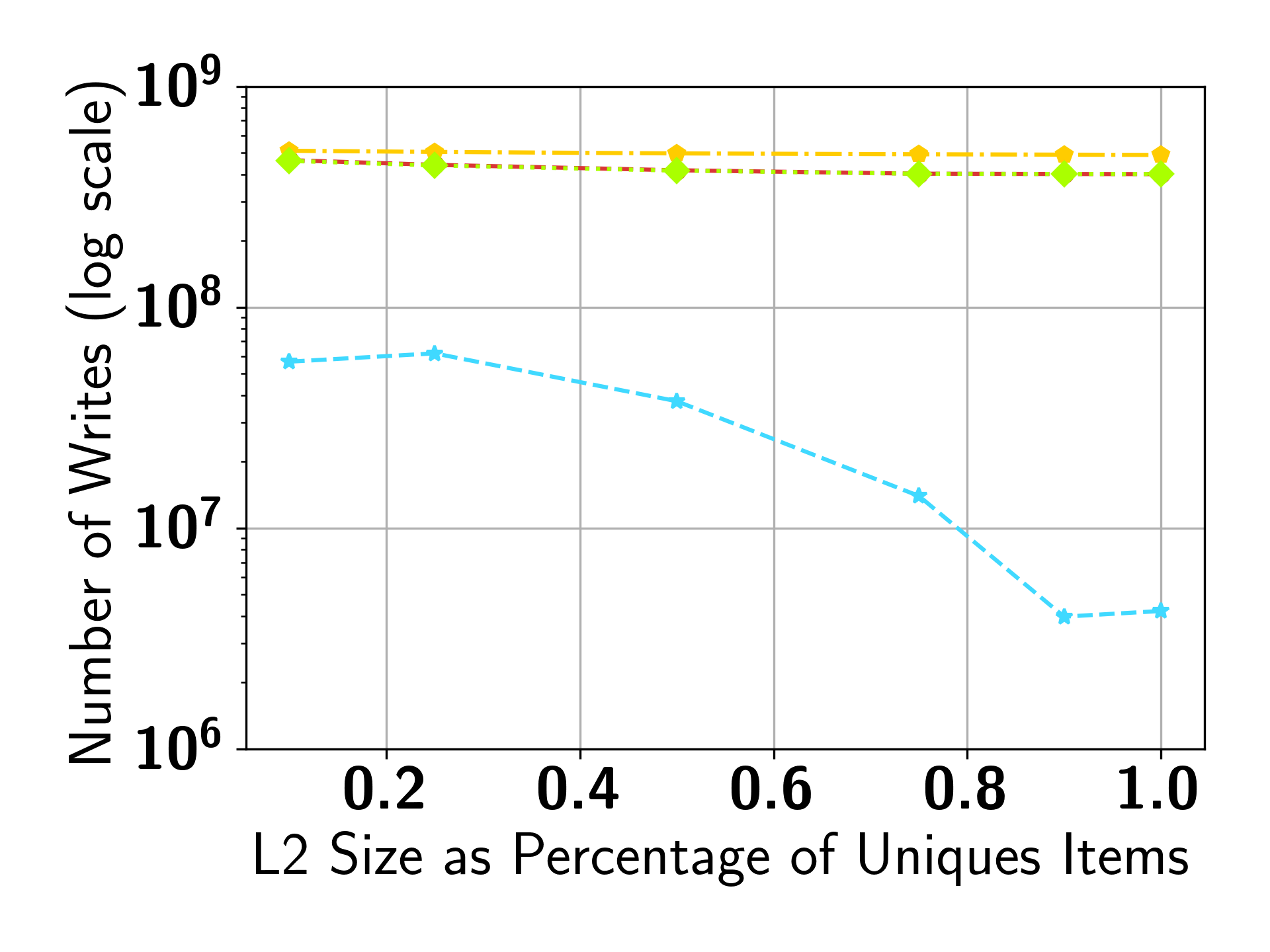}
	}
	\\
	\vspace{0.2cm}
	\centering CDN1 \\
	\vspace{-0.2cm}
	\subfloat[\normalfont{$L1:L2 = 1:10$ }]{
		\includegraphics[trim=0 0 0 10, clip, height=\Height]{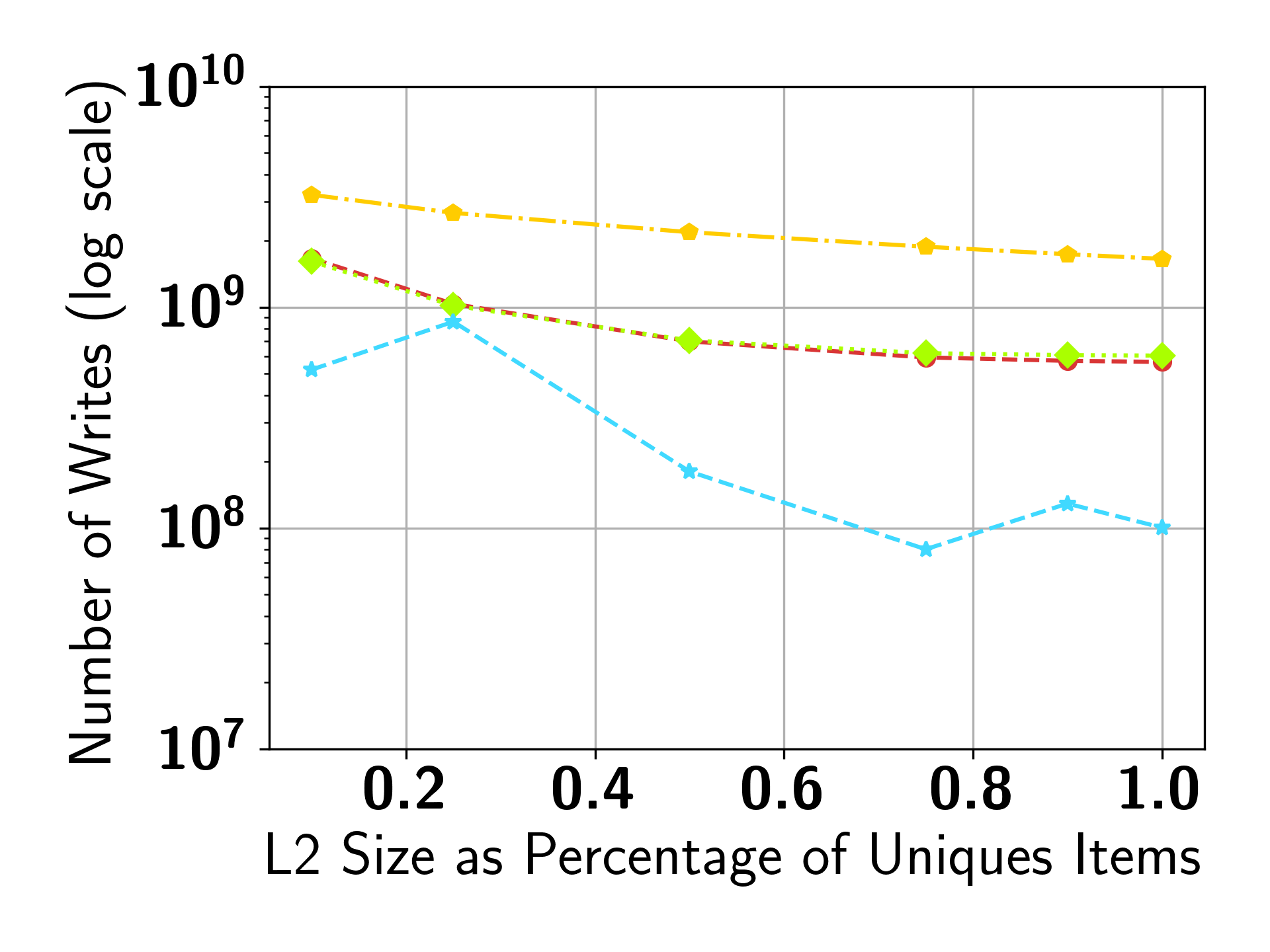}
	}
	\subfloat[\normalfont{$L1:L2 = 1:20$ }]{
		\includegraphics[trim=50 0 0 10, clip, height=\Height]{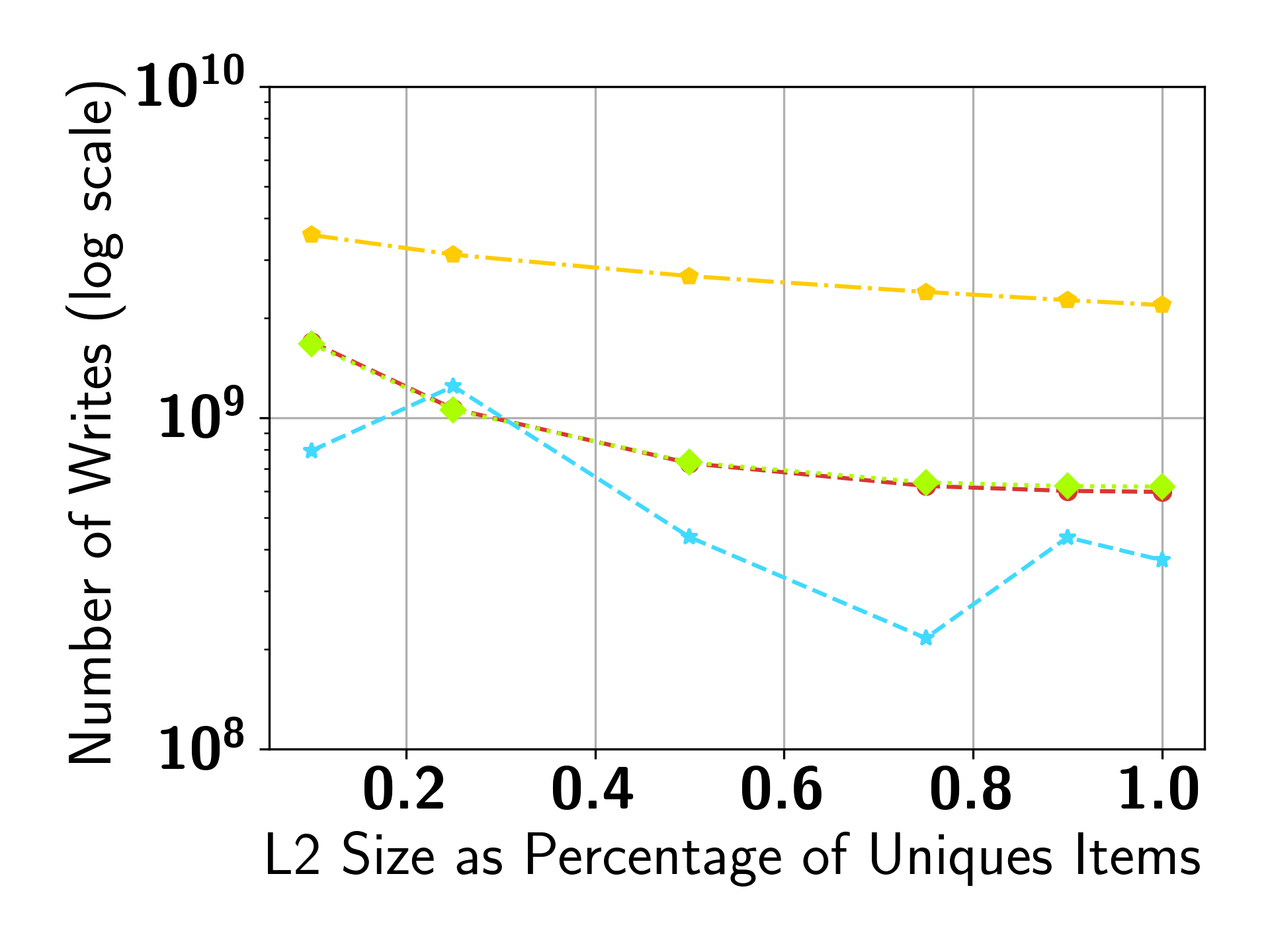}
	}
	\subfloat[\normalfont{$L1:L2 = 1:50$ }]{
		\includegraphics[trim=86 0 0 10, clip, height=\Height]{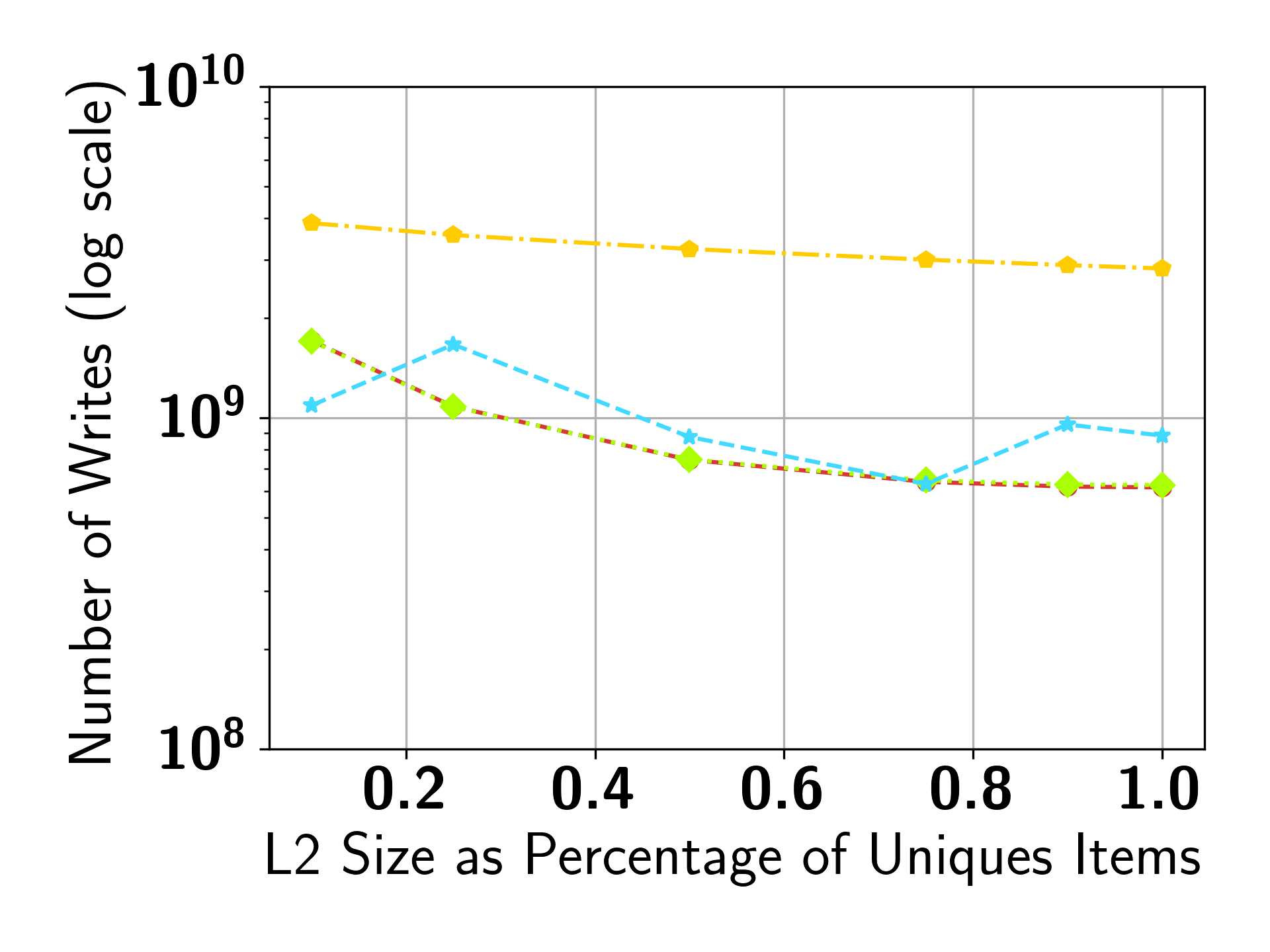}
	}
	\subfloat[\normalfont{$L1:L2 = 1:100$ }]{
		\includegraphics[trim=86 0 0 10, clip, height=\Height]{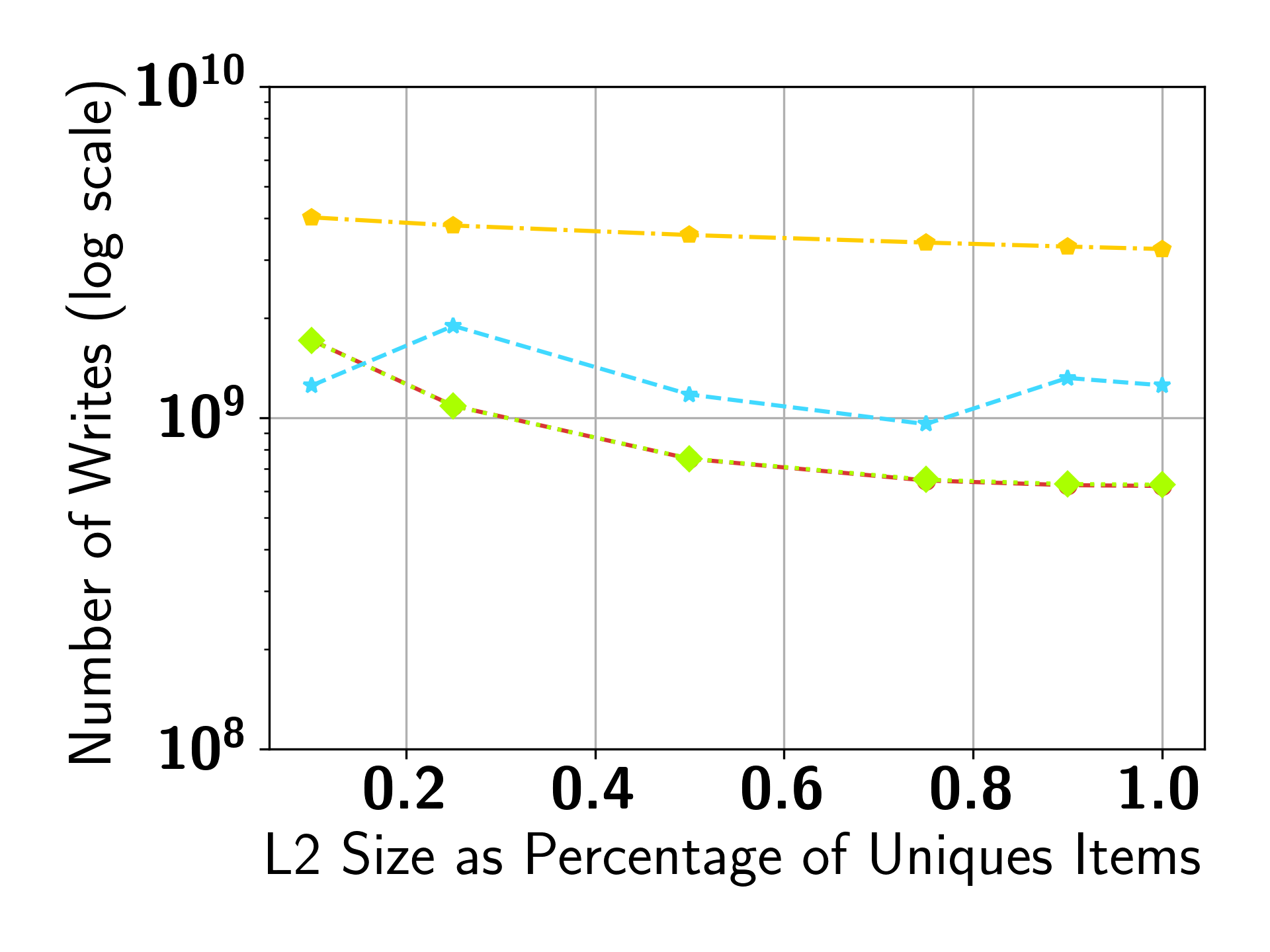}
	}
	\\
	\vspace{0.2cm}
	\centering TWITTER1 \\
	\vspace{-0.2cm}
	\subfloat[\normalfont{$L1:L2 = 1:10$ }]{
		\includegraphics[trim=0 0 0 10, clip, height=\Height]{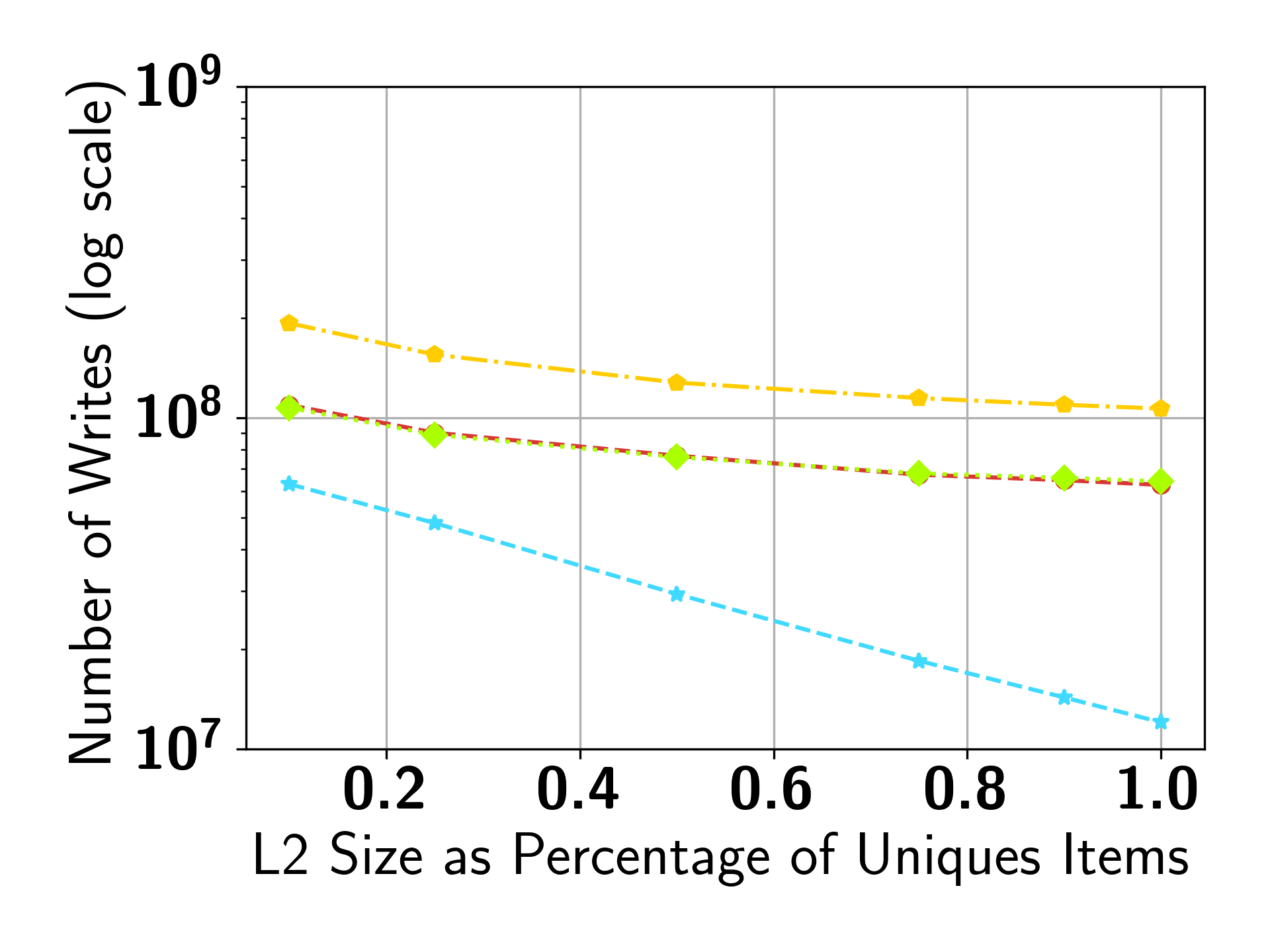}
	}
	\subfloat[\normalfont{$L1:L2 = 1:20$ }]{
		\includegraphics[trim=86 0 0 10, clip, height=\Height]{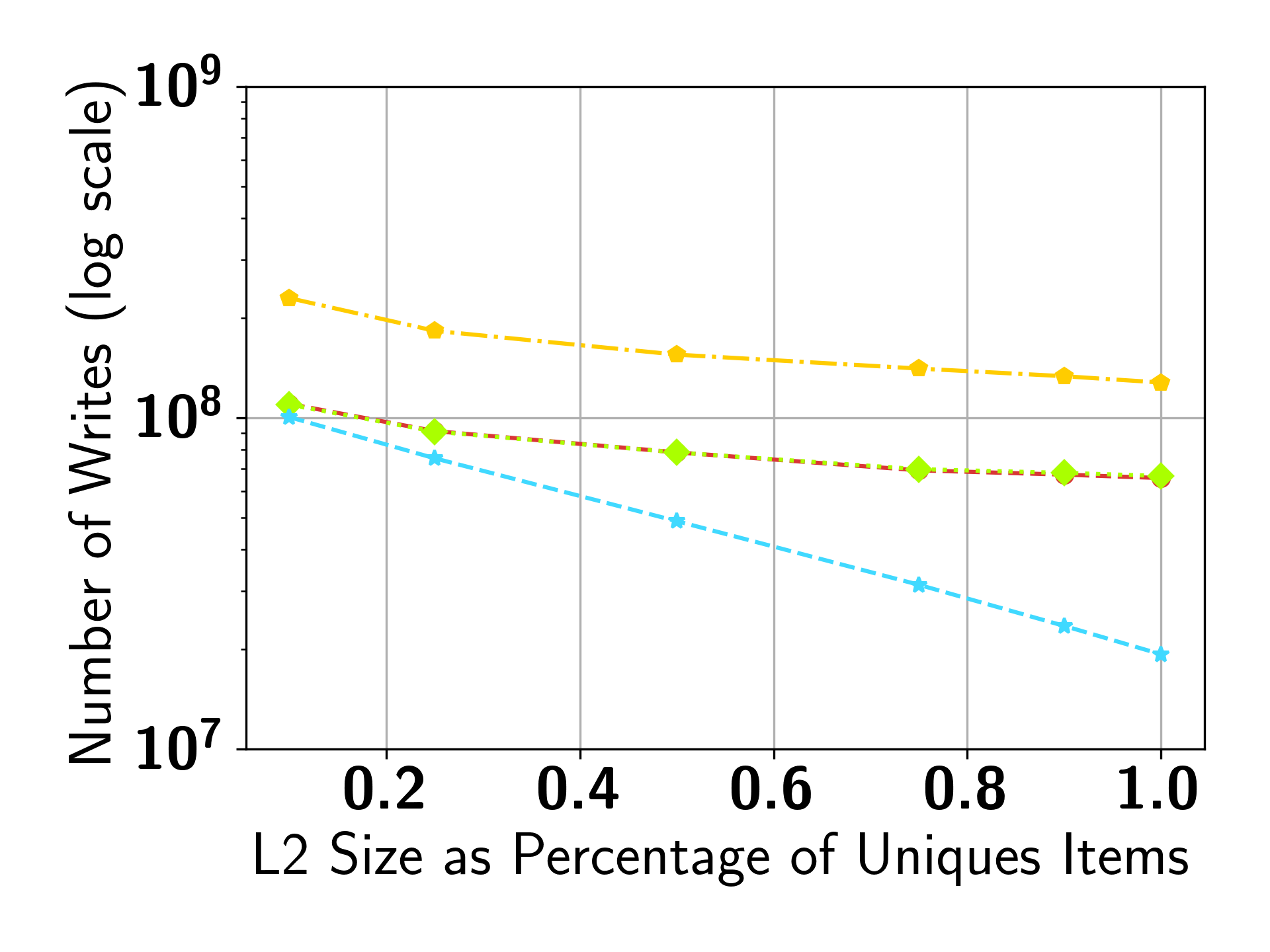}
	}
	\subfloat[\normalfont{$L1:L2 = 1:50$ }]{
		\includegraphics[trim=86 0 0 10, clip, height=\Height]{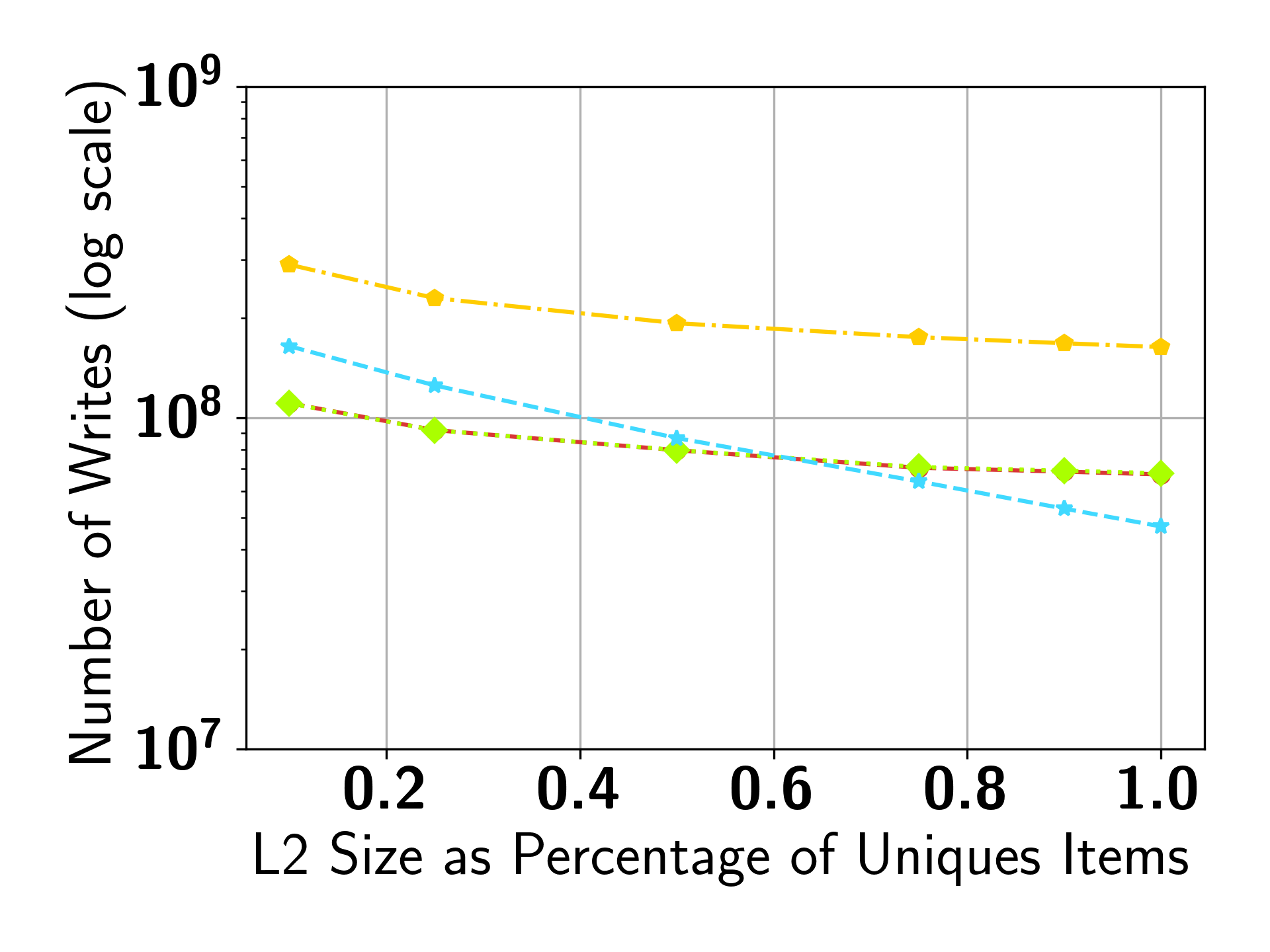}
	}
	\subfloat[\normalfont{$L1:L2 = 1:100$ }]{
		\includegraphics[trim=86 0 0 10, clip, height=\Height]{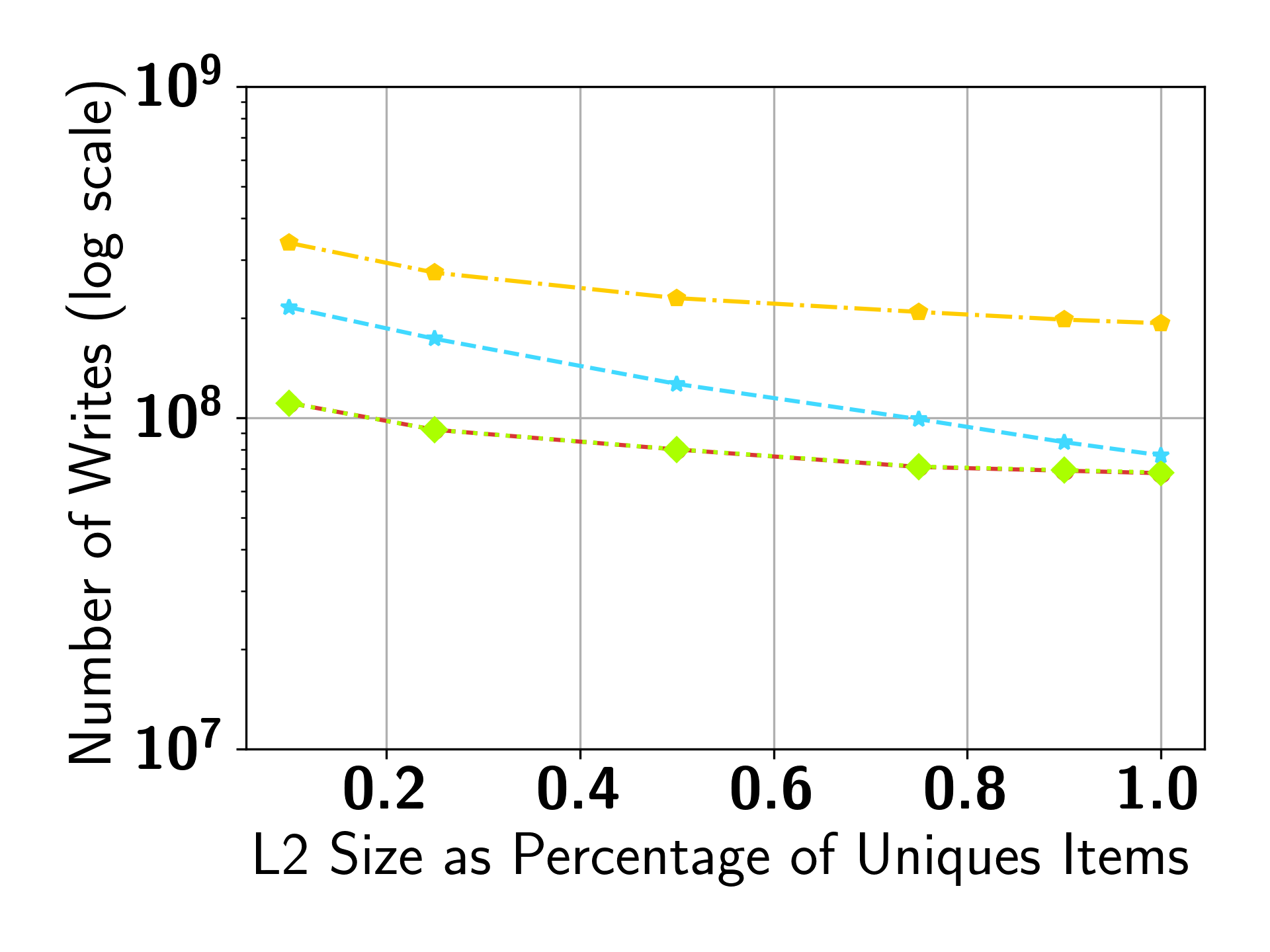}
	}
	\\
	\vspace{0.2cm}
	\centering TENCENT1 \\
	\vspace{-0.2cm}
	\subfloat[\normalfont{$L1:L2 = 1:10$ }]{
		\includegraphics[trim=0 0 0 10, clip, height=\Height]{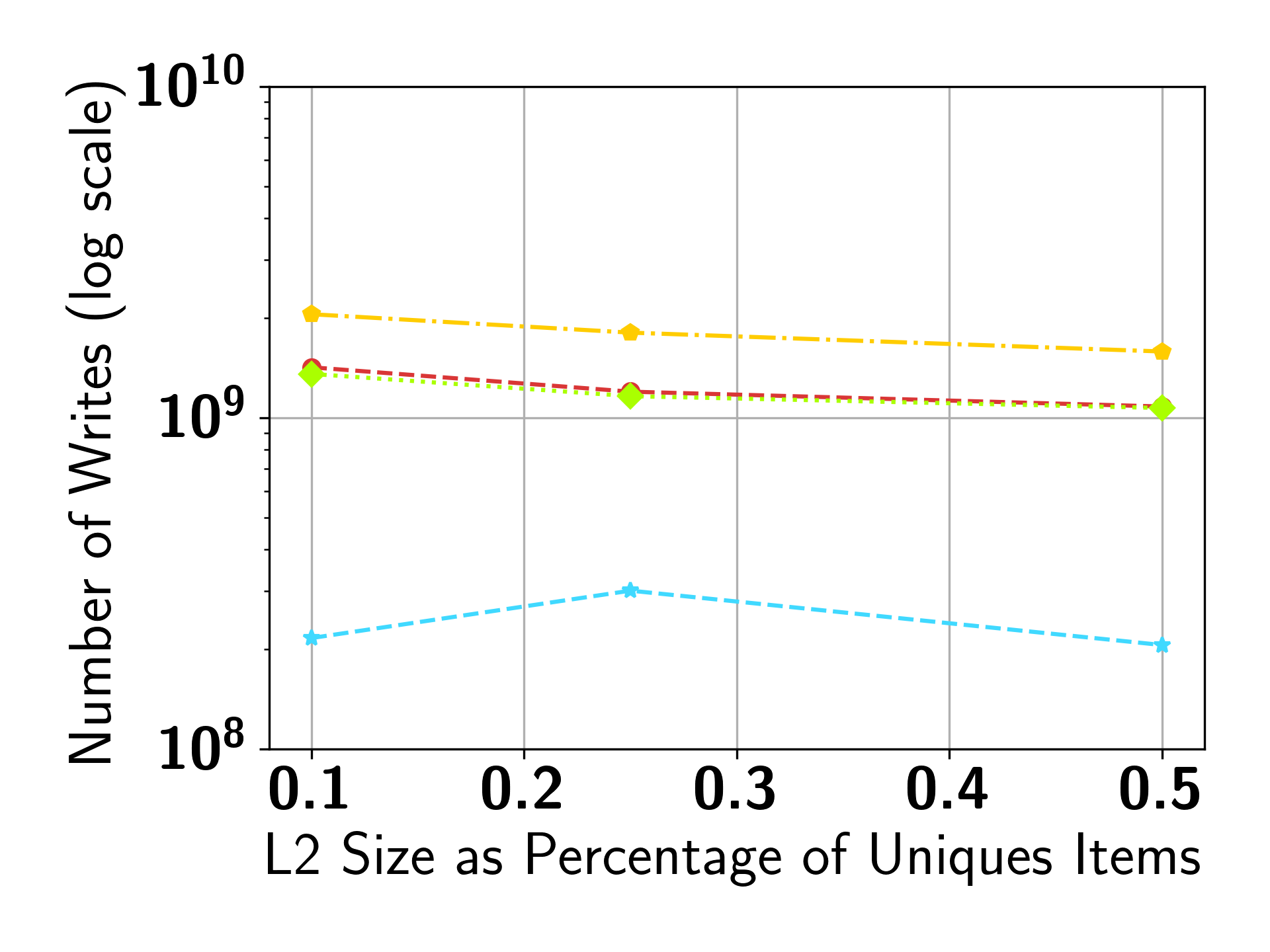}
	}
	\subfloat[\normalfont{$L1:L2 = 1:20$ }]{
		\includegraphics[trim=86 0 0 10, clip, height=\Height]{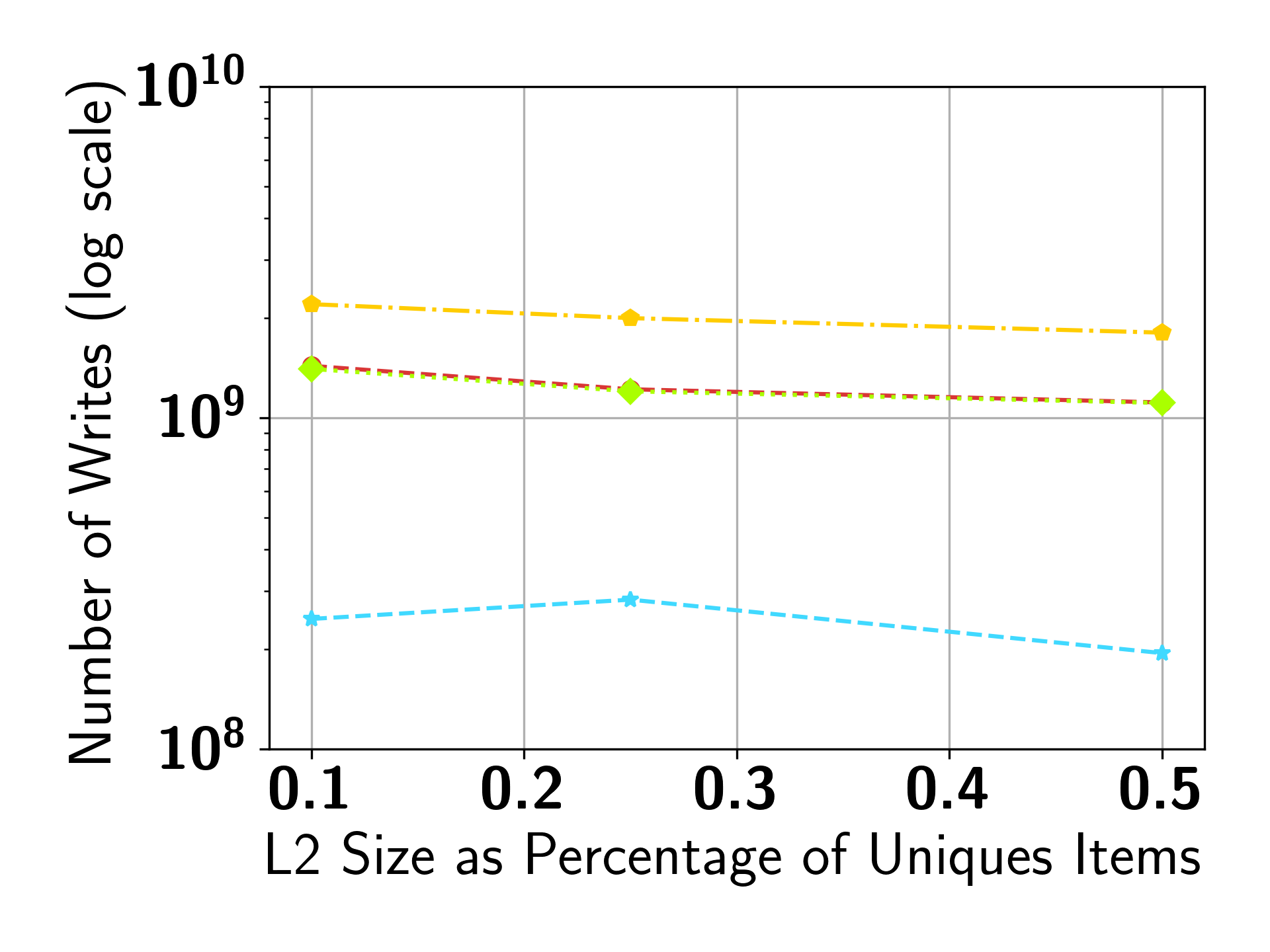}
	}
	\subfloat[\normalfont{$L1:L2 = 1:50$ }]{
		\includegraphics[trim=86 0 0 10, clip, height=\Height]{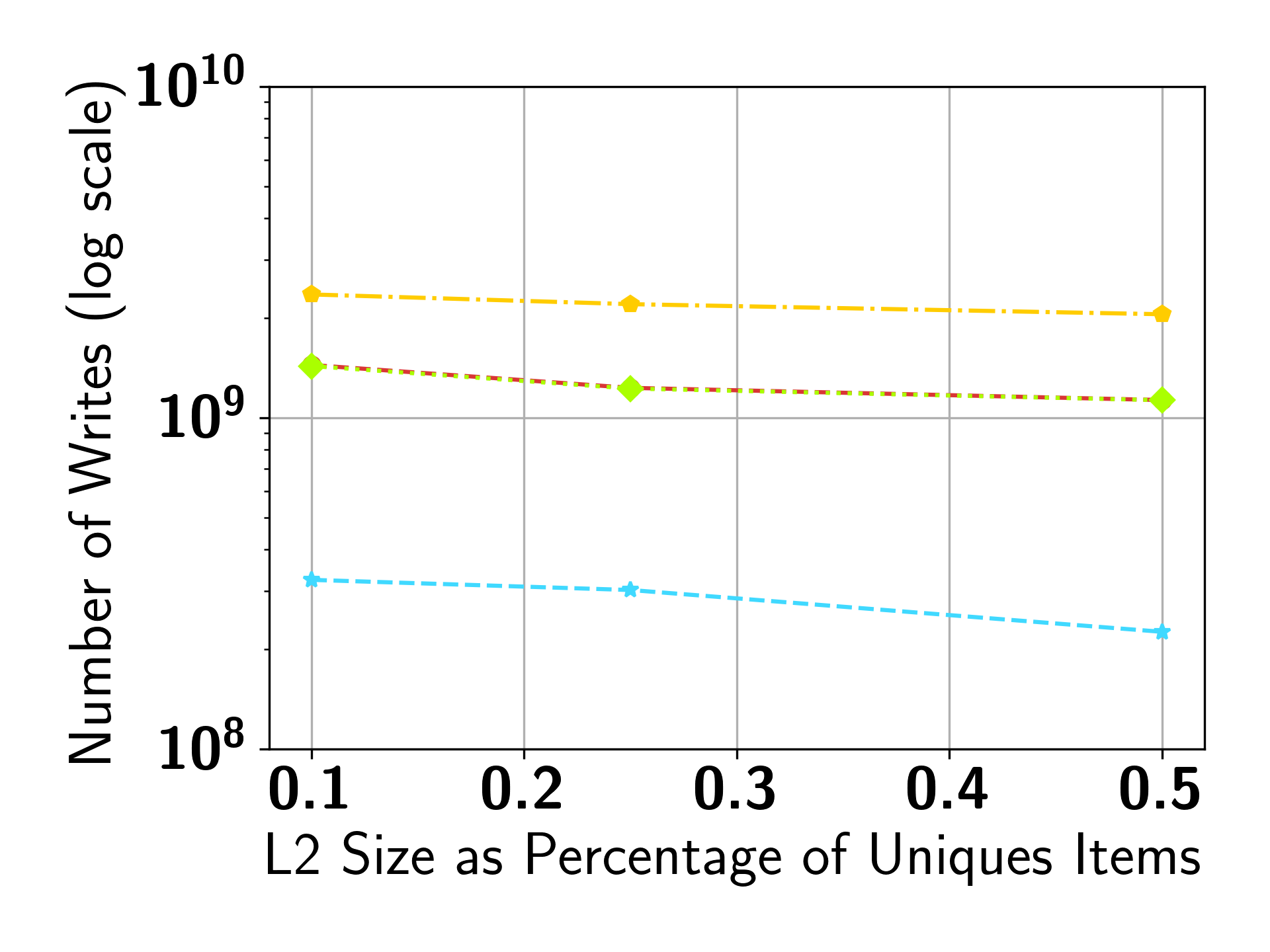}
	}
	\subfloat[\normalfont{$L1:L2 = 1:100$ }]{
		\includegraphics[trim=86 0 0 10, clip, height=\Height]{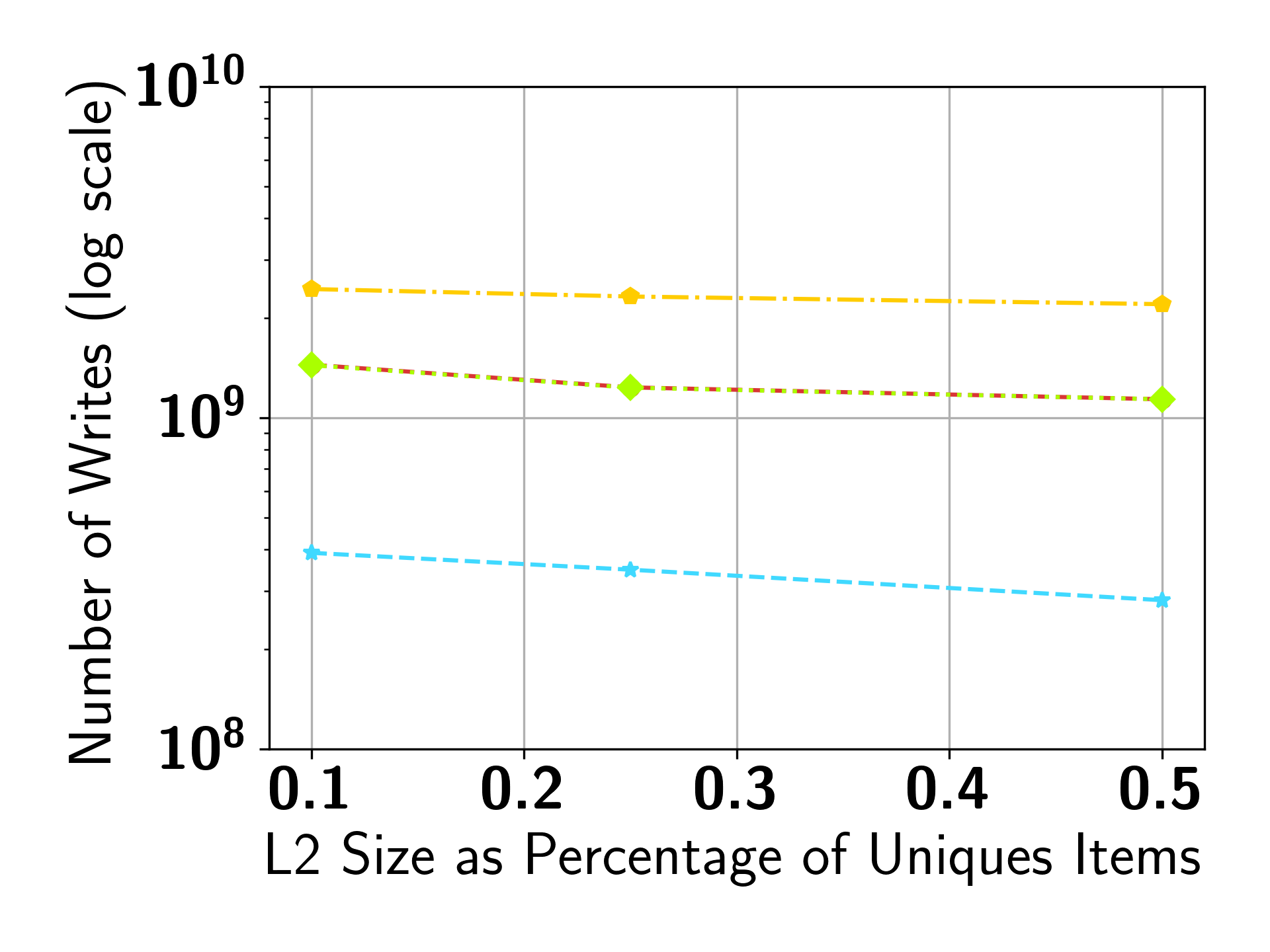}
	}
	\\
	\centering \includegraphics[trim=0 0 100 10, clip, height=1.3cm]{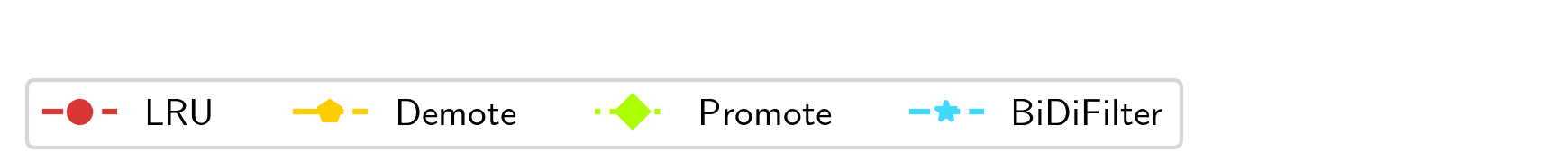}
	\caption{Number of writes to L2 for multiple traces and multiple ratios between L1 and L2}
	\label{eva:others:writes}
\end{figure*}

Here, we compare BiDiFilter with a $50\%$ window and admission on tie to other alternatives. 
The first two are versions of the traditional LRU modified to multilevel settings. 
One is Demote \cite{Demote} (also known as Global LRU), where we treat all the levels as a continuous one level; thus, on every miss and every level-two hit, an object is evicted from level-one and written into level-two. 
The second is a naive version to avoid level-two writes, where on a hit, we promote the item only inside its own level; we refer to this variant simply as LRU.
Additionally, we compare to Promote \cite{Promote}, a probabilistic approach targeted to avoid redundant writes on multilevel settings. 
We implement all these alternatives in JAVA as part of Caffeine Simulator \cite{CaffeineProject}. 
The results are presented in figures~\ref{eva:others:writes}~and~\ref{eva:others:latency}.

As expected, figure~\ref{eva:others:writes} shows that Demote always has the most level-two writes since it writes on any access except for level-one hits.
More surprising is that our naive modification to LRU saves writes quite similarly to the advanced Promote method. Nevertheless, it comes at the expense of hit ratio as we see in Figure~\ref{eva:others:latency} where Promote outperforms LRU in TWITTER1 and CDN1 traces while behaving similar on SYSTOR1 and TENCENT1.

Our solution, BiDiFilter, has done well on both metrics. Although we chose the less thrifty version (with admission on a tie),  we decreased the number of writes by one order of magnitude, i.e., by an $x10$ factor, on almost all cases. 
Only on small level-one caches in CDN1 and TWITTER1, we are sometimes less effective. 
Regarding latencies, we have the best results in CDN1 and TWITTER1 traces, improving over the competition by up to $20\%$. On the other hand, on SYSTOR1 and TENCENT1, we lag by a few percent where considering only the reads, but doing better if writes are included. 
The reason for this is that SYSTOR1 and TENCENT1 have a relatively low hit ratio (i.e., less than $60\%$), so frequent items are less beneficial, and recency plays a more prominent part.

\begin{figure*}[h]
	\centering SYSTOR1 \\ \small Hit-Ratio ranges from 7\%--23\% \\
	\subfloat[\normalfont{$L1:L2 = 1:10$ }]{ \begin{tabular}[b]{c}%
		\includegraphics[trim=0 0 0 10, clip, height=\Height]{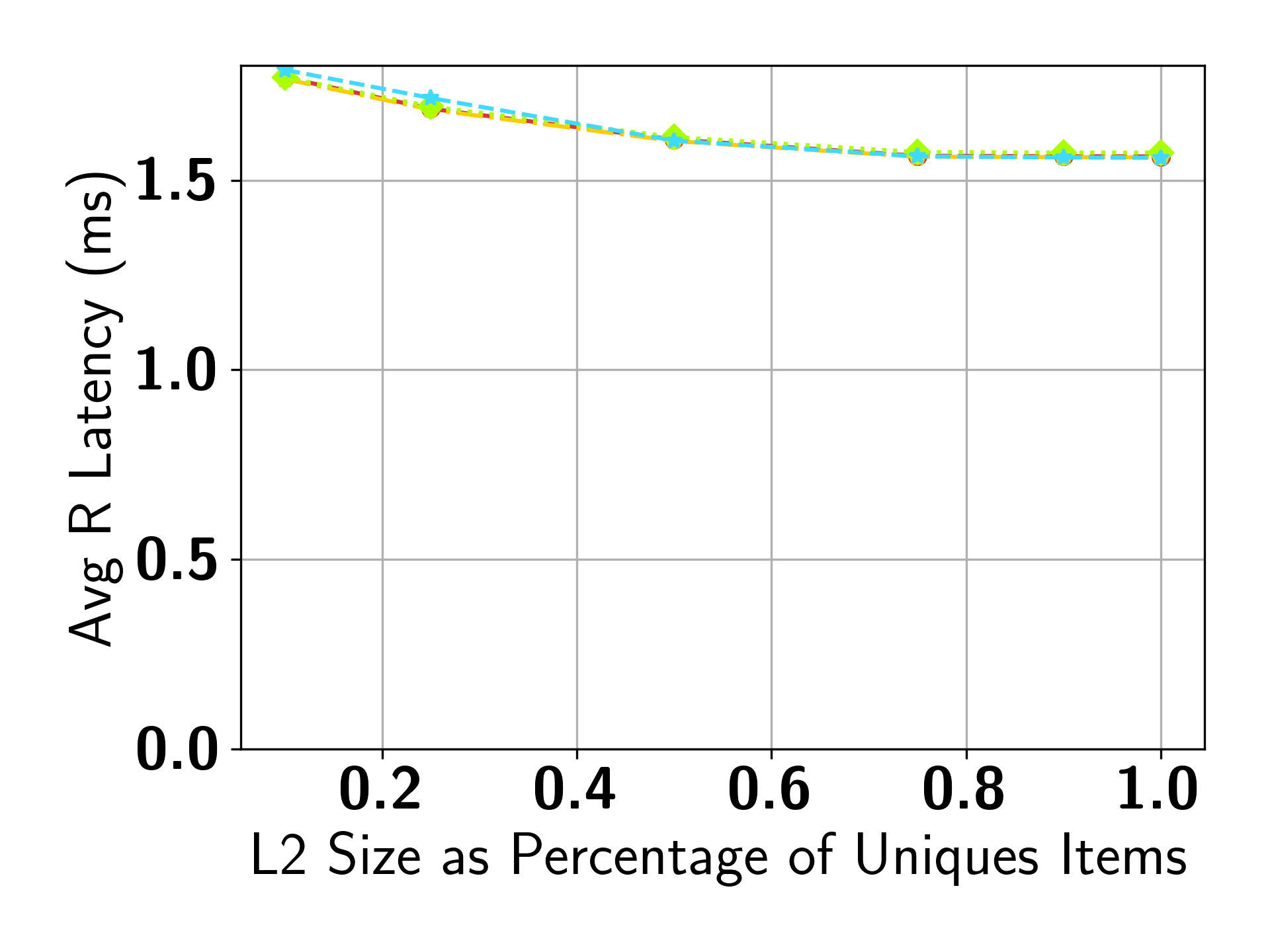} \\
		\includegraphics[trim=0 0 0 10, clip, height=\Height]{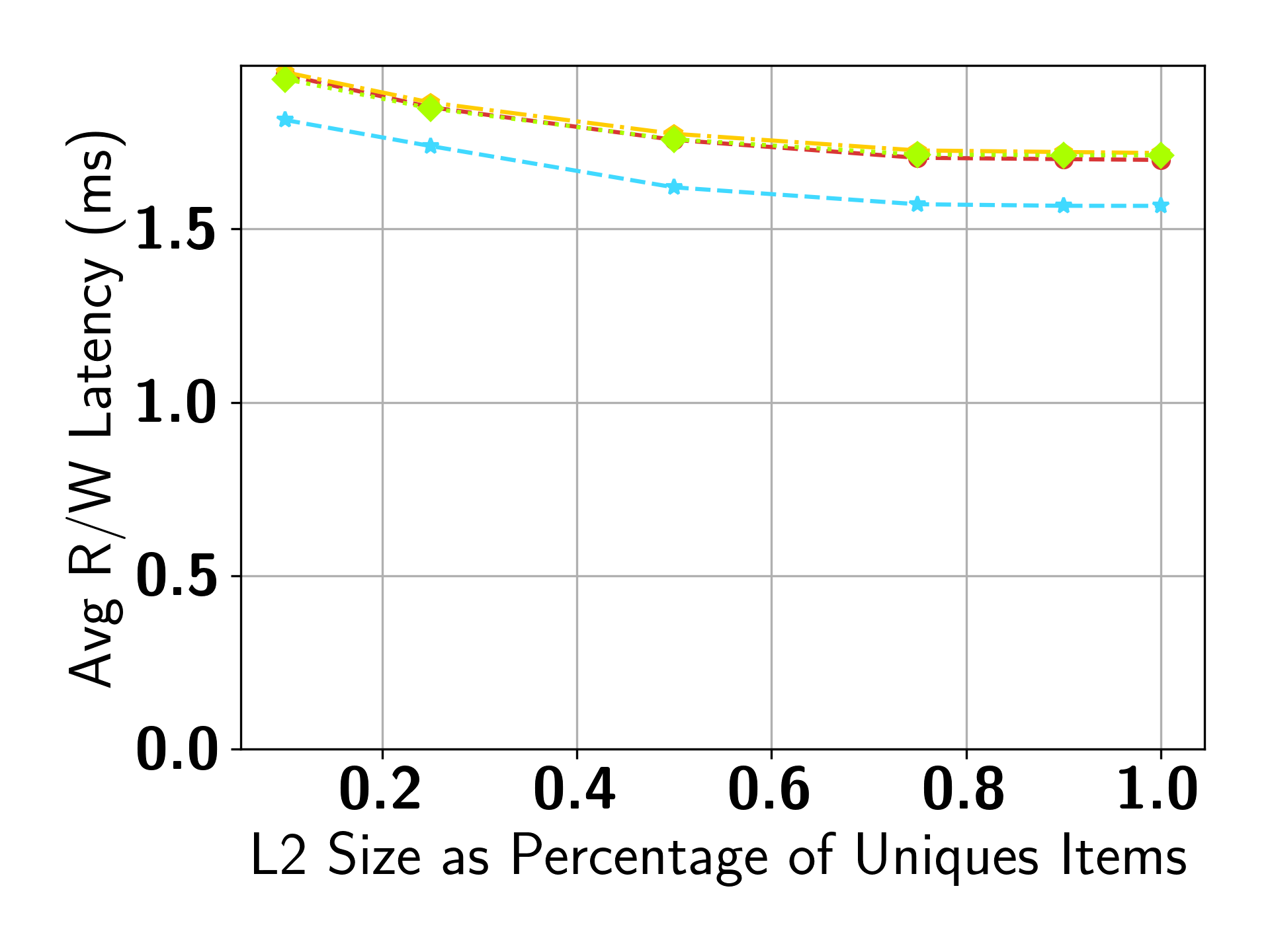}
	\end{tabular} \hspace{-0.5cm}
	}
	\subfloat[\normalfont{$L1:L2 = 1:20$ }]{ \begin{tabular}[b]{c}%
		\includegraphics[trim=86 0 0 10, clip, height=\Height]{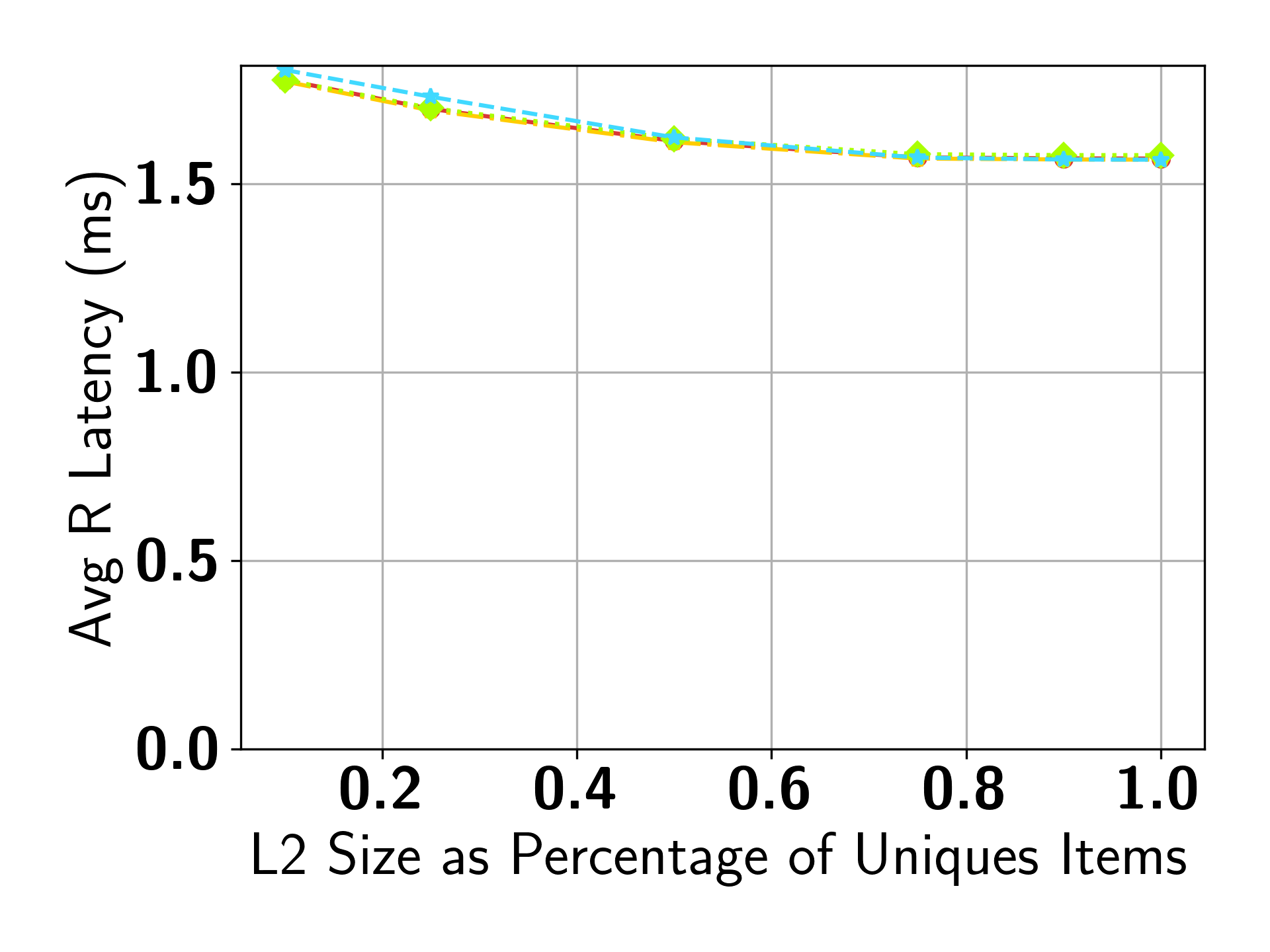} \\
		\includegraphics[trim=86 0 0 10, clip, height=\Height]{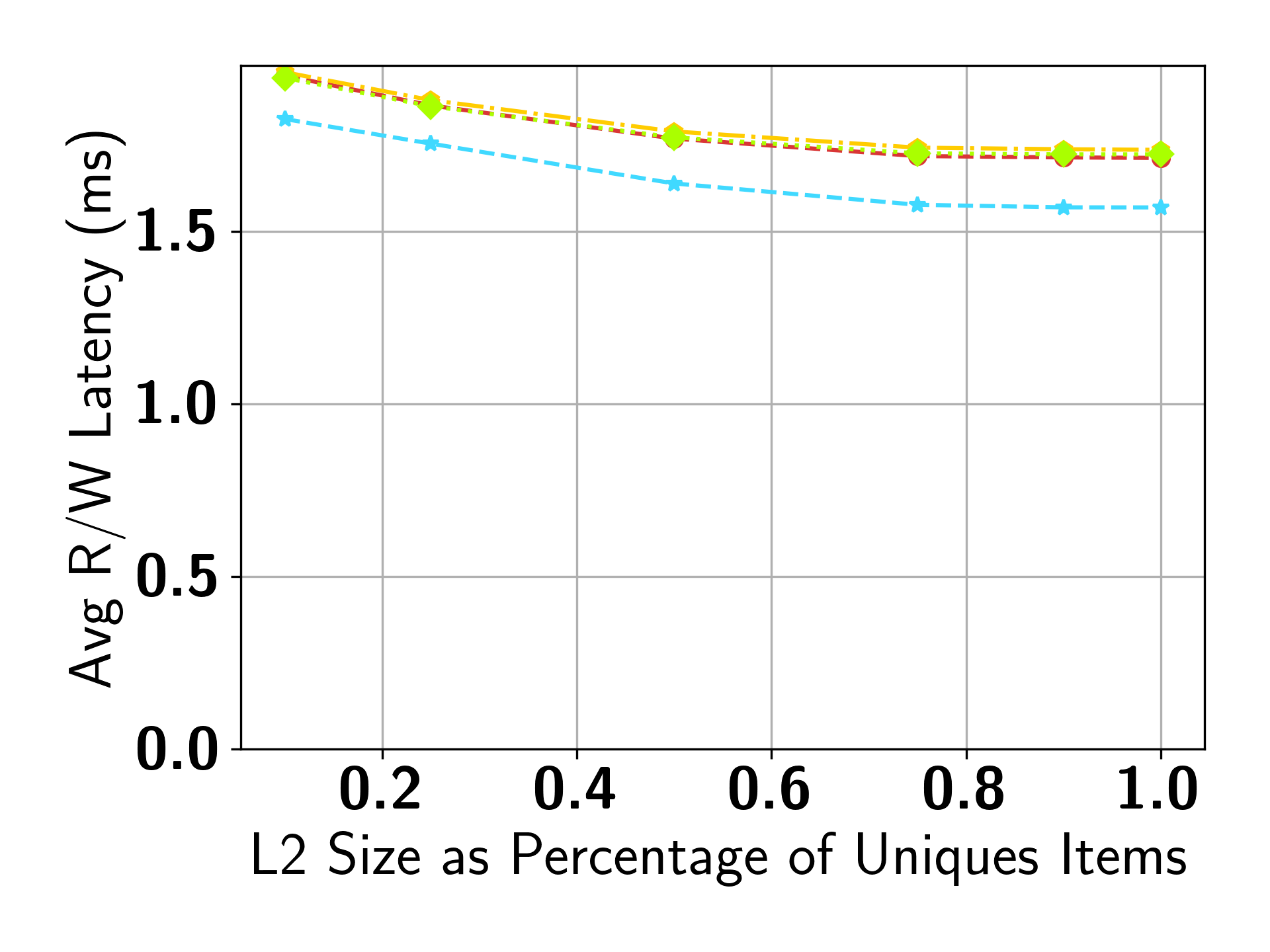}
	\end{tabular} \hspace{-0.5cm}
	}
	\subfloat[\normalfont{$L1:L2 = 1:50$ }]{ \begin{tabular}[b]{c}%
		\includegraphics[trim=86 0 0 10, clip, height=\Height]{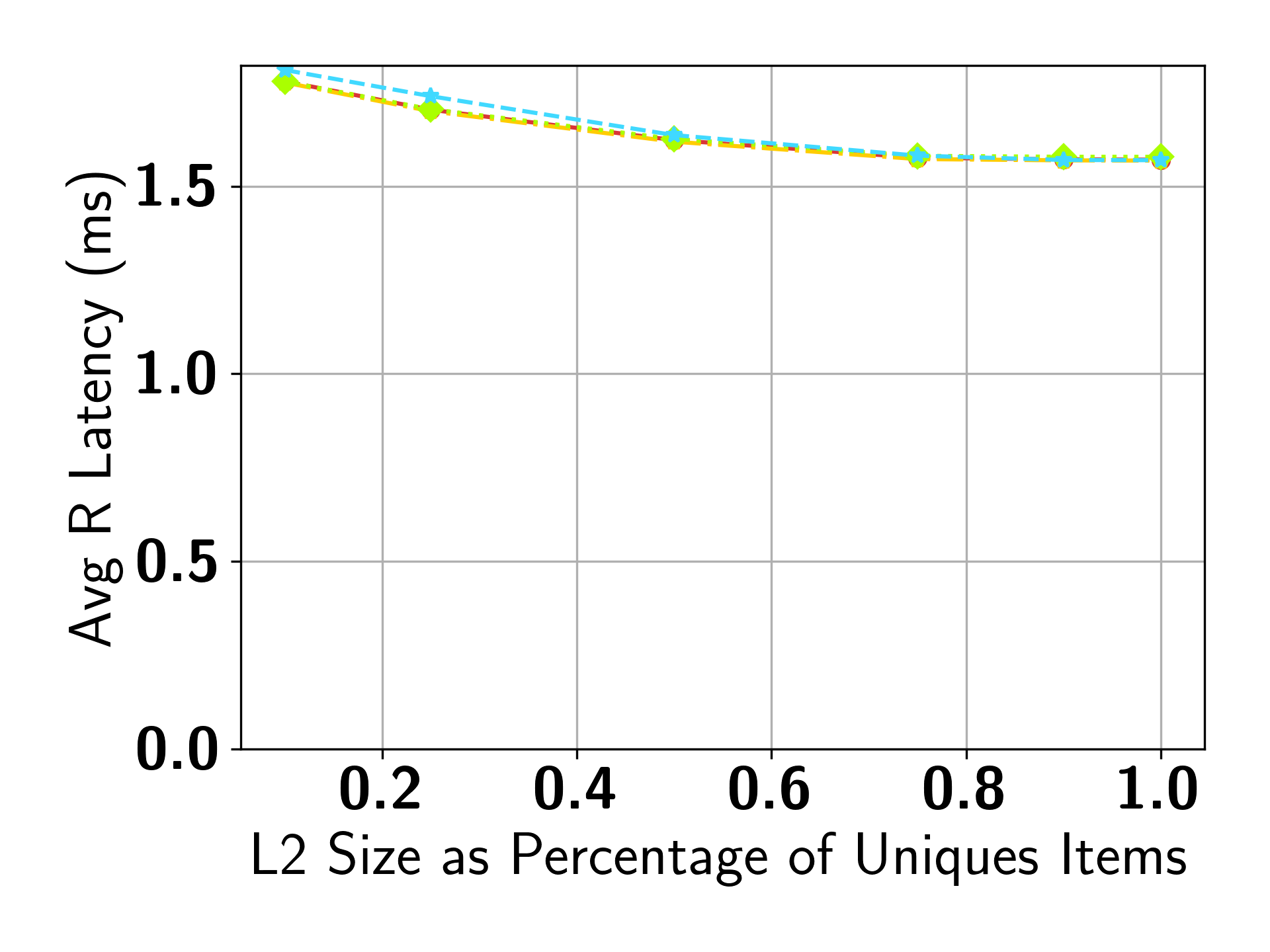} \\
		\includegraphics[trim=86 0 0 10, clip, height=\Height]{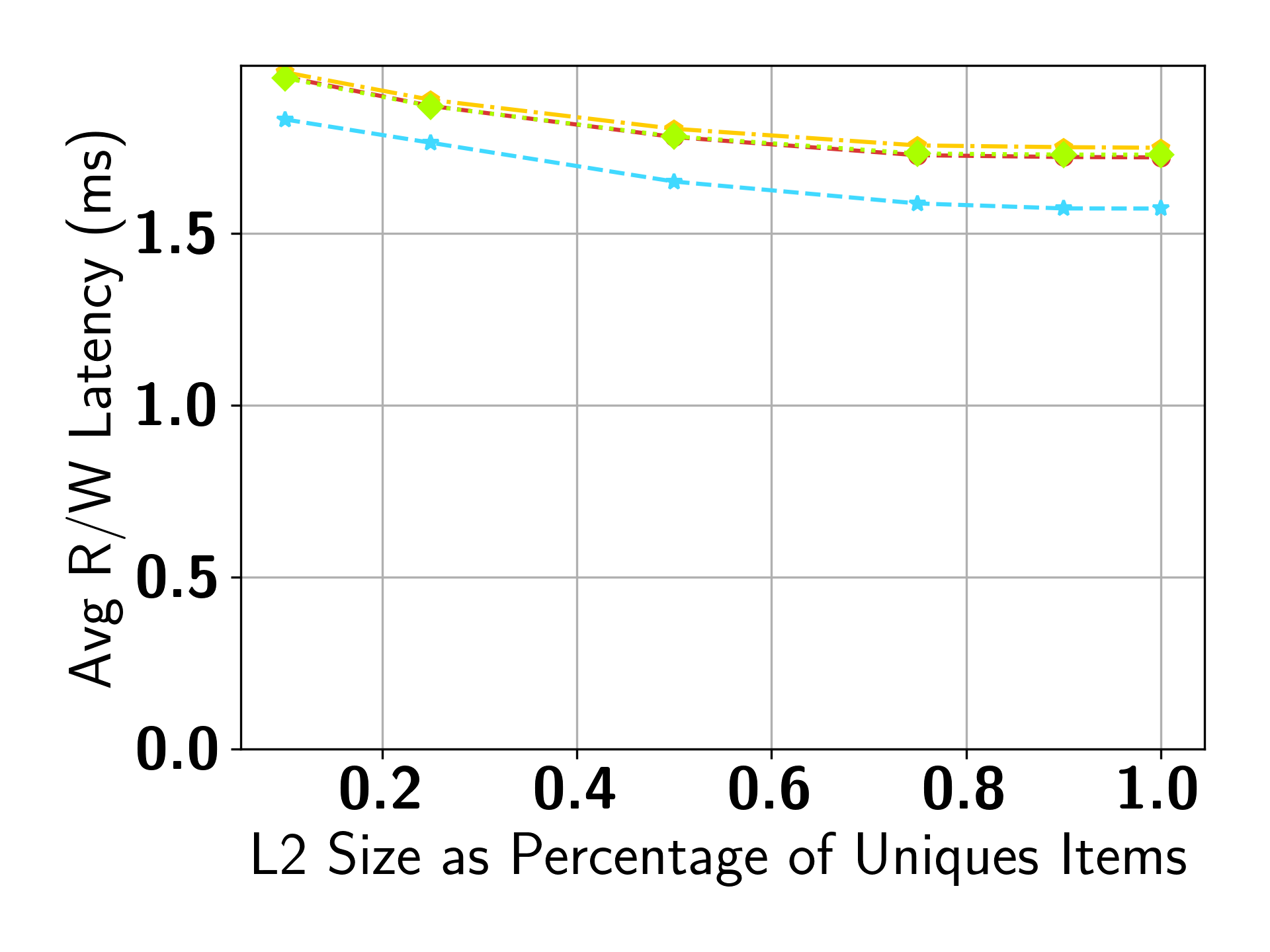}
	\end{tabular} \hspace{-0.5cm}
	}
	\subfloat[\normalfont{$L1:L2 = 1:100$ }]{ \begin{tabular}[b]{c}%
		\includegraphics[trim=86 0 0 10, clip, height=\Height]{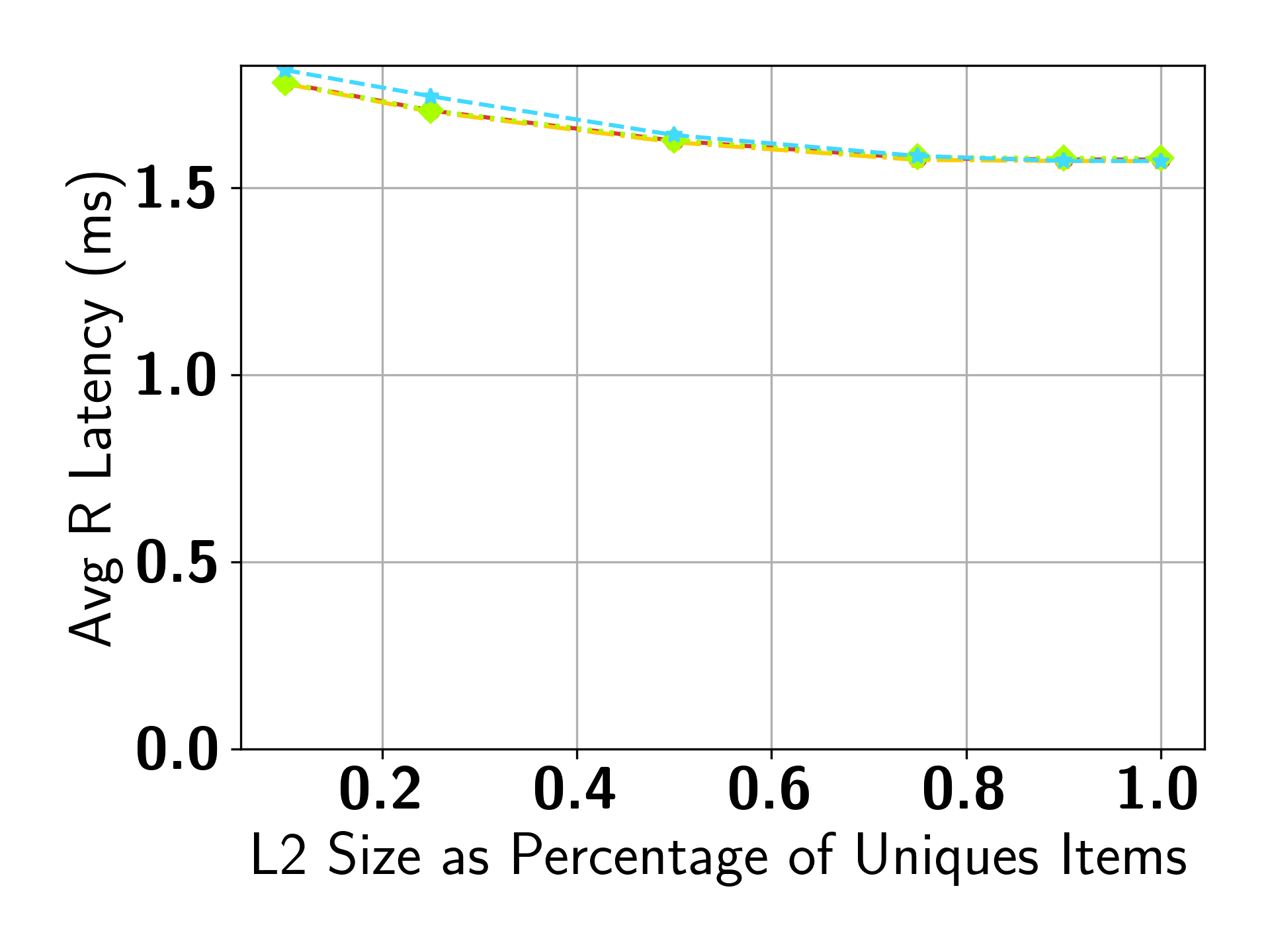} \\
		\includegraphics[trim=86 0 0 10, clip, height=\Height]{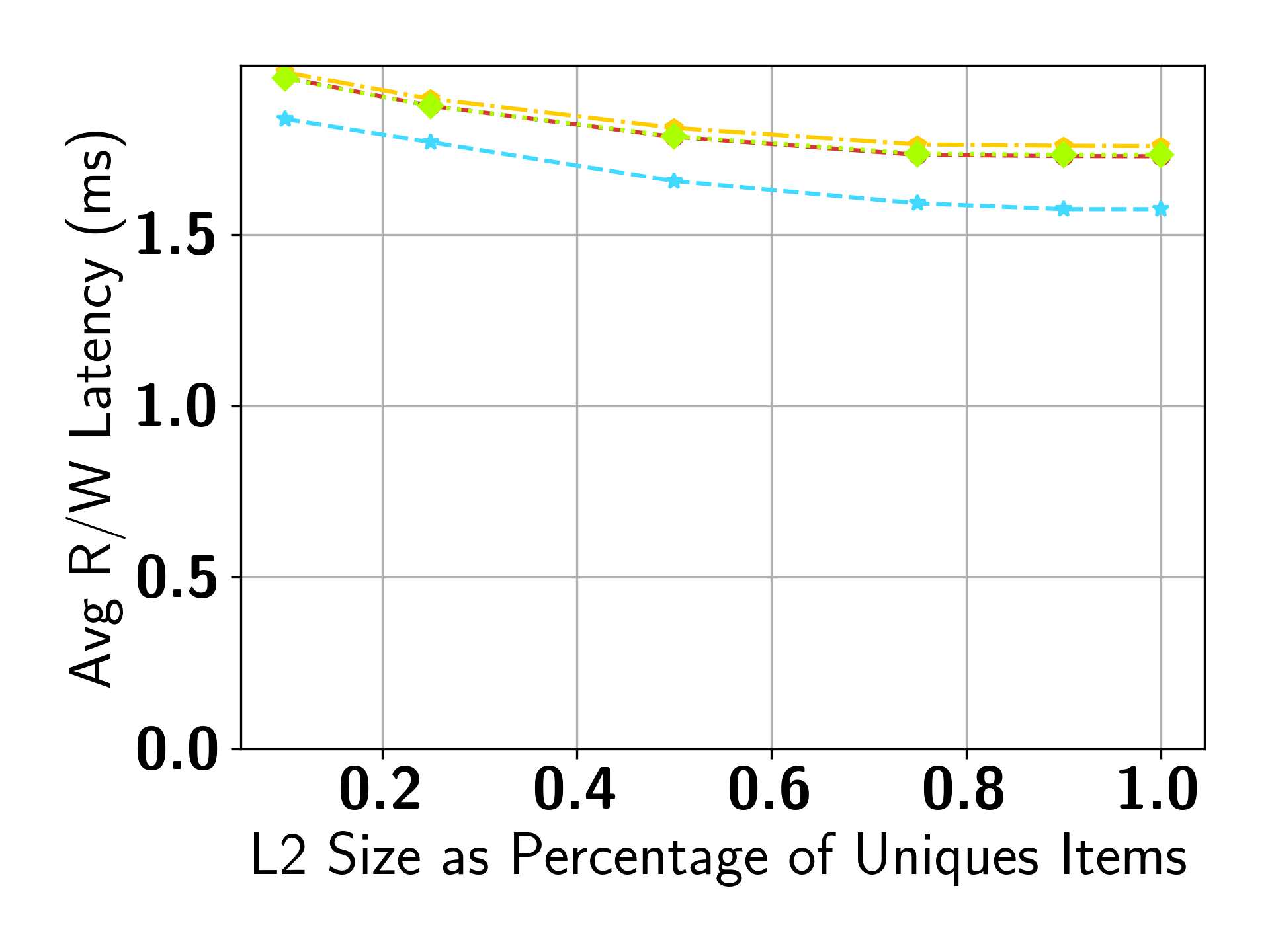}
	\end{tabular} \hspace{-0.5cm}
	}
	\\
	\vspace{0.2cm}
	\centering CDN1 \\ \small Hit-Ratio ranges from 20\%--86\% \\ 
	\subfloat[\normalfont{$L1:L2 = 1:10$ }]{ \begin{tabular}[b]{c}%
		\includegraphics[trim=0 0 0 10, clip, height=\Height]{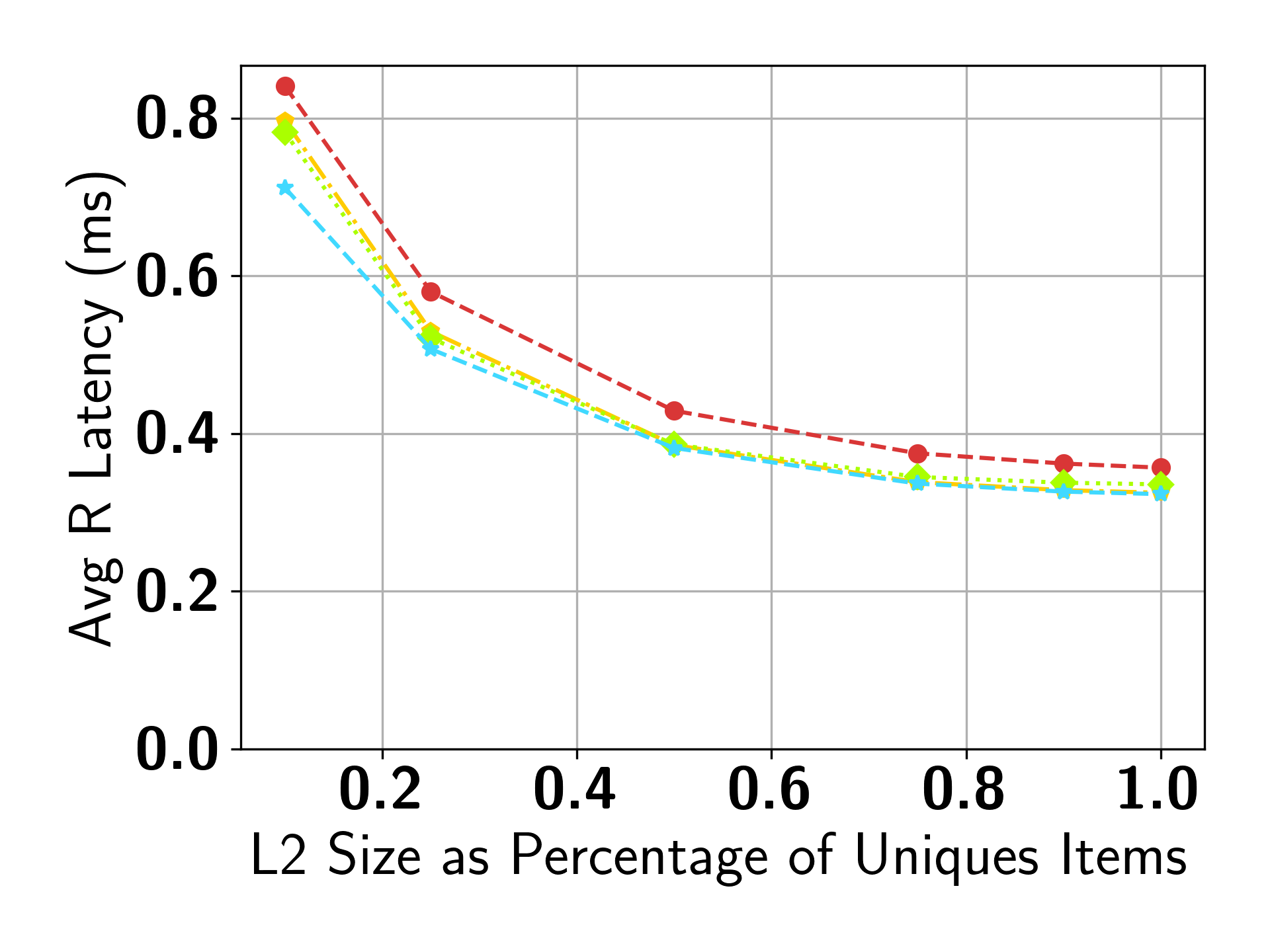} \\
		\includegraphics[trim=0 0 0 10, clip, height=\Height]{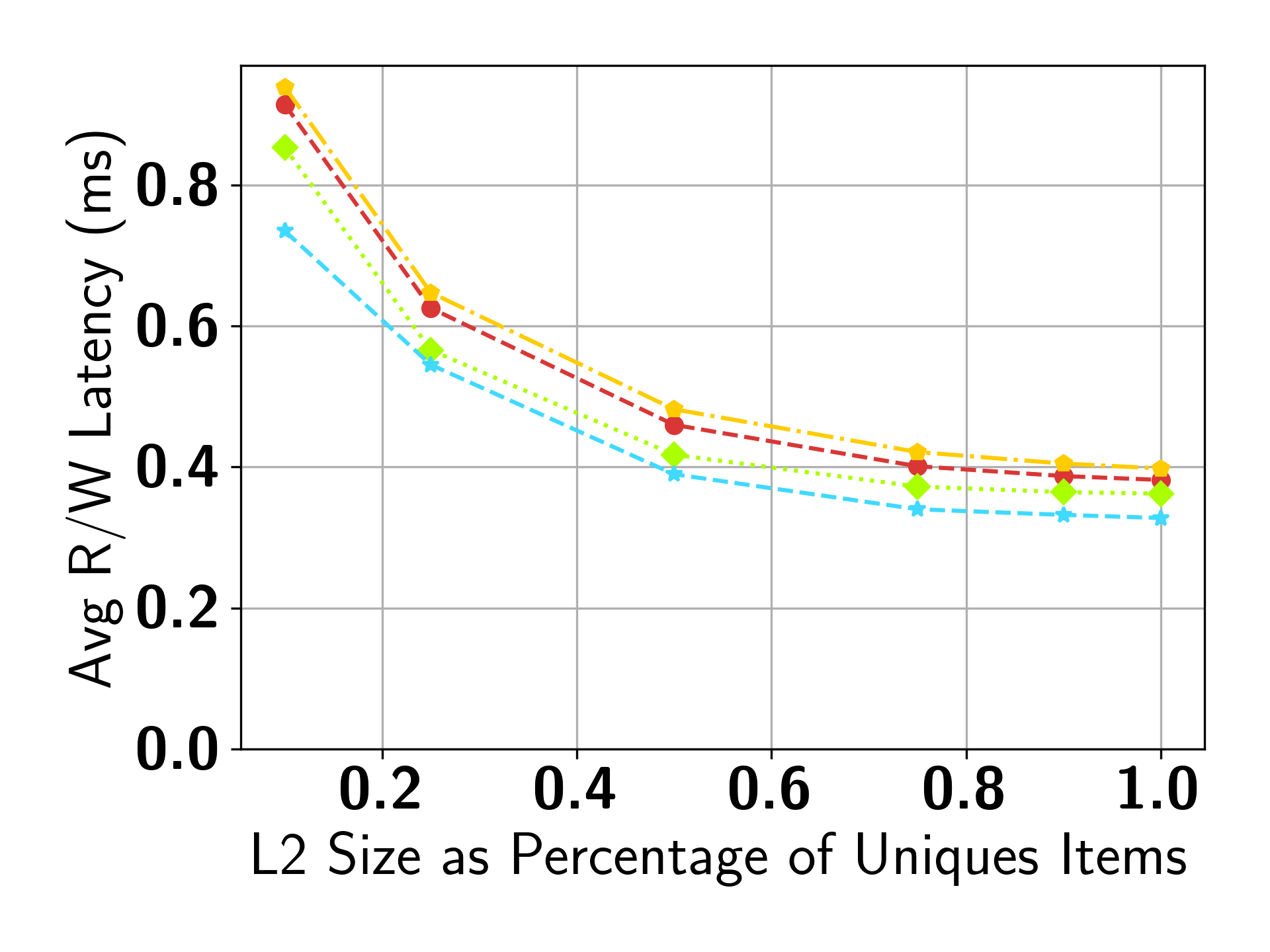}
	\end{tabular} \hspace{-0.5cm}
	}
	\subfloat[\normalfont{$L1:L2 = 1:20$ }]{ \begin{tabular}[b]{c}%
		\includegraphics[trim=86 0 0 10, clip, height=\Height]{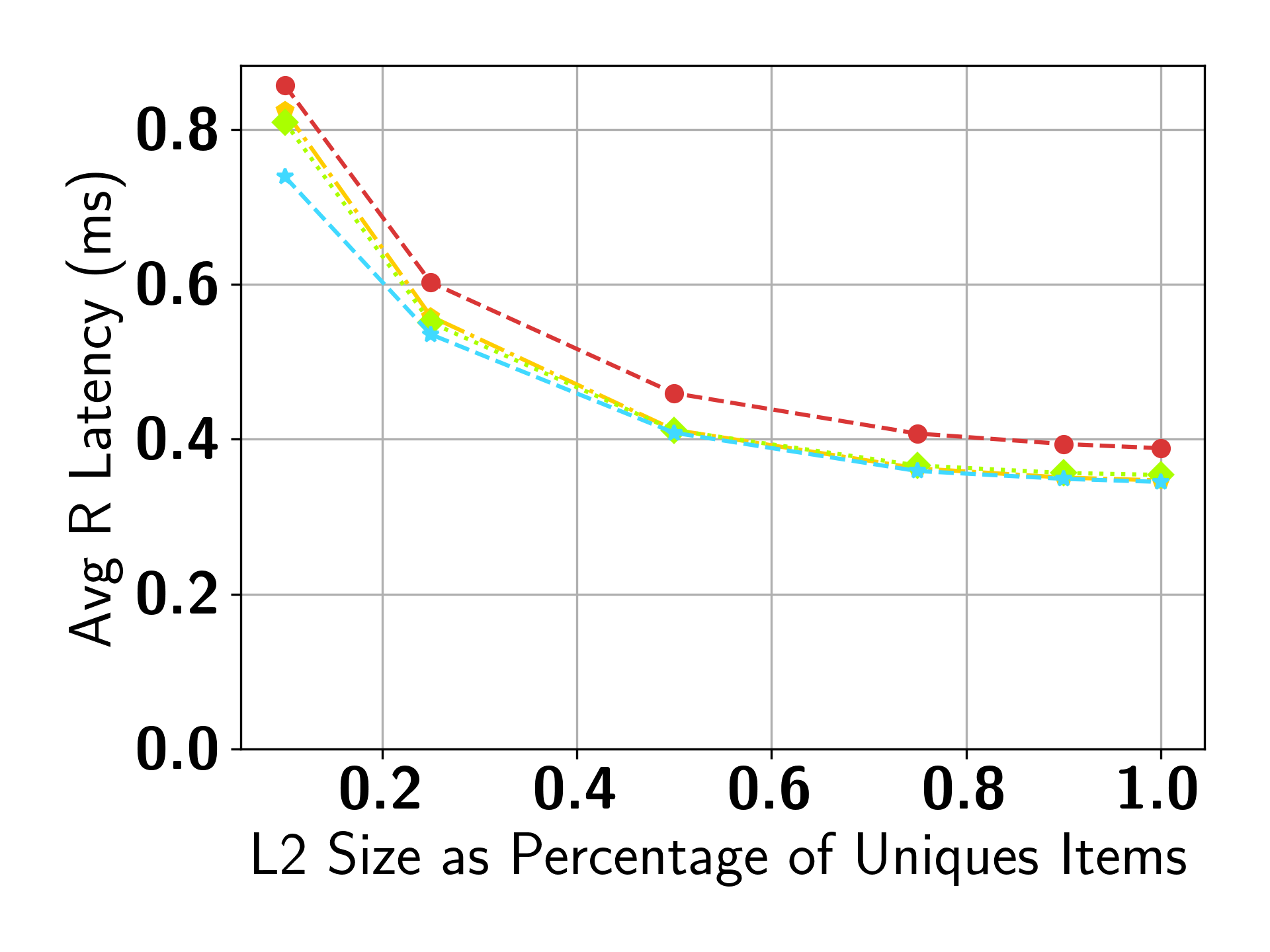} \\
		\includegraphics[trim=50 0 0 10, clip, height=\Height]{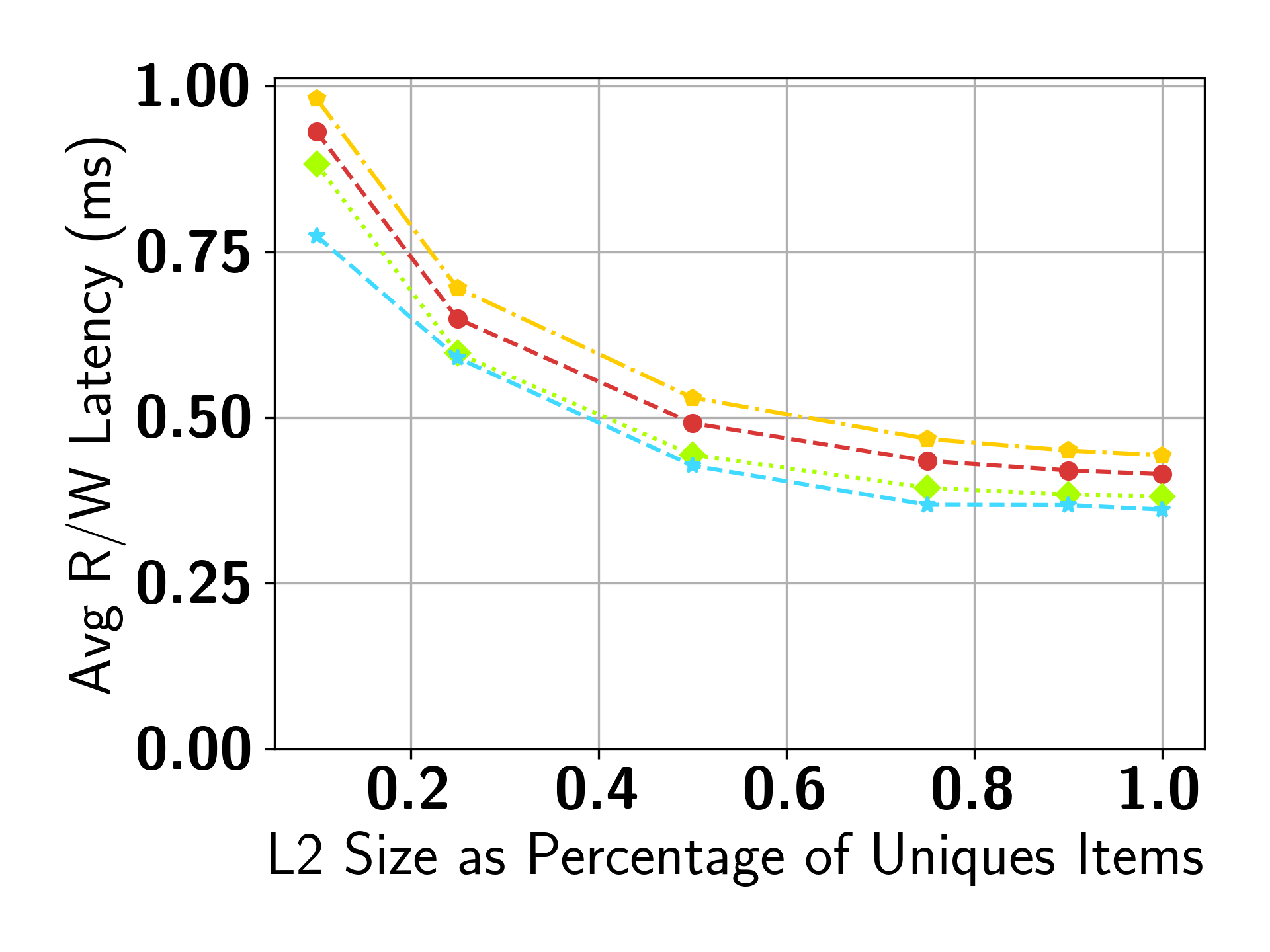}
	\end{tabular} \hspace{-0.5cm}
	}
	\subfloat[\normalfont{$L1:L2 = 1:50$ }]{ \begin{tabular}[b]{c}%
		\includegraphics[trim=86 0 0 10, clip, height=\Height]{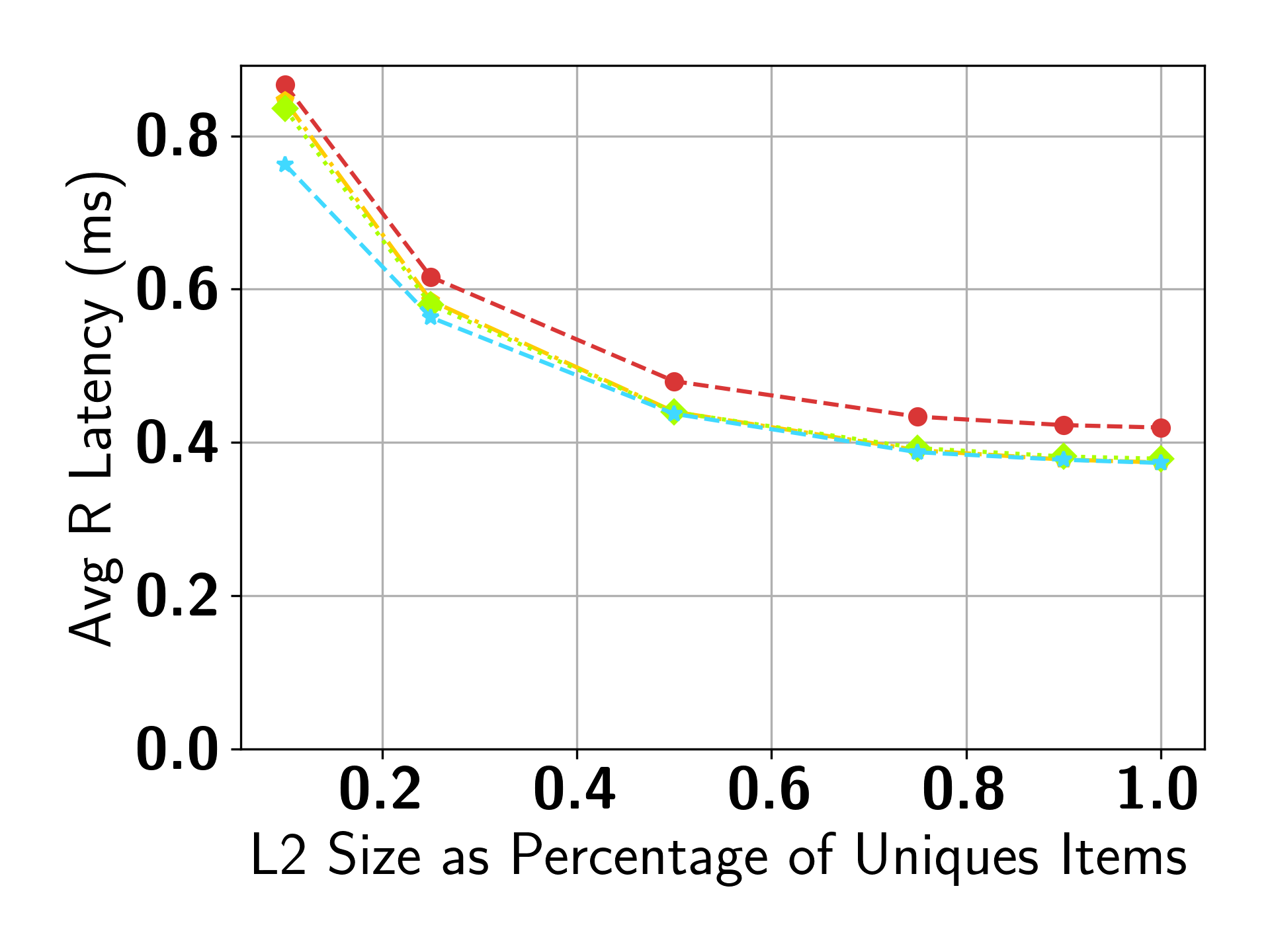} \\
		\includegraphics[trim=91 0 0 10, clip, height=\Height]{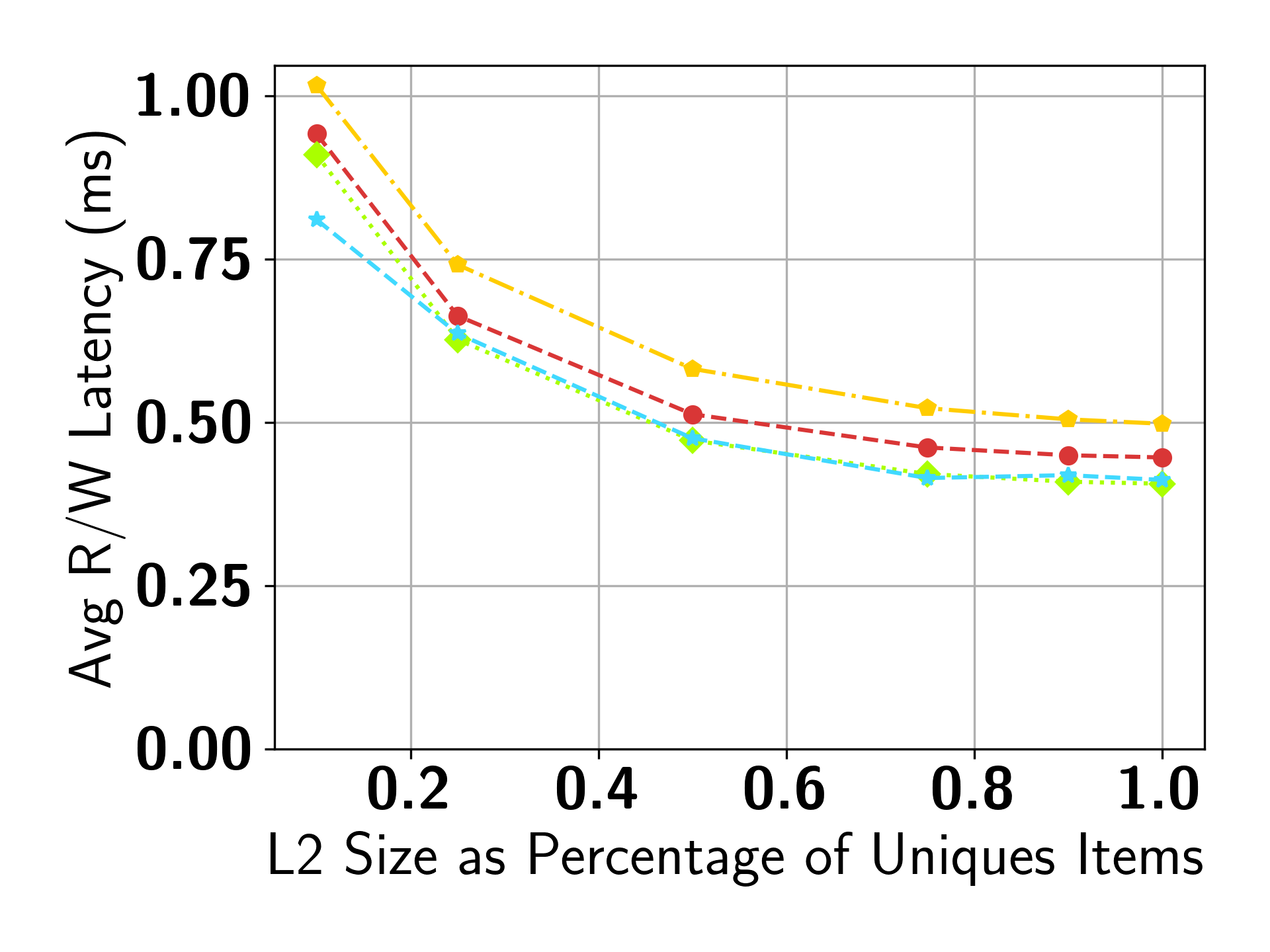}
	\end{tabular} \hspace{-0.5cm}
	}
	\subfloat[\normalfont{$L1:L2 = 1:100$ }]{ \begin{tabular}[b]{c}%
		\includegraphics[trim=86 0 0 10, clip, height=\Height]{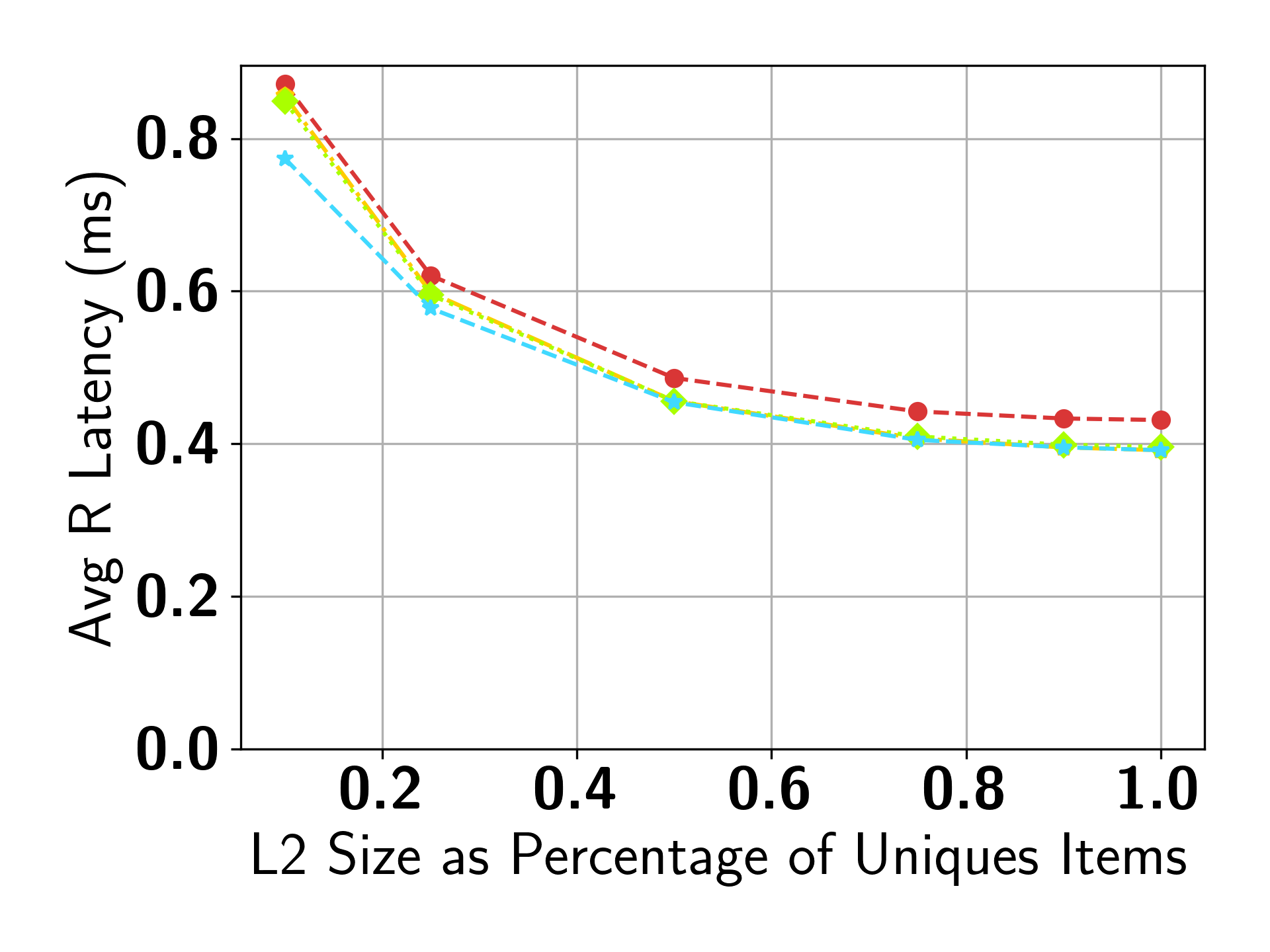} \\
		\includegraphics[trim=91 0 0 10, clip, height=\Height]{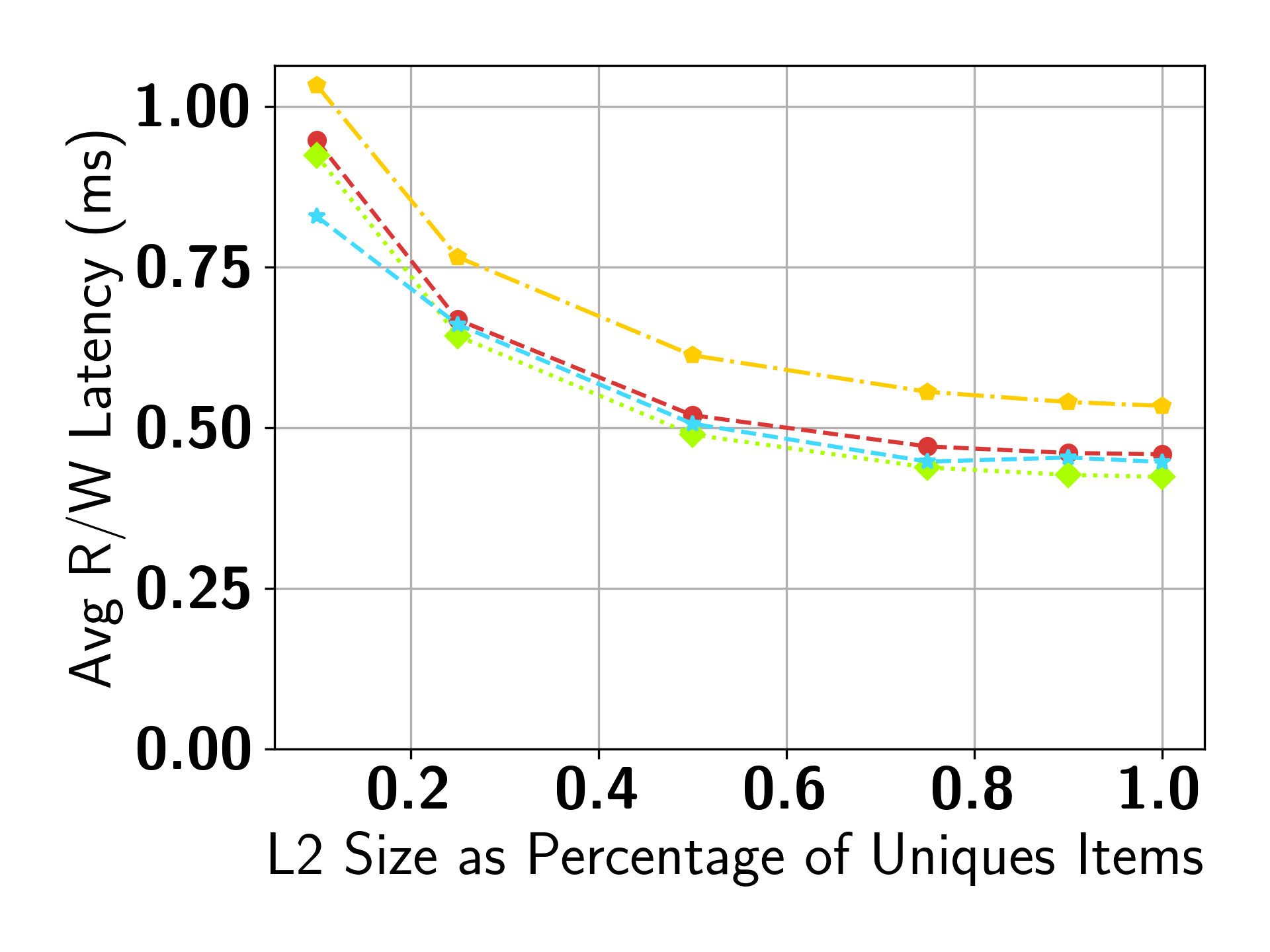}
	\end{tabular} \hspace{-0.5cm}
	}
	\\
	\centering \includegraphics[trim=0 0 100 10, clip, height=1.3cm]{Plots/others-legend-lines.png}
	\caption{Average latency per request, assuming latencies of 2ms, 200us and 100ns for disk, L2, and L1 respectively, for multiple traces and multiple ratios between L1 and L2}
	\label{eva:others:latency}
\end{figure*}
\begin{figure*}[h]
\ContinuedFloat
	\centering TWITTER1 \\ \small Hit-Ratio ranges from 65\%--93\% \\
	\subfloat[\normalfont{$L1:L2 = 1:10$ }]{ \begin{tabular}[b]{c}%
		\includegraphics[trim=0 0 0 10, clip, height=\Height]{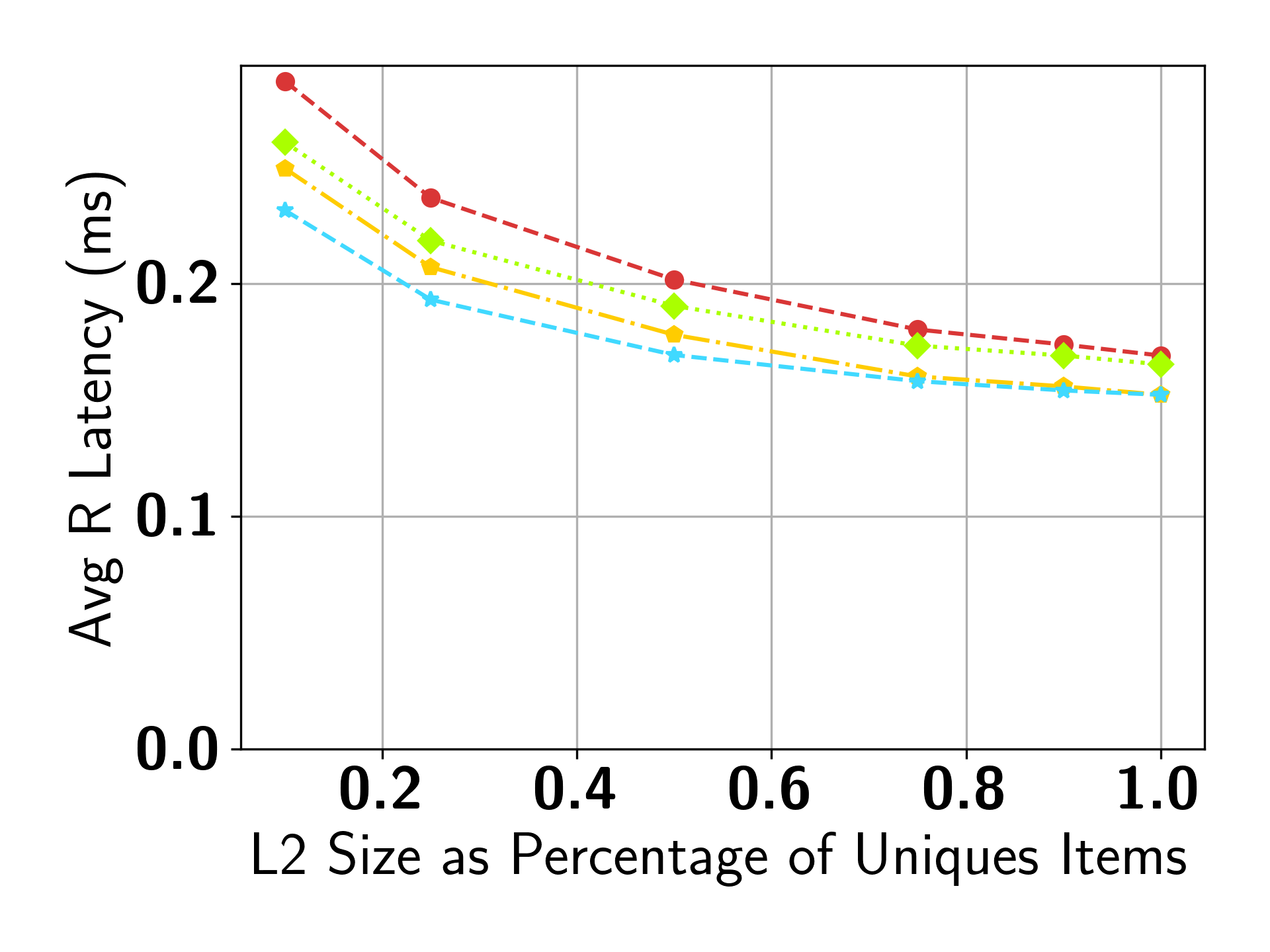} \\
		\includegraphics[trim=0 0 0 10, clip, height=\Height]{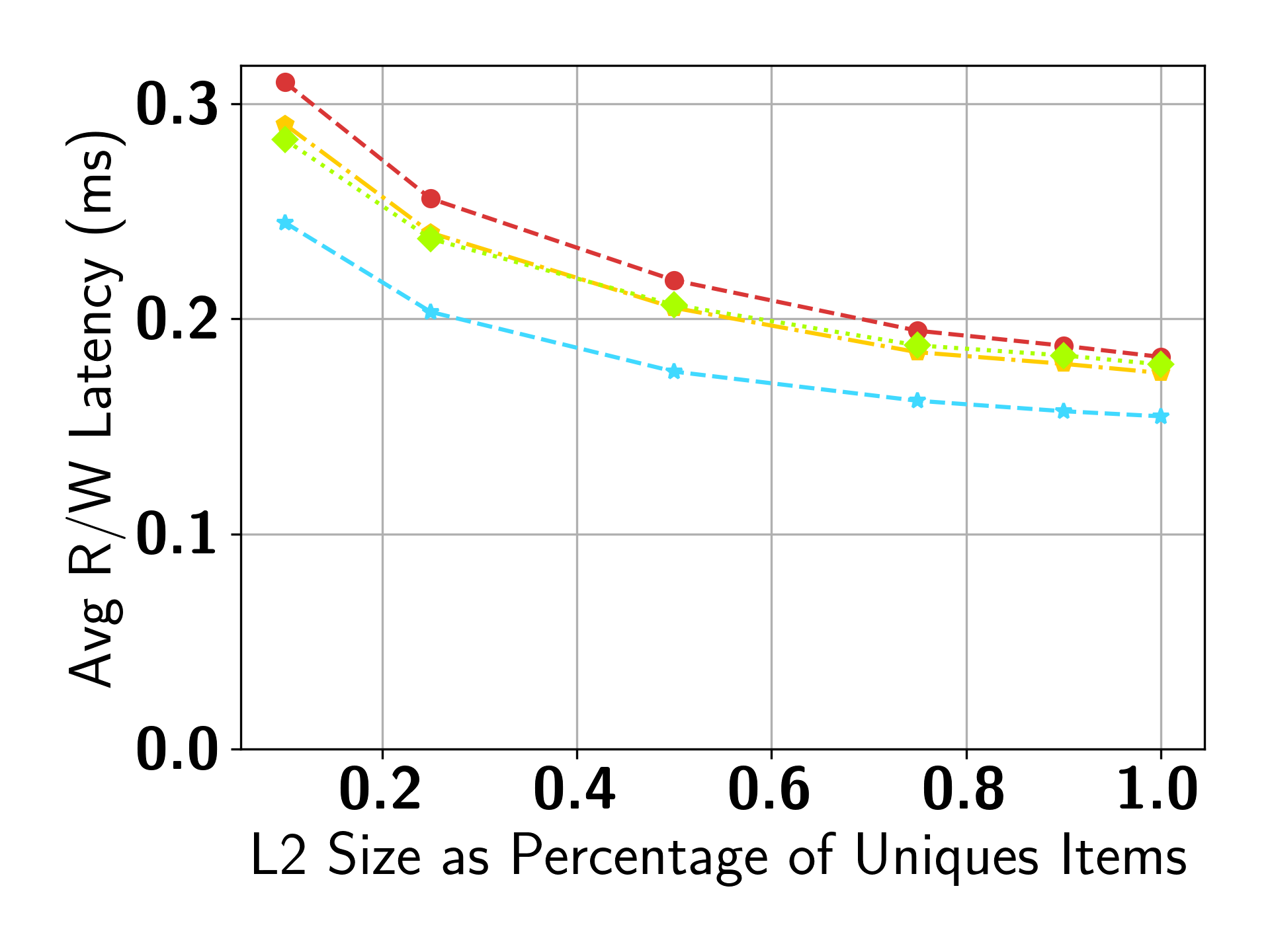}
	\end{tabular} \hspace{-0.5cm}
	}
	\subfloat[\normalfont{$L1:L2 = 1:20$ }]{ \begin{tabular}[b]{c}%
		\includegraphics[trim=86 0 0 10, clip, height=\Height]{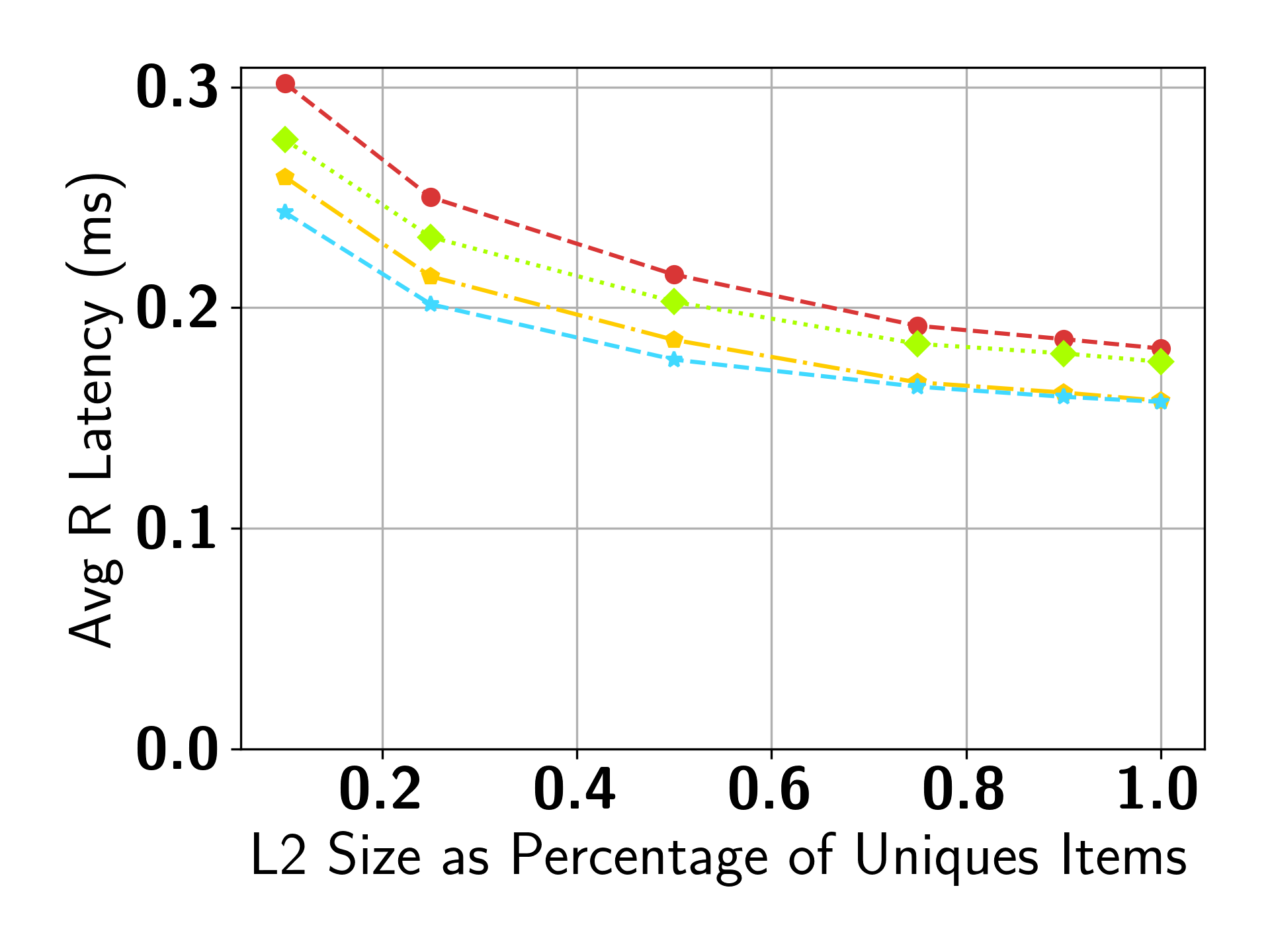} \\
		\includegraphics[trim=86 0 0 10, clip, height=\Height]{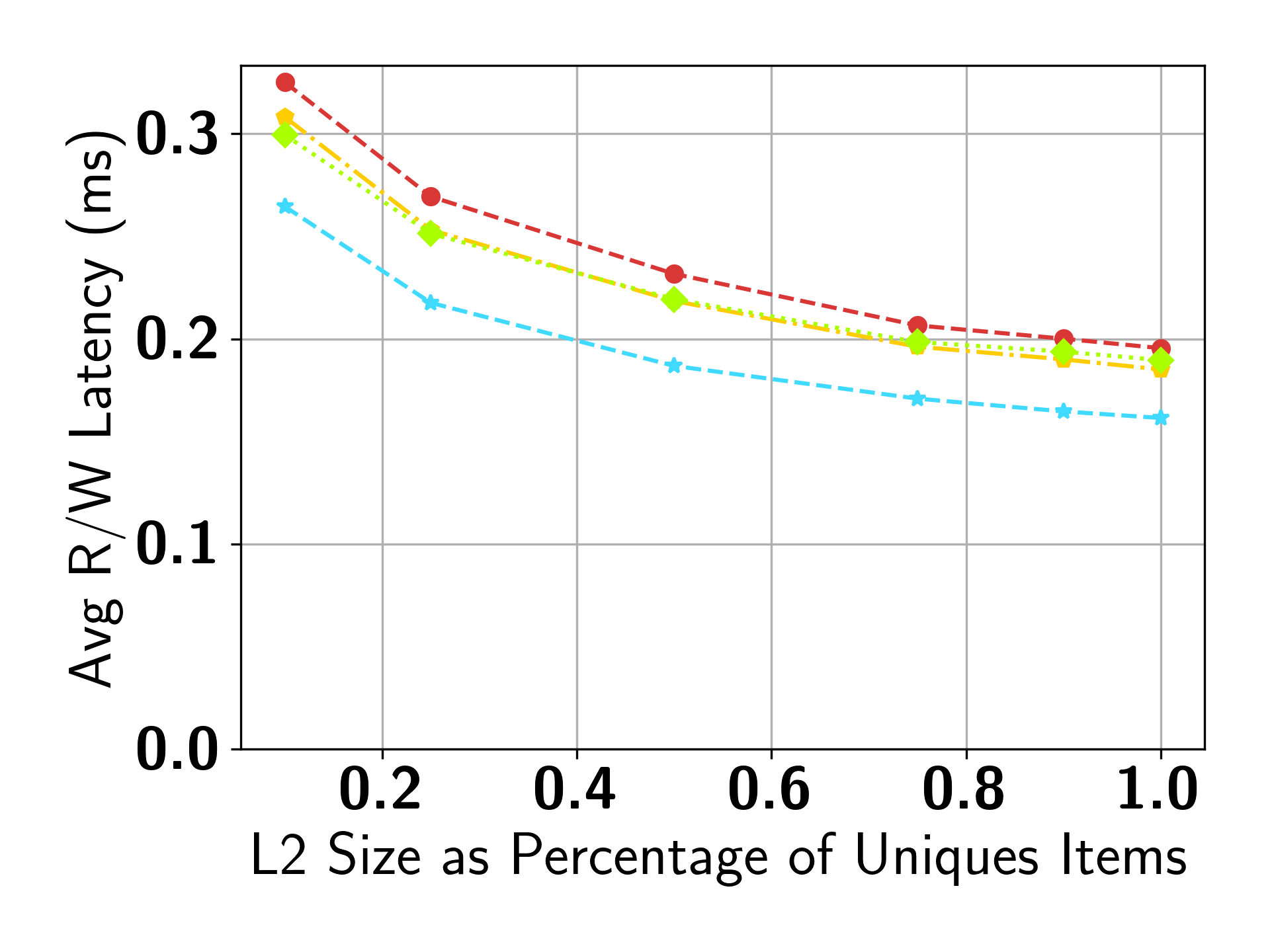}
	\end{tabular} \hspace{-0.5cm}
	}
	\subfloat[\normalfont{$L1:L2 = 1:50$ }]{ \begin{tabular}[b]{c}%
		\includegraphics[trim=86 0 0 10, clip, height=\Height]{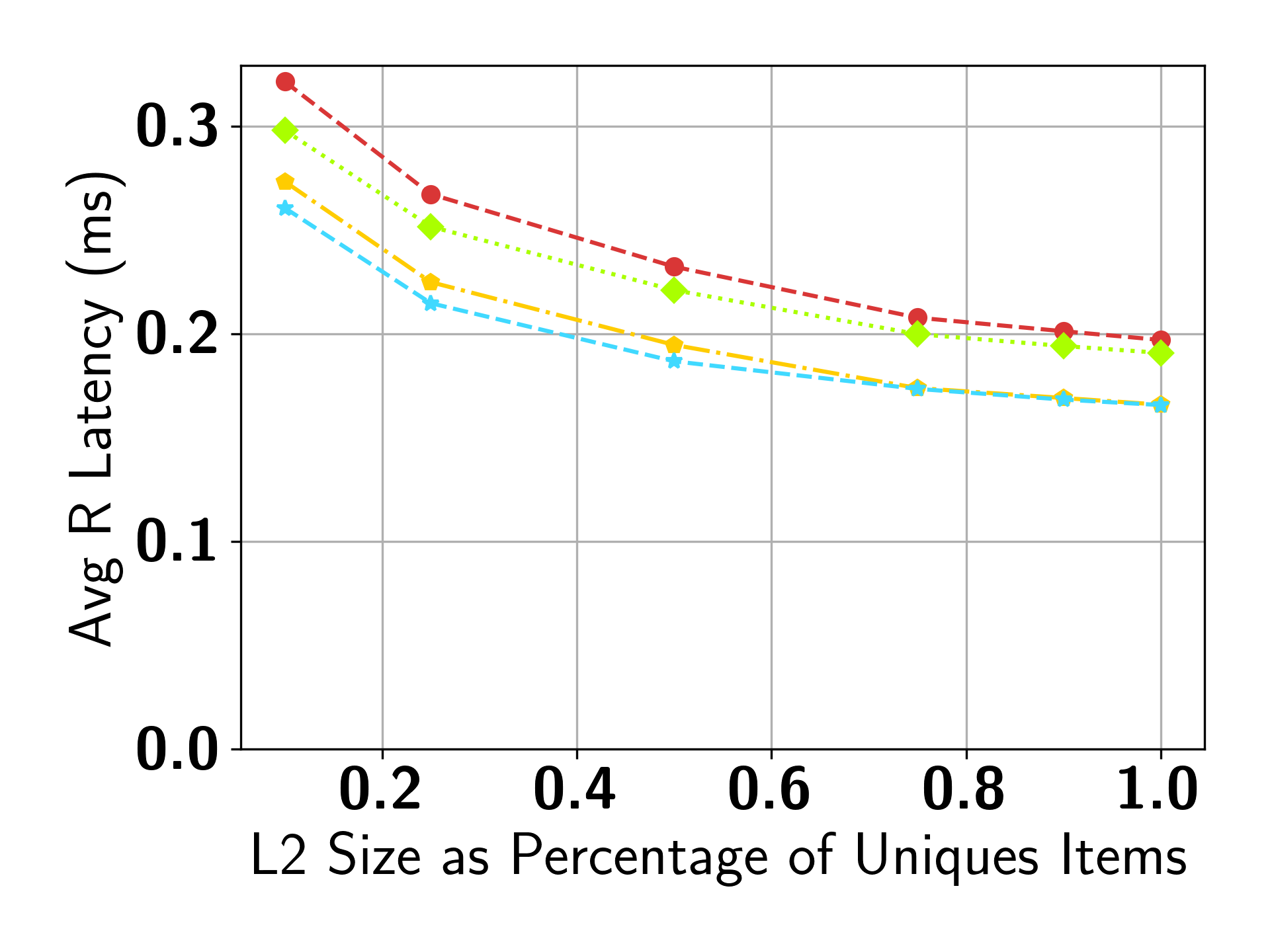} \\
		\includegraphics[trim=86 0 0 10, clip, height=\Height]{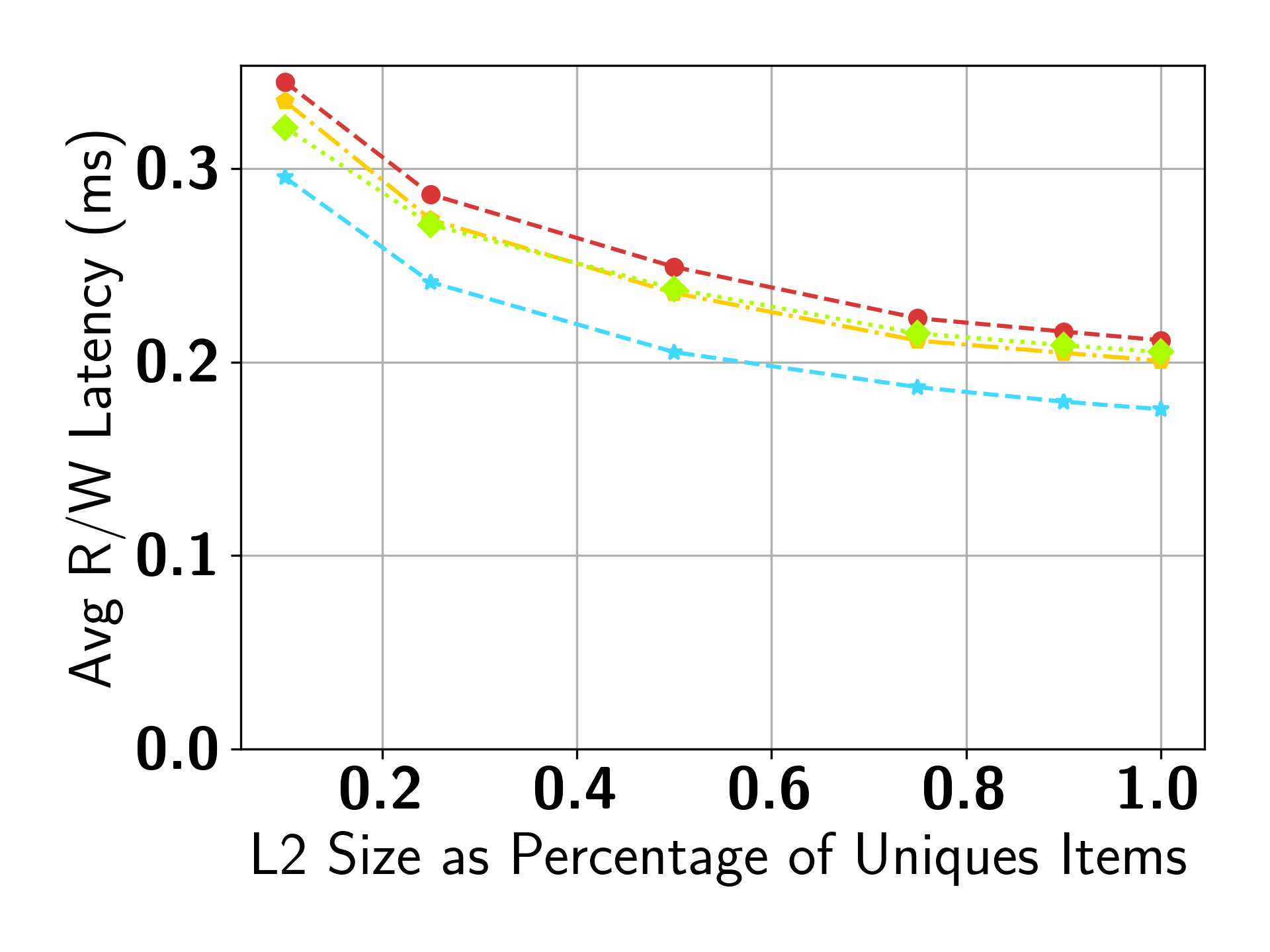}
	\end{tabular} \hspace{-0.5cm}
	}
	\subfloat[\normalfont{$L1:L2 = 1:100$ }]{ \begin{tabular}[b]{c}%
		\includegraphics[trim=86 0 0 10, clip, height=\Height]{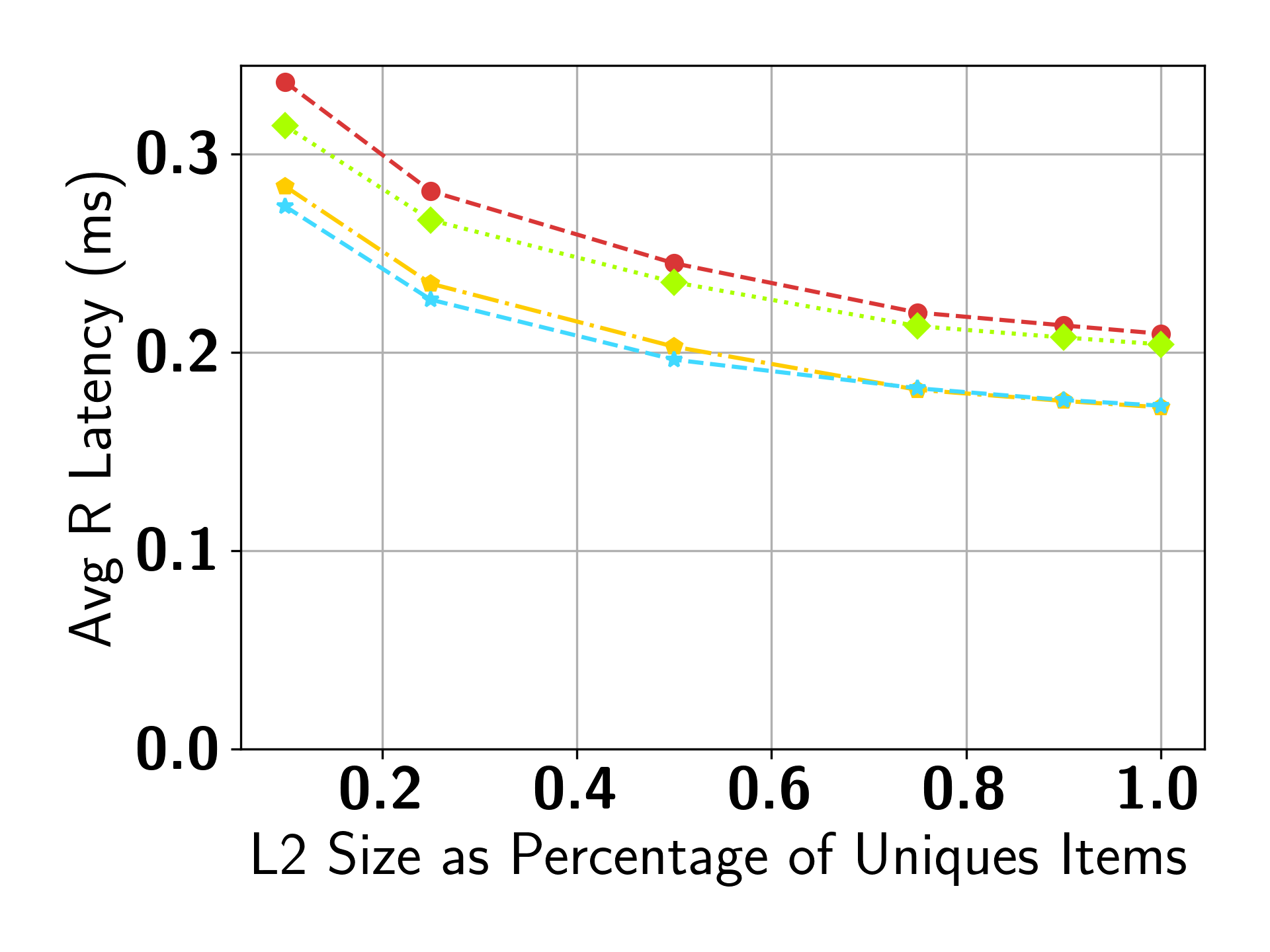} \\
		\includegraphics[trim=86 0 0 10, clip, height=\Height]{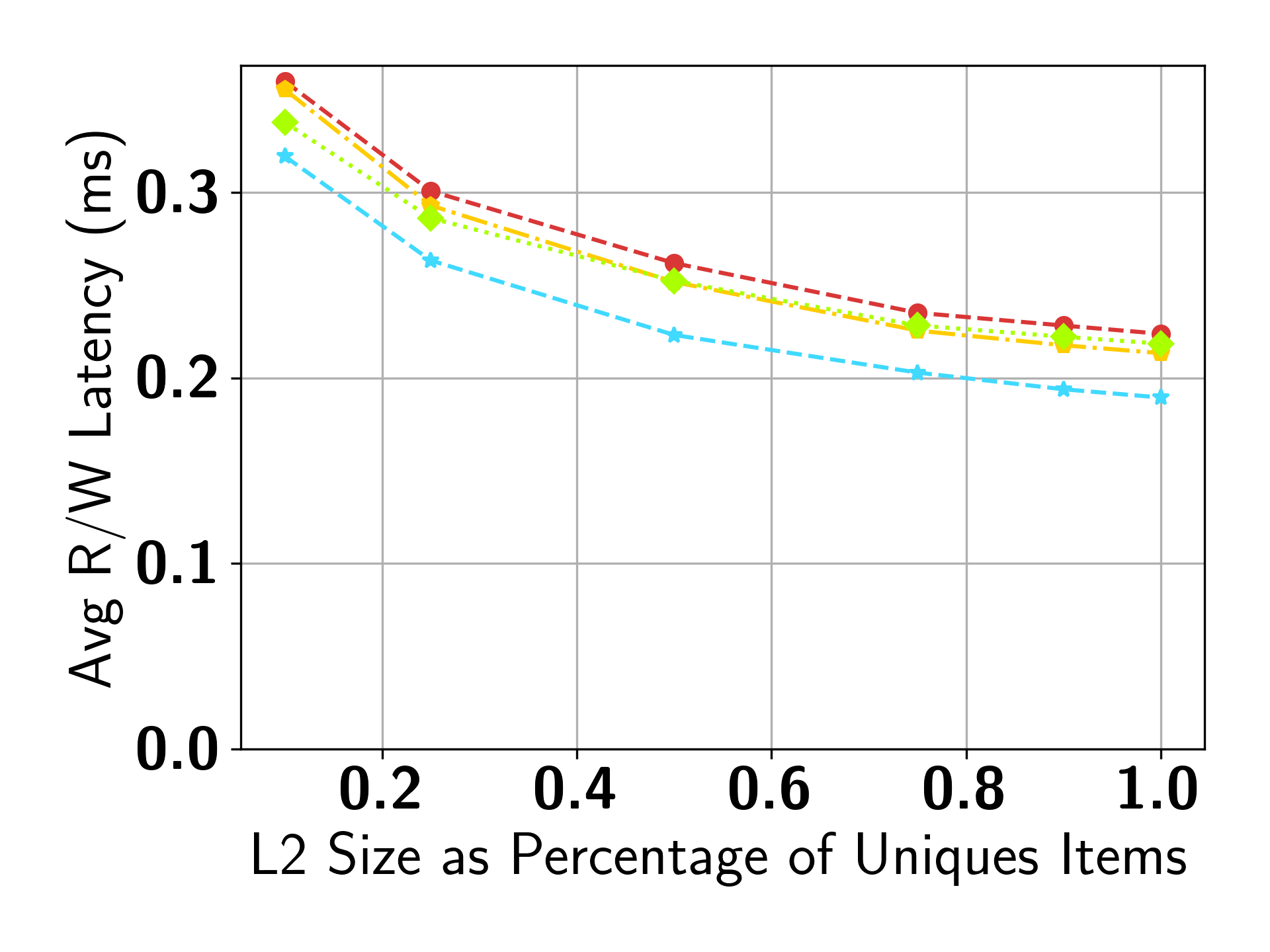}
	\end{tabular} \hspace{-0.5cm}
	}
	\\
	\vspace{0.2cm}
	\centering TENCENT1 \\ \small Hit-Ratio ranges from 18\%--59\% \\
	\subfloat[\normalfont{$L1:L2 = 1:10$ }]{ \begin{tabular}[b]{c}%
		\includegraphics[trim=0 0 0 10, clip, height=\Height]{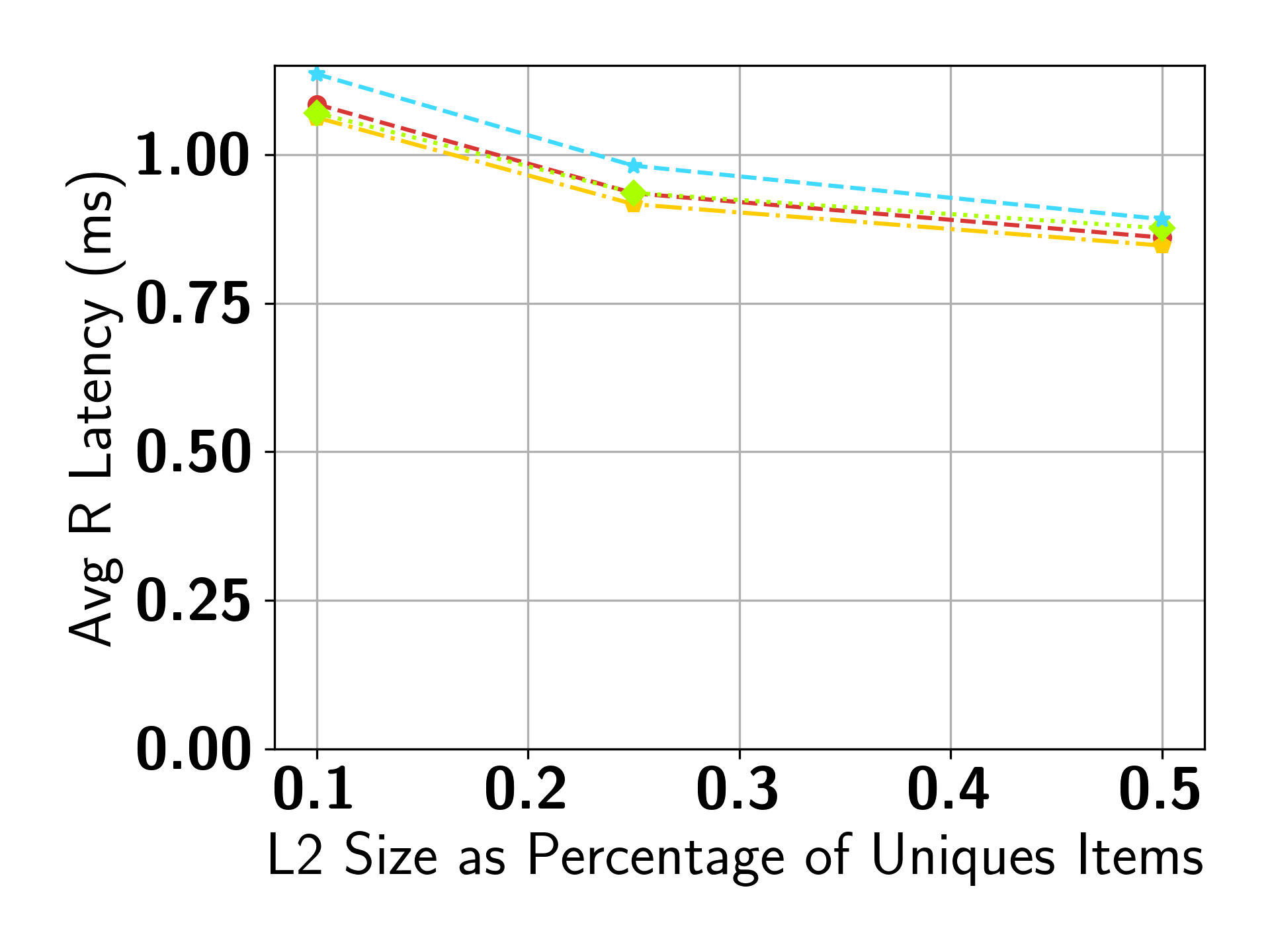} \\
		\includegraphics[trim=0 0 0 10, clip, height=\Height]{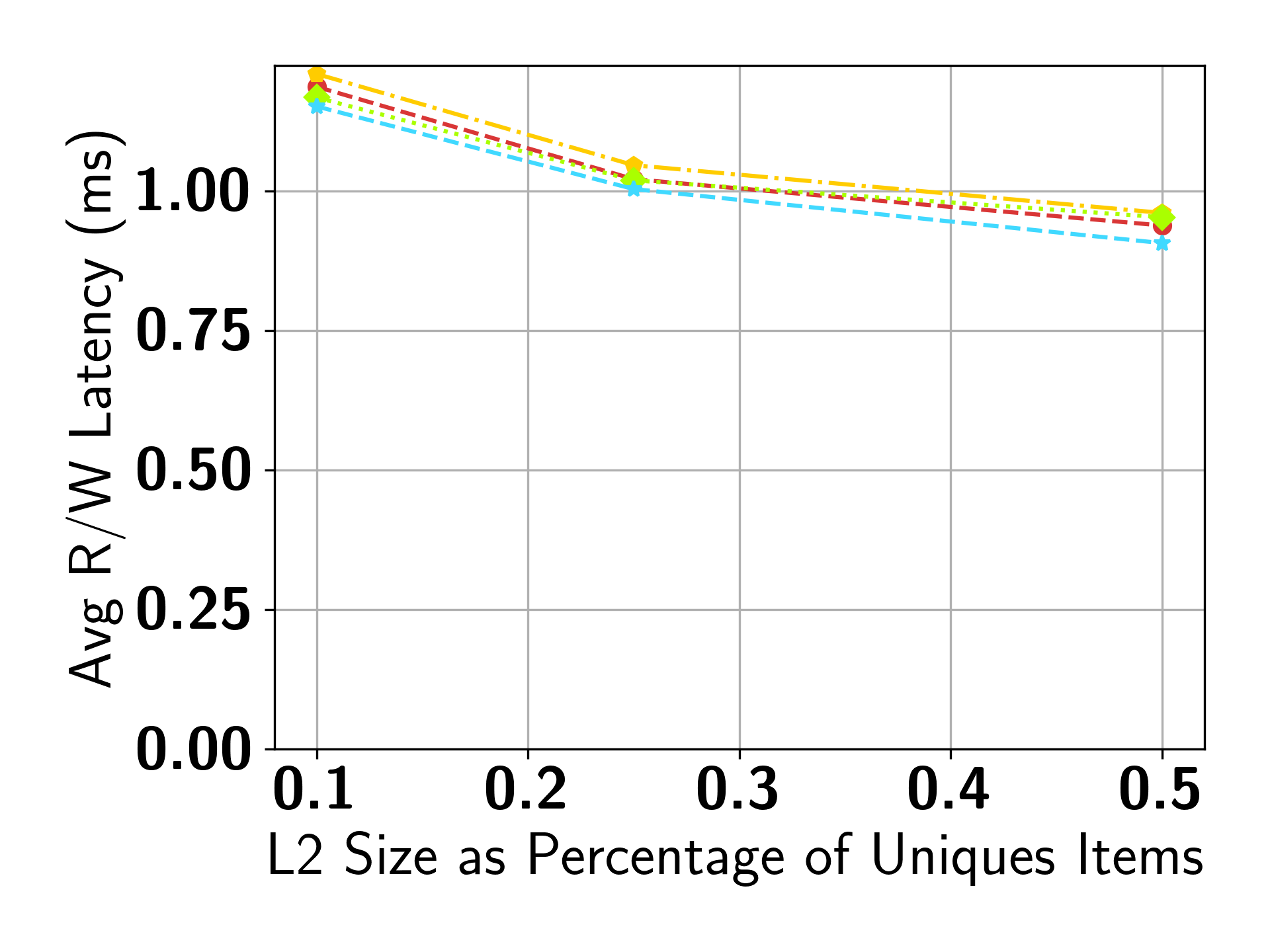}
	\end{tabular} \hspace{-0.5cm}
	}
	\subfloat[\normalfont{$L1:L2 = 1:20$ }]{ \begin{tabular}[b]{c}%
		\includegraphics[trim=91 0 0 10, clip, height=\Height]{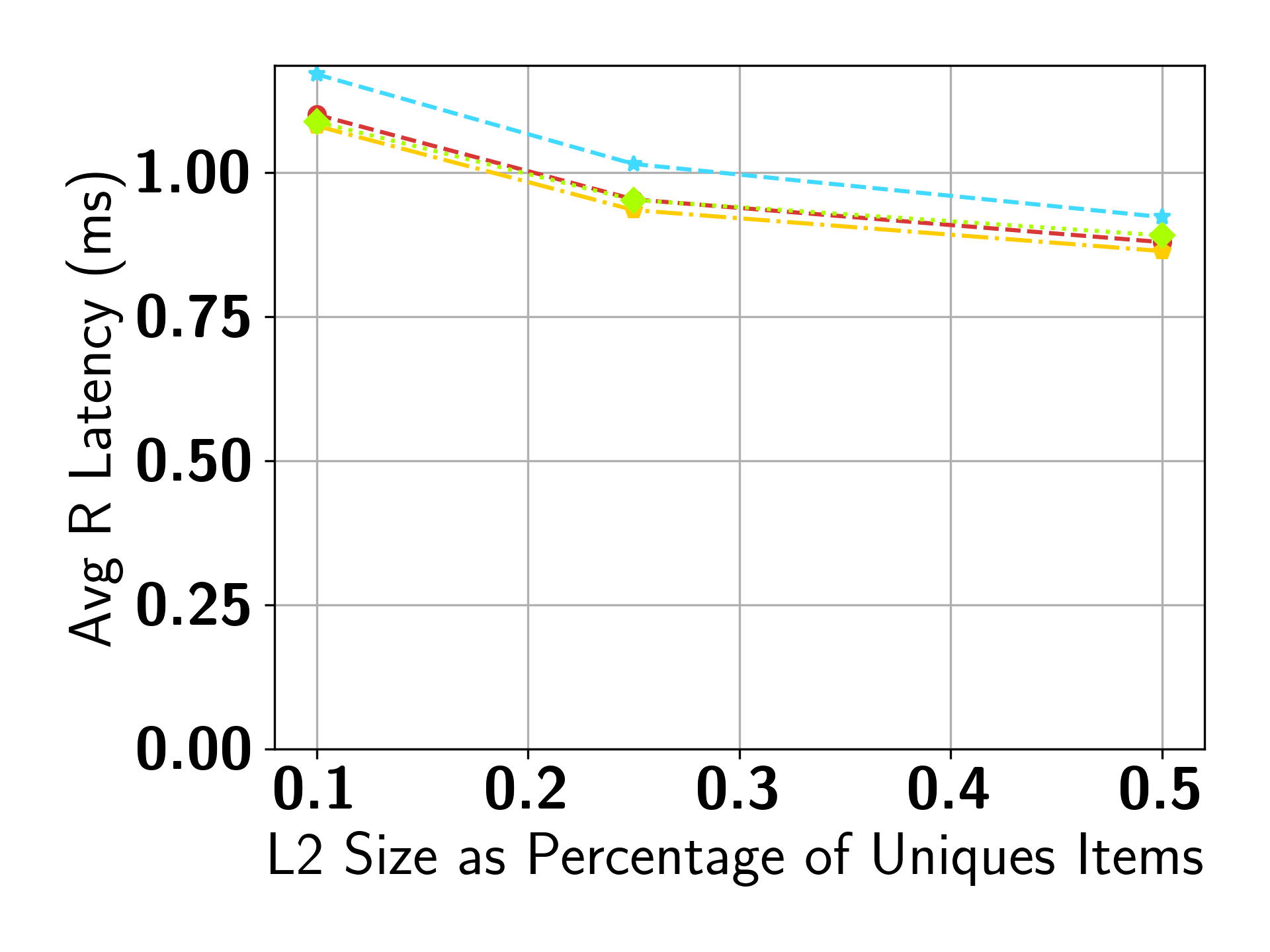} \\
		\includegraphics[trim=50 0 0 10, clip, height=\Height]{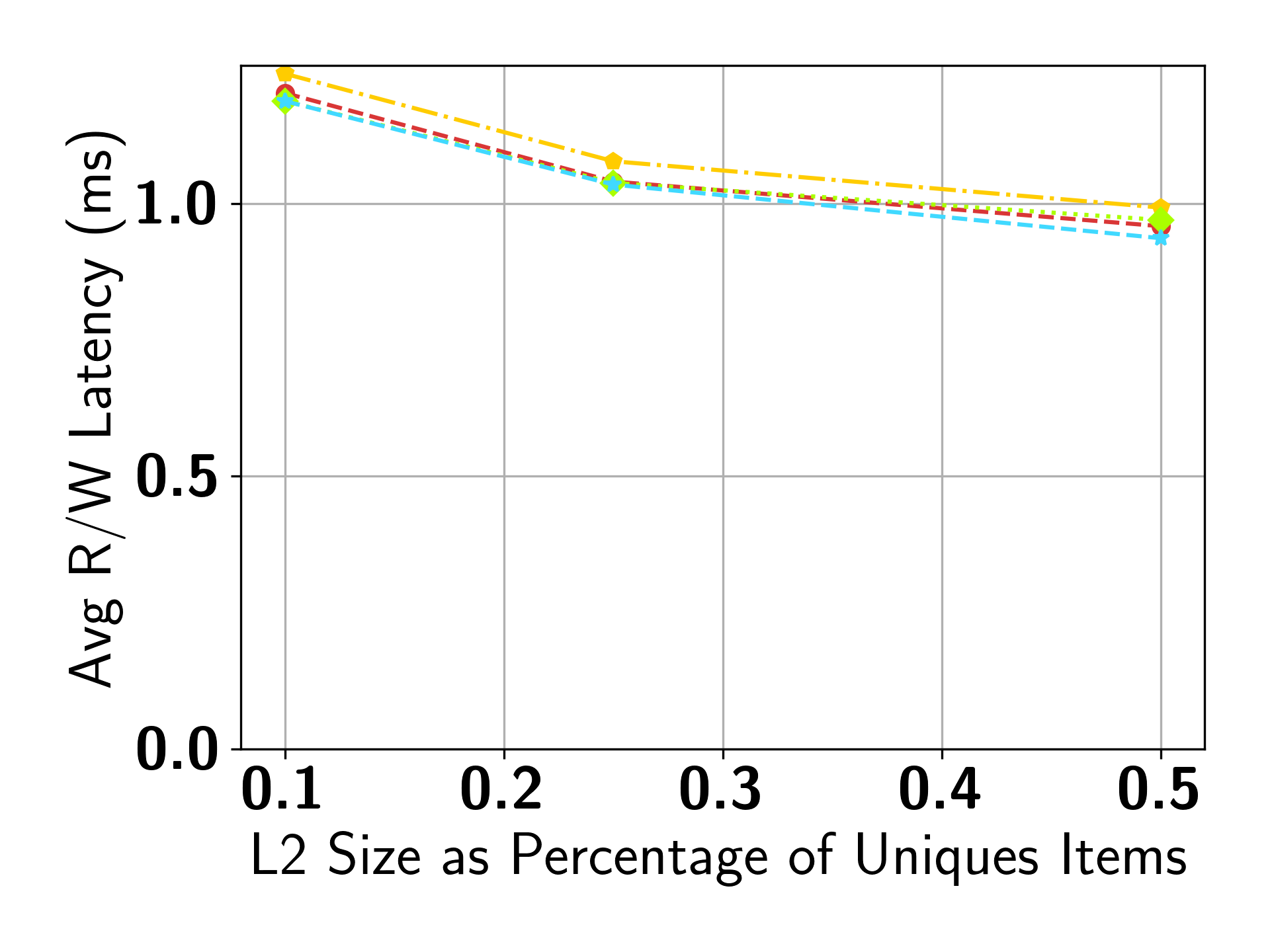}
	\end{tabular} \hspace{-0.5cm}
	}
	\subfloat[\normalfont{$L1:L2 = 1:50$ }]{ \begin{tabular}[b]{c}%
		\includegraphics[trim=91 0 0 10, clip, height=\Height]{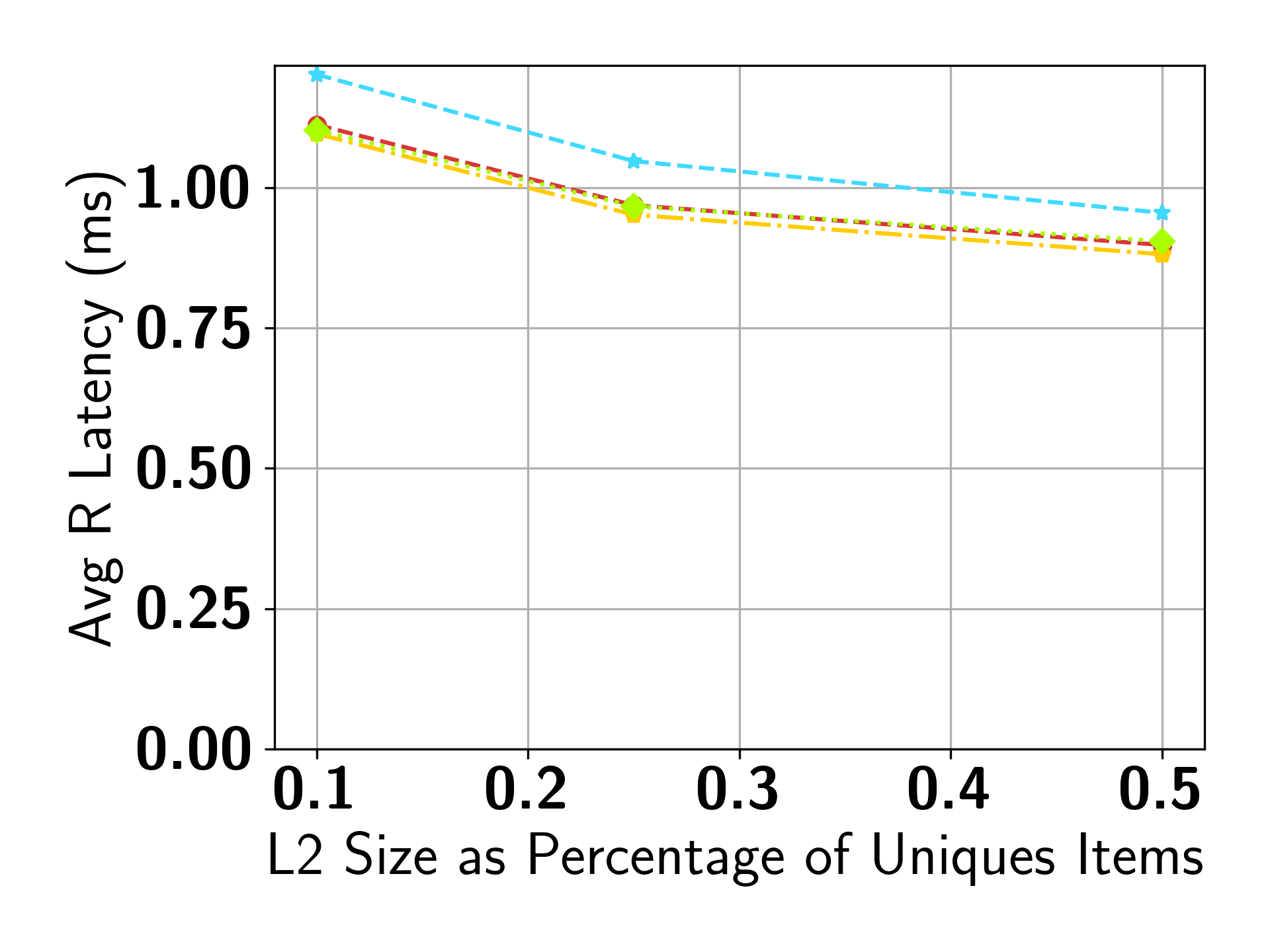} \\
		\includegraphics[trim=86 0 0 10, clip, height=\Height]{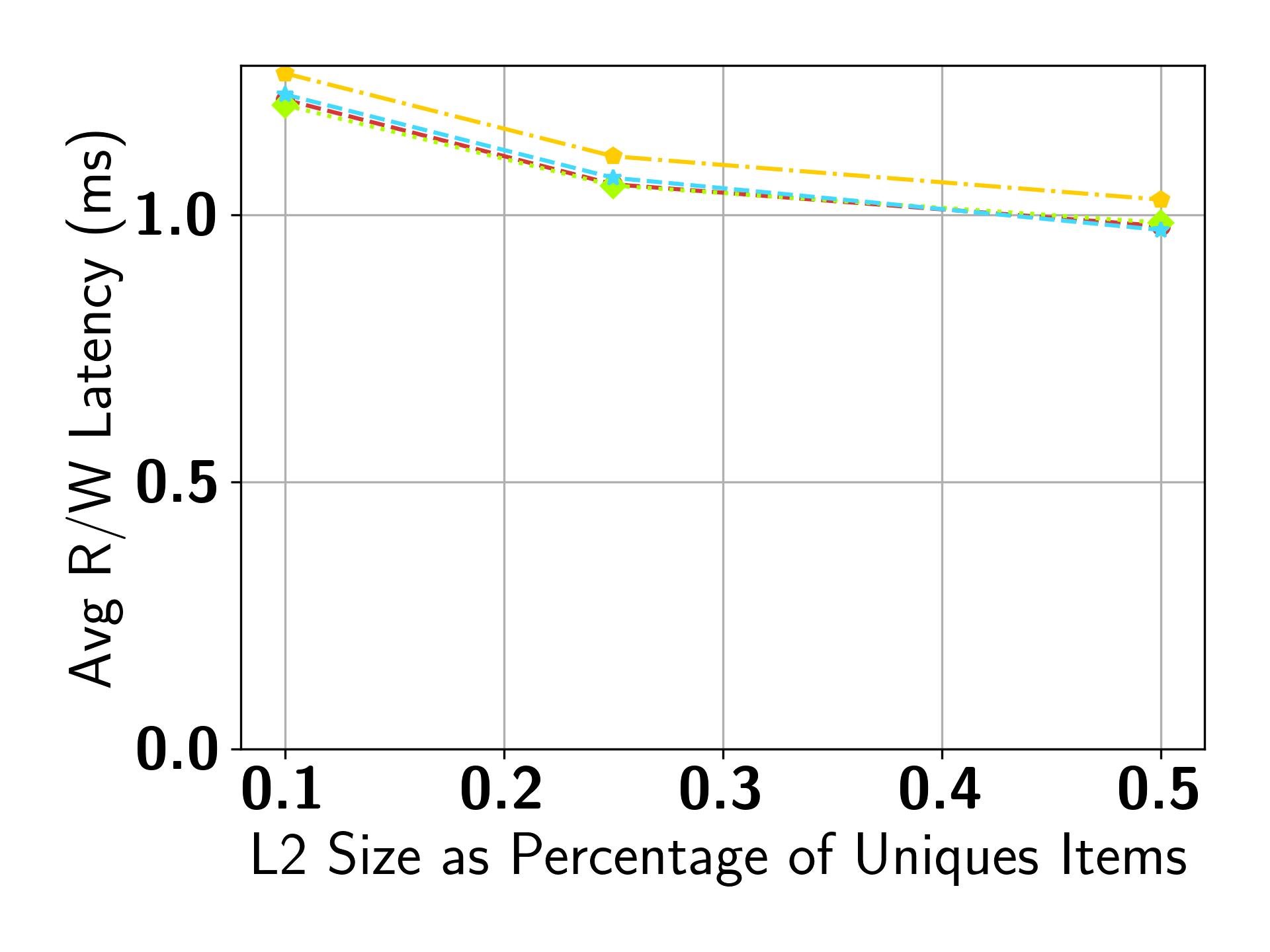}
	\end{tabular} \hspace{-0.5cm} 
	}
	\subfloat[\normalfont{$L1:L2 = 1:100$ }]{ \begin{tabular}[b]{c}%
		\includegraphics[trim=91 0 0 10, clip, height=\Height]{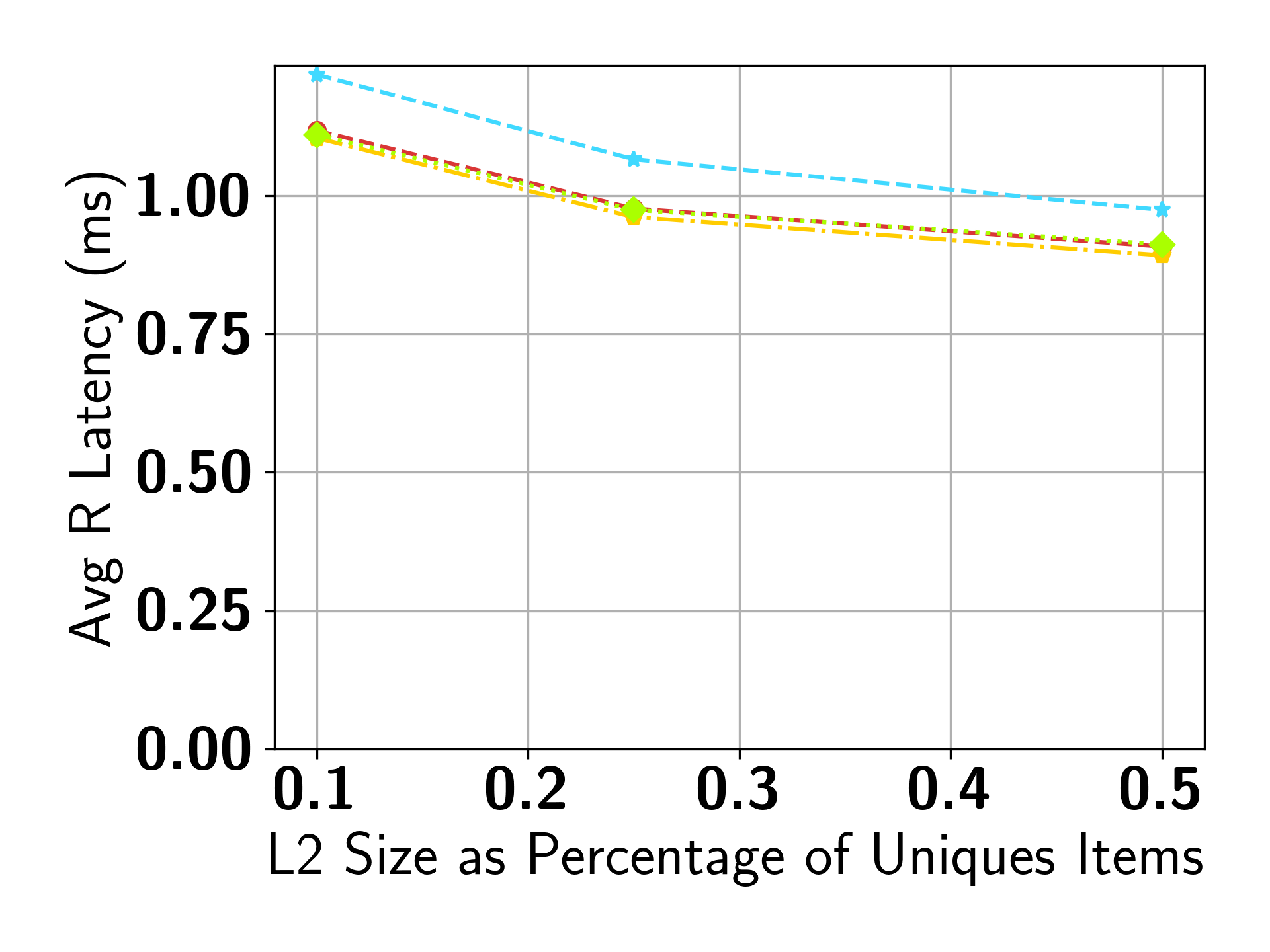} \\
		\includegraphics[trim=86 0 0 10, clip, height=\Height]{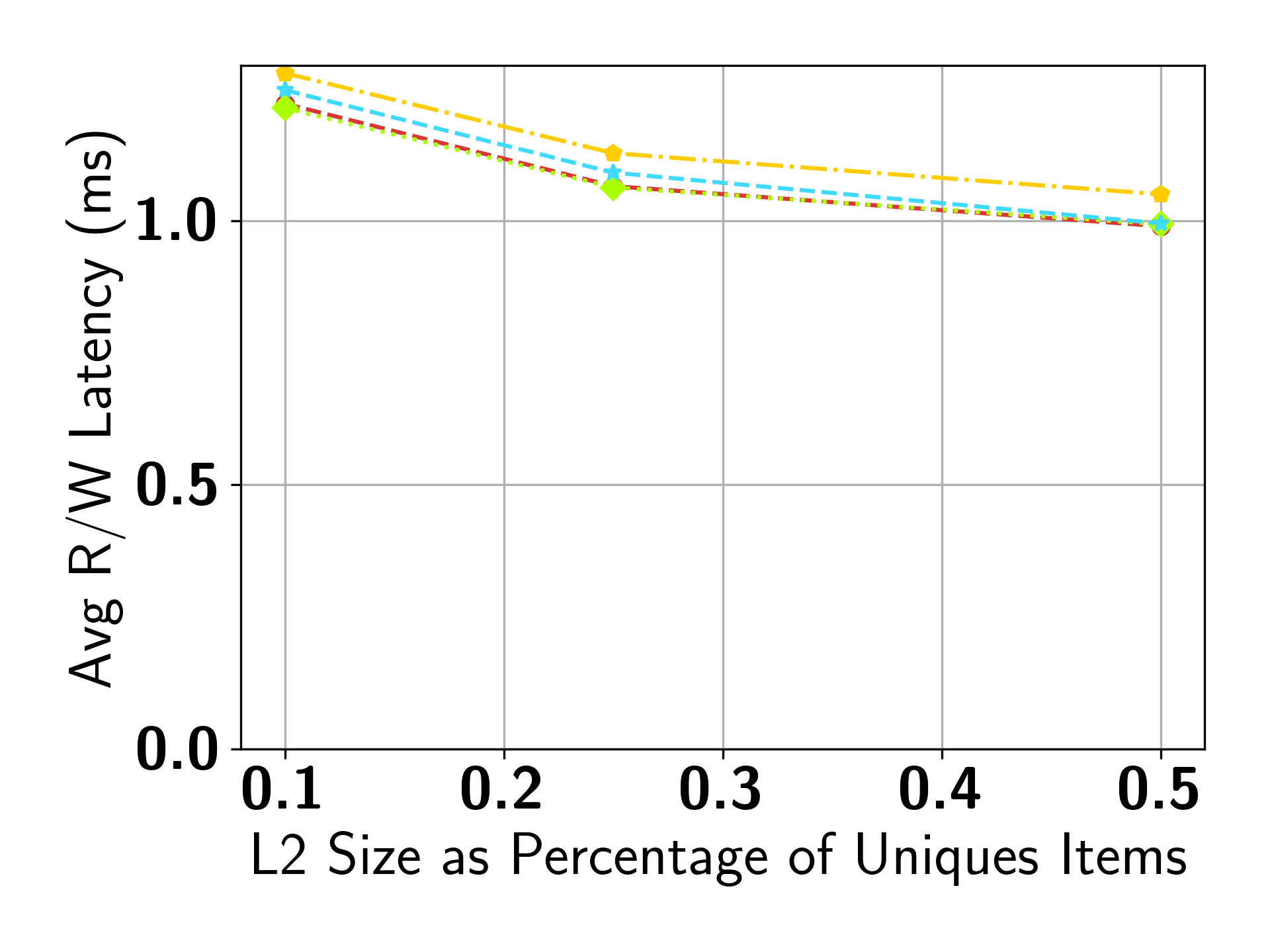}
	\end{tabular} \hspace{-0.5cm}
	}
	\\
	\centering \includegraphics[trim=0 0 100 10, clip, height=1.3cm]{Plots/others-legend-lines.png}
	\caption{Average latency per request, assuming latencies of 2ms, 200us and 100ns for disk, L2, and L1 respectively, for multiple traces and multiple ratios between L1 and L2}
\end{figure*}

%% file: discussion.tex
\section{Discussion}
\label{multi:sec:discussion}
This work explored the use of an aging frequency sketch as a bidirectional filter between levels in a hybrid cache environment. 
We saw a clear benefit in terms of write saving by at least one order of magnitude in almost all cases compared to competing approaches.
We also observed latency improvements on many workloads. 

Here we used LRU and SLRU to manage the internal cache levels. 
Using the same framework with a simpler internal cache, such as FIFO~\cite{FIFO}, or advanced ones like ARC~\cite{ARC}, is left for future work and could provide additional benefit. 
We also believe that combing improvements such an adaptivity mechanism~\cite{AdaptiveTinyLFU} and size-aware admission~\cite{size-aware} should improve the hit ratio and latency, especially where it currently~lags.

As described in~\cite{desperately}, comprehensive performance evaluation of multilevel caching is a complex and daunting task. 
We hope that new tools and methods will be developed and released in the coming future, and will provide further analysis and results, showing the potential benefits of using our~scheme.

